\documentclass[longauth]{aa} 

\usepackage{graphicx}
\usepackage{rotating}
\usepackage{longtable}
\usepackage{txfonts}
\begin{document}

   \title{Panchromatic evolution of three luminous red novae}

   \subtitle{Forbidden hugs in pandemic times -- IV}

   \author{A. Pastorello\inst{1}\fnmsep\thanks{andrea.pastorello@inaf.it}
              \and
          G.~Valerin\inst{1,2}
          \and
          M.~Fraser\inst{3}
           \and
          A.~Reguitti\inst{4,5,1}
          \and
          N.~Elias-Rosa\inst{1,6}
         \and
          A.~V.~Filippenko\inst{7,8}          
         \and
          C. Rojas-Bravo\inst{9}
         \and
          L.~Tartaglia\inst{1}
            \and       
          T.~M.~Reynolds\inst{10,11}
         \and
         S.~Valenti\inst{12}
         \and
         J.~E. Andrews\inst{13}
         \and
          C.~Ashall\inst{14}
             \and
          K.~A.~Bostroem\inst{15}
           \and
          T.~G.~Brink\inst{7}
            \and
         J.~Burke\inst{16,17}   
           \and
         Y.-Z.~Cai\inst{18,19,20}
         \and
         E.~Cappellaro\inst{1}
          \and
         D.~A.~Coulter\inst{9}
          \and
         R.~Dastidar\inst{4,5}   
          \and
         K.~W.~Davis\inst{9}
          \and  
         G.~Dimitriadis\inst{21}   
          \and
         A.~Fiore\inst{22,23}
          \and
         R.~J.~Foley\inst{9}
           \and 
          D.~Fugazza\inst{24}
          \and
          L.~Galbany\inst{6,25}
          \and
          A.~Gangopadhyay\inst{26,27}
          \and
          S.~Geier\inst{28,29}
         \and
          C.~P.~Guti\'errez\inst{30,10}
         \and
          J.~Haislip\inst{31}
         \and
         D.~Hiramatsu\inst{16,17,32,33}
         \and
         S.~Holmbo\inst{34}
         \and
         D.~A. Howell\inst{16,17}
         \and 
         E.~Y.~Hsiao\inst{35}
         \and
         T.~Hung\inst{9}
         \and 
          S.~W.~Jha\inst{36}
         \and 
         E.~Kankare\inst{10,37}
          \and
          E.~Karamehmetoglu\inst{34}
          \and
          C.~D. Kilpatrick\inst{38}
          \and 
          R.~Kotak\inst{10}
          \and
          V.~Kouprianov\inst{31}
          \and
          T.~Kravtsov\inst{10}
          \and
          S.~Kumar\inst{35}
          \and
         Z.-T.~Li\inst{39,40}
         \and
         M.~J.~Lundquist\inst{41}
         \and
         P.~Lundqvist\inst{42}
          \and
         K.~Matilainen\inst{10} 
          \and
          P.~A.~Mazzali\inst{43,44}
         \and
         C.~McCully\inst{16}
         \and
          K.~Misra\inst{26}
          \and
           A.~Morales-Garoffolo\inst{45}
          \and
          S.~Moran\inst{10}
           \and
          N.~Morrell\inst{46}
         \and
         M.~Newsome\inst{16,17}
         \and
         E.~Padilla~Gonzalez\inst{16,17}
         \and
         Y.-C.~Pan\inst{47}
         \and
         C.~Pellegrino\inst{16,17}
           \and 
          M.~M.~Phillips\inst{46}
           \and
          G.~Pignata\inst{4,5}
          \and
          A.~L.~Piro\inst{48}
          \and
          D.~E.~Reichart\inst{31}
          \and
          A.~Rest\inst{49,50}
           \and
          I.~Salmaso\inst{1,2} 
           \and  
          D.~J.~Sand\inst{51}
           \and 
          M.~R.~Siebert\inst{9}
          \and
          S.~J.~Smartt\inst{52}
           \and
          K.~W.~Smith\inst{52}
         \and
          S.~Srivastav\inst{52}
          \and     
          M.~D.~Stritzinger\inst{34}
          \and
          K.~Taggart\inst{9}
          \and
          S.~Tinyanont\inst{9}
          \and
          S.-Y.~Yan\inst{20}
          \and
           L.~Wang\inst{53}
          \and
          X.-F.~Wang\inst{20,54} 
          \and
          S.~C.~Williams\inst{30,10}
          \and
          S.~Wyatt\inst{51}  
          \and
          T.-M.~Zhang\inst{39,40}
          \and
          T.~de~Boer\inst{55}
          \and
          K.~Chambers\inst{55}
          \and
          H.~Gao\inst{55} 
          \and
          E.~Magnier\inst{55}  
}

   \institute{INAF - Osservatorio Astronomico di Padova, Vicolo dell'Osservatorio 5, I-35122 Padova, Italy\\ \email{andrea.pastorello@inaf.it}
         \and
    Universit\'a degli Studi di Padova, Dipartimento di Fisica e Astronomia, Vicolo dell’Osservatorio 2, 35122 Padova, Italy 
          \and
             School of Physics, O’Brien Centre for Science North, University College Dublin, Belfield, Dublin 4, Ireland        
          \and
           Instituto de astrofísica, Facultad de Ciencias Exactas, Universidad Andres Bello, Fern\'andez Concha 700, Las Condes, Santiago, Chile
          \and
             Millennium Institute of Astrophysics (MAS), Nuncio Monsenor S\'otero Sanz 100, Providencia, Santiago, 8320000, Chile
             \and
             Institute of Space Sciences (ICE, CSIC), Campus UAB, Carrer de Can Magrans s/n, E-08193, Barcelona, Spain
            \and
          Department of Astronomy, University of California, Berkeley, CA 94720-3411, USA
         \and
         Miller Institute for Basic Research in Science, University of California, Berkeley, CA 94720, USA
          \and
             Department of Astronomy and Astrophysics, University of California, Santa Cruz, CA 95064, USA
          \and
             Tuorla Observatory, Department of Physics and Astronomy, University of Turku, FI-20014 Turku, Finland
        \and
            Cosmic Dawn Center (DAWN), Niels Bohr Institute, University of Copenhagen, Jagtvej 128, 2200 K\o{}benhavn N, Denmark
        \and
           Department of Physics and Astronomy, University of California, Davis, 1 Shields Avenue, Davis, CA 95616-5270, USA
        \and
          Gemini Observatory, 670 North A'ohoku Place, Hilo, HI 96720-2700, USA 
         \and
          Department of Physics, Virginia Tech, 850 West Campus Drive, Blacksburg, VA 24061, USA
         \and
          DIRAC Institute, Department of Astronomy, University of Washington, 3910 15th Avenue NE, Seattle, WA 98195-0002, USA   
            \and
          Las Cumbres Observatory, 6740 Cortona Dr. Suite 102, Goleta, CA 93117, USA
            \and
          Department of Physics, University of California, Santa Barbara, Santa Barbara, CA 93106, USA
            \and
          Yunnan Observatories, Chinese Academy of Sciences, Kunming 650216, China
            \and
          Key Laboratory for the Structure and Evolution of Celestial Objects, Chinese Academy of Sciences, Kunming 650216, China
            \and
             Physics Department and Tsinghua Center for Astrophysics (THCA), Tsinghua University, Beijing 100084, China
         \and
          School of Physics, Trinity College Dublin, The University of Dublin, Dublin 2, Ireland
         \and
         European Centre for Theoretical Studies in Nuclear Physics and Related Areas (ECT$^\star$), Fondazione Bruno Kessler, I-38123, Trento, Italy
         \and
        INFN-TIFPA, Trento Institute for Fundamental Physics and Applications, Via Sommarive 14, I-38123 Trento, Italy
         \and
        INAF – Osservatorio Astronomico di Brera, via E. Bianchi 46 I-23807, Merate, Italy
         \and
         Institut d’Estudis Espacials de Catalunya (IEEC), E-08034, Barcelona, Spain
         \and
         Aryabhatta Research Institute of observational sciencES, Manora Peak, Nainital 263 002, India
         \and
         Hiroshima Astrophysical Science Centre, Hiroshima University, 1-3-1 Kagamiyama, Higashi-Hiroshima, Hiroshima 739-8526, Japan
         \and
         Gran Telescopio Canarias (GRANTECAN), Cuesta de San Jos\'e s/n, 38712 Bre\~na Baja, La Palma, Spain
         \and
        Instituto de Astrof\'isica de Canarias, V\'ia L\'actea s/n, 38200 La Laguna, Tenerife, Spain 
          \and
        Finnish Centre for Astronomy with ESO (FINCA), University of Turku, FI-20014 Turku, Finland
            \and
         Department of Physics and Astronomy, University of North Carolina, 120 East Cameron Avenue, Chapel Hill, NC 27599, USA
         \and
         Center for Astrophysics \textbar{} Harvard \& Smithsonian, 60 Garden Street, Cambridge, MA 02138-1516, USA
         \and
          The NSF AI Institute for Artificial Intelligence and Fundamental Interactions
         \and
             Department of Physics and Astronomy, Aarhus University, Ny Munkegade 120, 8000 Aarhus C, Denmark
            \and
         Department of Physics, Florida State University, 77 Chieftan Way, Tallahassee, FL 32306, USA
         \and
        Department of Physics and Astronomy, Rutgers, The State University of New Jersey, 136 Frelinghuysen Road, Piscataway, NJ 08854, USA
          \and
         Turku Collegium for Science, Medicine and Technology, University of Turku, FI-20014 Turku, Finland
          \and
         Center for Interdisciplinary Exploration and Research in Astrophysics (CIERA), Northwestern University, Evanston, IL 60208, USA
          \and
         Key Laboratory of Optical Astronomy, National Astronomical Observatories, Chinese Academy of Sciences, Beijing 100101, China
          \and 
         School of Astronomy and Space Science, University of Chinese Academy of Sciences, Beijing 101408, China
          \and
          W.~M.~Keck Observatory, 65-1120 Ma-malahoa Highway, Kamuela, HI 96743-8431, USA
          \and 
             The Oskar Klein Centre, Department of Astronomy, Stockholm University, AlbaNova, SE-10691 Stockholm, Sweden
          \and
             Astrophysics Research Institute, Liverpool John Moores University, ic2, 146 Brownlow Hill, Liverpool L3 5RF, UK
          \and
             Max-Planck Institut f\"ur Astrophysik, Karl-Schwarzschild-Str. 1, D-85741 Garching, Germany
          \and
            Department of Applied Physics, University of C\'adiz, Campus of Puerto Real, 11510 C\'adiz, Spain
          \and
         Las Campanas Observatory, Carnegie Observatories, Casilla 601, La Serena, Chile
          \and
          Graduate Institute of Astronomy, National Central University, 300 Zhongda Road, Zhongli, Taoyuan 32001, Taiwan
          \and
         The Observatories of the Carnegie Institution for Science, 813 Santa Barbara St., Pasadena, CA 91101, USA
         \and
         Department of Physics and Astronomy, Johns Hopkins University, 3400 North Charles Street, Baltimore, MD 21218, USA 
         \and
         Space Telescope Science Institute, 3700 San Martin Drive, Baltimore, MD 21218, USA
         \and
         Steward Observatory, University of Arizona, 933 North Cherry Avenue, Tucson, AZ 85721-0065, USA
         \and
             Astrophysics Research Centre, School of Mathematics and Physics, Queen’s University Belfast, BT7 1NN, UK
         \and
          Chinese Academy of Sciences, South America Center for Astronomy, National Astronomical Observatories, CAS, Beijing 100101, China
         \and
            Beijing Planetarium, Beijing Academy of Science and Technology, Beijing 100044, China
          \and
          Institute for Astronomy, University of Hawaii, 2680 Woodlawn Drive, Honolulu, HI 96822, USA\\
             }

   \date{Received August 03, 2022; accepted Month dd, 202X}

  \abstract
   {We present photometric and spectroscopic data on three extragalactic luminous red novae (LRNe): \object{AT~2018bwo}, \object{AT~2021afy}, and \object{AT~2021blu}. 
\object{AT~2018bwo} was discovered in \object{NGC~45} (at about 6.8~Mpc) a few weeks after the outburst onset. During 
the monitoring period, the transient reached a peak luminosity of $10^{40}$~erg~s$^{-1}$. \object{AT~2021afy}, hosted by \object{UGC~10043} ($\sim 49.2$~Mpc), 
showed a double-peaked light curve, with the two peaks reaching a similar luminosity of $2.1 (\pm 0.6) \times 10^{41}$~erg~s$^{-1}$. Finally, for \object{AT~2021blu} in \object{UGC~5829} 
($\sim 8.6$~Mpc), the pre-outburst phase was well-monitored by several photometric surveys, and the object showed a slow luminosity rise before the outburst. The light curve of \object{AT~2021blu} was 
sampled with an unprecedented cadence until the object disappeared behind the Sun, and it was then recovered at late phases. The light curve of LRN \object{AT~2021blu} shows a double peak, 
with a prominent early maximum reaching a luminosity of $6.5 \times 10^{40}$~erg~s$^{-1}$, which is half of that of \object{AT~2021afy}. 
The spectra of \object{AT~2021afy} and \object{AT~2021blu} display the expected evolution for LRNe: a blue continuum dominated by 
prominent Balmer lines in emission during the first peak, and a redder continuum consistent with that of a K-type star with narrow absorption metal lines during the second, 
broad maximum. The spectra of \object{AT~2018bwo} are markedly different, with a very red continuum dominated by broad molecular features in absorption. As these spectra closely 
resemble those of LRNe after the second peak, \object{AT~2018bwo} was probably discovered at the very late evolutionary stages. This would explain its fast evolution and 
the spectral properties compatible with that of an M-type star. From the analysis of deep frames of the LRN sites years before the outburst, and considerations of the
light curves, the quiescent progenitor systems of the three LRNe were likely massive, with primaries ranging from about 13~M$_\odot$ for \object{AT~2018bwo}, to $14_{-1}^{+4}$~M$_\odot$ 
for \object{AT~2021blu}, and over 40~M$_\odot$ for \object{AT~2021afy}.}

   \maketitle

\section{Introduction} \label{sect:intro}

Luminous red novae (LRNe) are optical transients that are thought to result from a close binary interaction leading 
to the ejection of a common envelope, eventually followed by the coalescence of the stellar cores \citep[e.g.][]{iva17}. LRNe 
span an enormous range of luminosities, but they have a surprisingly similar spectral evolution.
About five orders of magnitude separate the peak luminosity of the faintest Galactic objects such as \object{OGLE~2002-BLG-360} \citep{tyl13} 
and \object{V1309~Sco} \citep{mas10,tyl11} from bright extragalactic events \citep[$0 \gtrsim M_V \gtrsim -15$~mag; see, e.g. the sample presented by][]{pasto19a}. 
The latter objects, with intermediate luminosity between those of classical novae and core-collapse supernovae, are  'gap transients' \citep[][]{kas12,pasto19,cai22a}. 
LRNe have structured light curves, with a phase of slowly rising luminosity lasting months to years, followed  
by a major outburst. The outburst is usually characterised by a short-duration early peak, during which the object has a blue colour, followed by a plateau or a second 
broad peak with a redder colour. While a LRN during the slow pre-outburst brightening has not been spectroscopically monitored yet, spectra during the early peak show a 
blue continuum with prominent lines in emission of the Balmer series, similar to those of other gap transients. At later phases (during the plateau or the second peak), the spectral continuum of LRNe 
becomes progressively redder, the Balmer lines become weaker, and many narrow absorption lines of metals appear. At very late phases, the optical
spectrum resembles that of intermediate to late M-type stars, dominated by prominent absorption bands of TiO and VO \citep{mar99,kim06,bar07,bar14,tyl15}.

While not all of the physical processes leading to LRN outbursts have been fully understood \citep{iva13a,iva13b}, significant progress has been made
 in the last decade from the observational side. In particular, follow-up observations of \object{V1309~Sco} revealed the signature of unstable 
mass transfer in a binary system when the primary filled its Roche lobe. 
The process may led to the ejection of a common envelope and the loss of angular momentum of the binary \citep{pej16a,pej17,mcl18,mcl20}.

Regardless of the mass, a binary stellar system after the ejection of the common envelope can 
either evolve to a new stable and closer binary configuration \citep[][and references therein]{jon20}, or to a final coalescence \citep{tyl06}, 
as is what happened for \object{V1309~Sco} \citep{tyl11}.
The merging event is the most popular scenario to explain the LRN observables \citep[e.g.][]{kam21}.
The double-peaked light curves and most of the observational properties of LRN outbursts are fairly well explained by gas outflow following 
the coalescence of the two stellar cores \citep[e.g.][for V838~Mon]{tyl06}, and subsequent shock interaction with the outer envelope. 
This scenario has been successfully modelled by a number of authors \citep{sha10,nan14,met17,mcl17}.
However, the merger scenario for LRNe has been occasionally challenged by late-time observations of the remnant \citep[e.g. in the case of V838~Mon;][]{gor20}.
Regardless of the fate of the system (merger or survived binary), the mass accretion onto an equatorial disk may power polar jets colliding 
with the slow-moving envelope,  which may account for the properties of LRNe \citep{kas16,sok16a,sok16b,sok20,sok21}.

From an observational point of view,  most LRNe display an early blue peak in the light curve resulting from
the outflow of hot material ejected in the merging process. However, in some LRNe the initial blue colour is not detected \citep[e.g. in \object{AT~2015dl} 
and \object{AT~2020hat};][]{bla17,pasto21a}. This can be due to an observational bias, as the early blue peak is a short-duration event. 
Alternatively, the lack of an initial blue phase can be due to an expanded, red giant (or supergiant) primary star. 

In this paper, we report extensive datasets for three LRNe. First, we present new data for AT~2018bwo which complement those 
released by \citet{bla21}.  \object{AT~2018bwo} is an object whose observations do not show evidence of an early blue phase, but its 
explosion epoch is poorly constrained. Furthermore, 
we present optical and near-infrared (NIR) data for two LRNe discovered in 2021: \object{AT~2021afy} and \object{AT~2021blu}. 
In the case of the latter, Sloan $g$- and $r$-band light curves were also presented by \citet{sora22}.
In contrast to the monitoring campaigns of other LRNe in our programme \citep{pasto21a,pasto21b}, the follow-up campaigns of these 
two objects were not significantly affected by the COVID-19 pandemic restrictions as to our access to observational facilities.
This study is complemented by a companion paper on another LRN monitored during the same period,
AT~2021biy \citep{cai22}.

We provide the basic information for the three transients and their host galaxies in  Sect. \ref{Sect:hosts}. The
photometric data are presented in Sect. \ref{sect:photo}; the evolution of their physical parameters 
(bolometric luminosity, photospheric radius, and temperature) is illustrated in Sect. \ref{sect:TLR}; and their spectral evolution 
is described in Sect. \ref{sect:spec}. The nature of the progenitors, the mechanisms producing LRN outbursts,
and the updated version of the correlations presented by \citet{pasto19a} and  \citet{pasto21a} are discussed in Sect. \ref{Sect:discussion}.
A brief summary follows in Sect. \ref{Sect:conclusions}.

\section{\object{AT~2018bwo},  \object{AT~2021afy},  \object{AT~2021blu}, and their host galaxies} \label{Sect:hosts}

\object{AT~2018bwo}\footnote{The object is also known as DLT18x, ATLAS18qgb, and Gaia18blv.} was discovered by the DLT40 survey \citep{tar18}
on 2018 May 22.93 \citep[UT dates are used throughout this paper;][]{val18}.\footnote{As mentioned by \citet{bla21}, the discovery unfiltered magnitude reported by \protect\citet{val18}, 
16.44 (AB mag scale), is incorrect; see Sect. 
\ref{sect:2018bwo_lc}.} Its coordinates are $\alpha = 00^h 14^m 01\fs69$ and $\delta = -23\degr11\arcmin35\farcs21$ (J2000). \citet{cla18} noted the similarity
with the spectrum of an F-type star and, also taking into account the faint absolute magnitude of the object, proposed an LRN classification for \object{AT~2018bwo}.

\begin{figure}
   \centering
   {\includegraphics[angle=0,width=8.3cm]{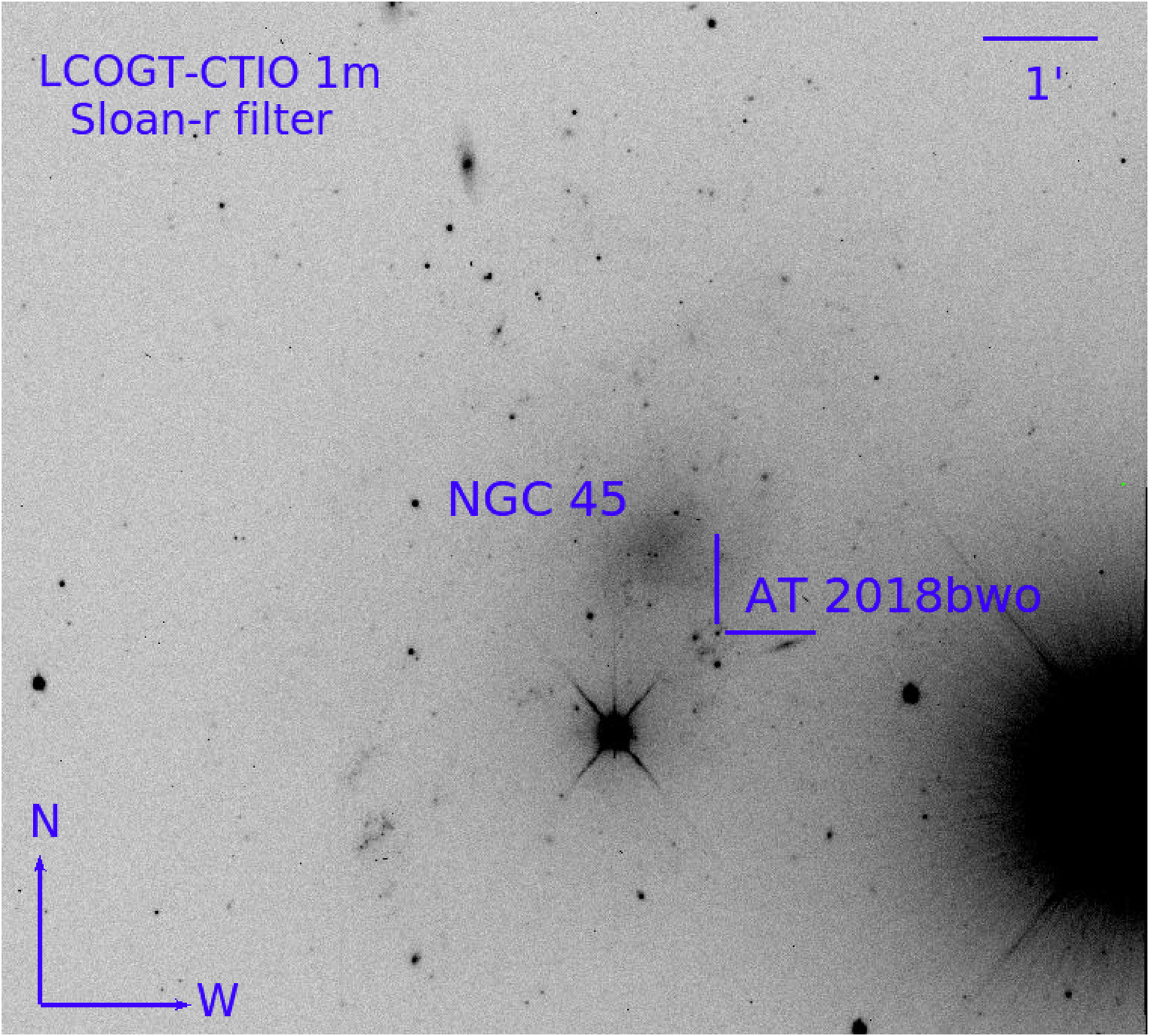}
\includegraphics[angle=0,width=8.35cm]{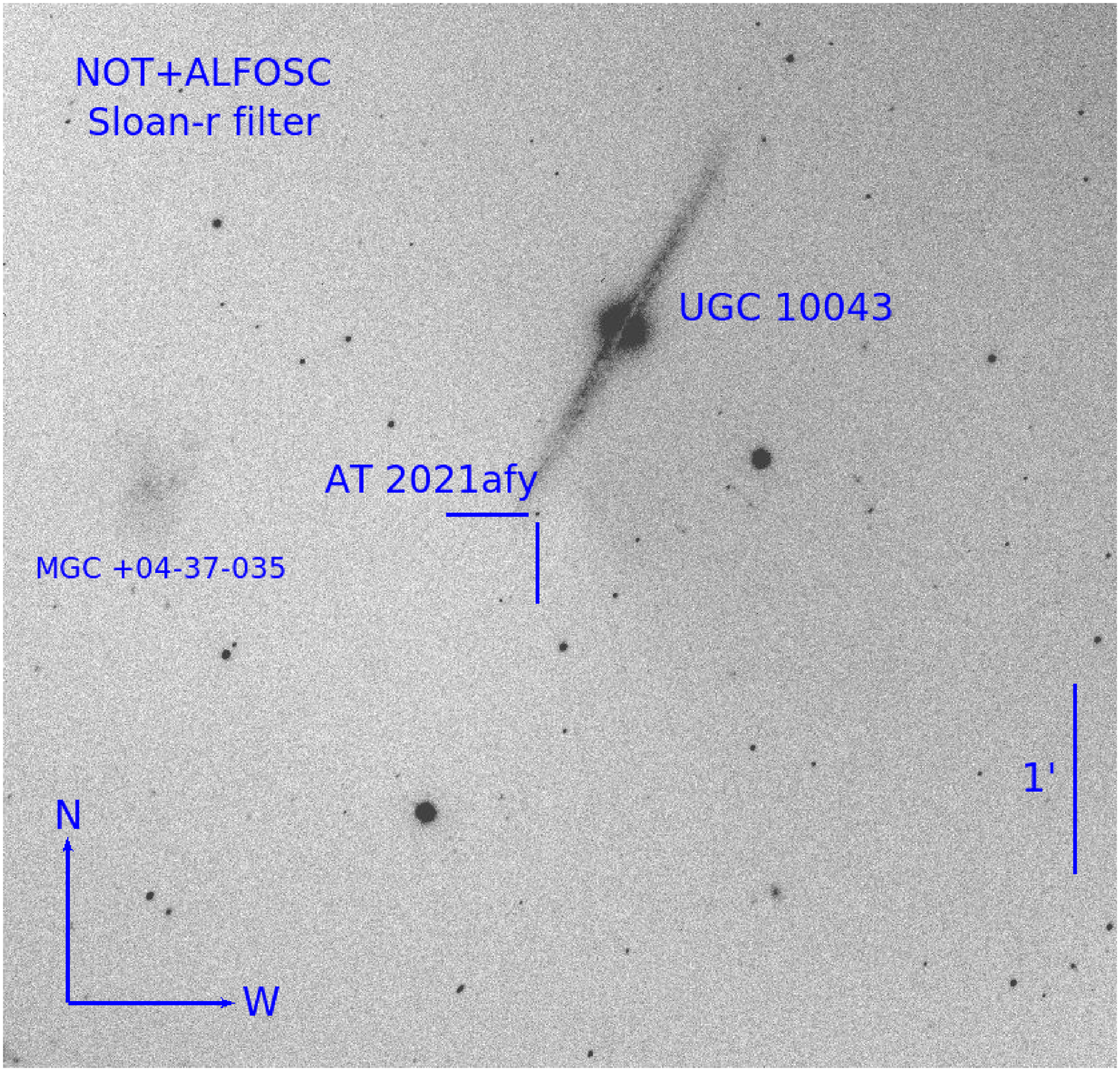}
\includegraphics[angle=0,width=8.35cm]{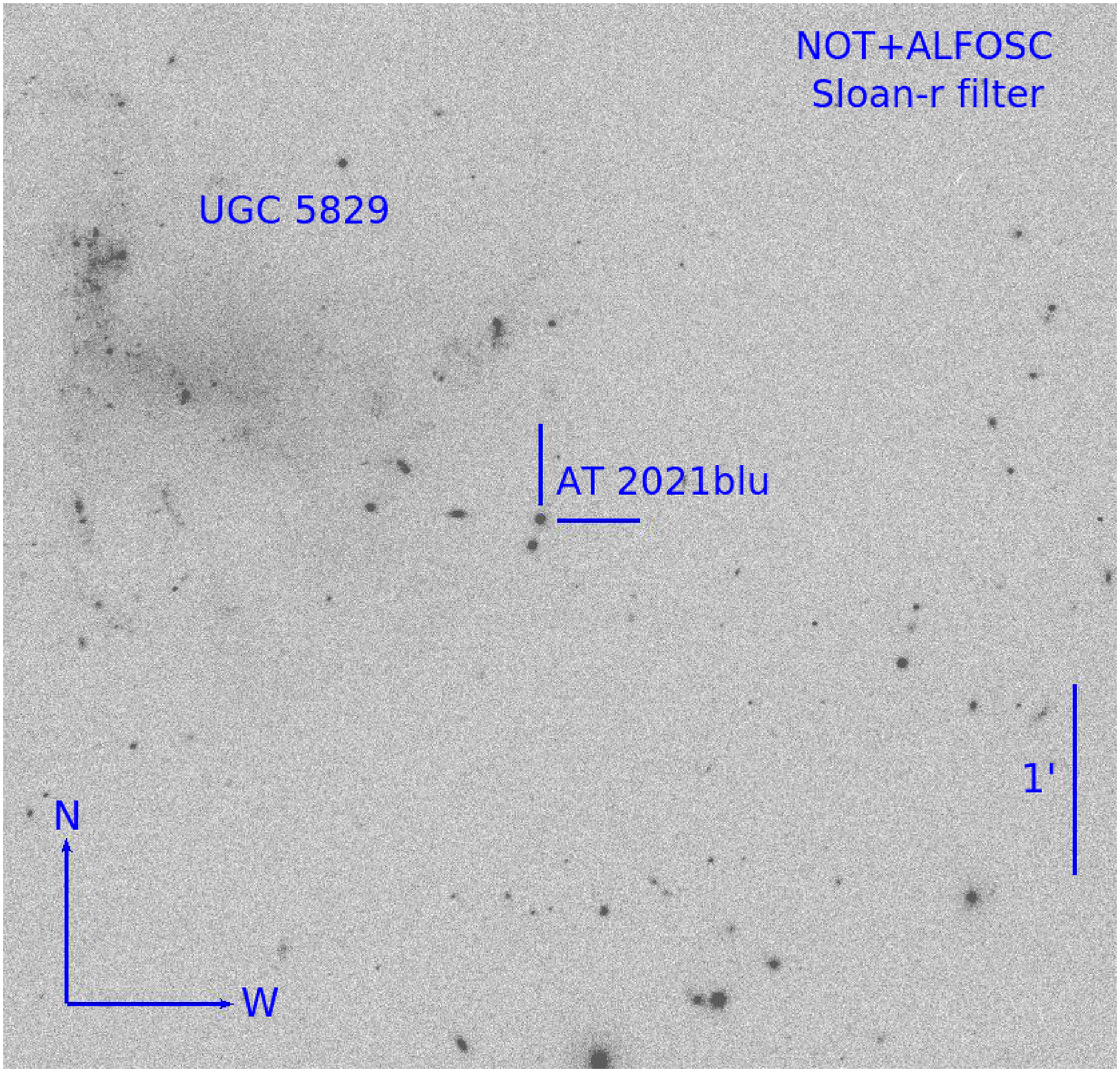}
}
      \caption{Sites of \object{AT~2018bwo} (top), \object{AT~2021afy} (middle), and \object{AT~2021blu} (bottom), with their host galaxies.}
         \label{Fig:21afy_21blu_FC}
   \end{figure}

The host galaxy, \object{NGC~45}, is a nearly face-on SABd spiral. Although the object lies in the outskirts of
\object{NGC~45} ($31\farcs7$ W and $39\farcs7$ S of the host-galaxy nucleus), it is very close to contaminating background sources (Fig. \ref{Fig:21afy_21blu_FC}, top panel). 
While at odds with other methods (``sosie'' galaxies\footnote{See, e.g., \protect\citet{bot85} for a description of the method.}, Tully-Fisher, kinematic), for the host-galaxy distance ($d$) 
we adopt the most recent value based on the tip of the red giant branch method \citep{sab18}, $d = 6.79\pm0.49$~Mpc, corresponding to a distance modulus of $\mu = 29.16 \pm 0.36$~mag. 
This value of $\mu$ is similar to that adopted by \citet{bla21}, $\mu = 29.11 \pm 0.10$~mag.

A (modest) average reddening within the host galaxy was estimated by \citet{mora07}, $E(B-V) = 0.04$~mag. \citet{bla21} adopted an even lower value ($E(B-V) = 0.01$~mag)
based on the spectral energy distribution (SED) of the LRN progenitor.
The peripheral location of \object{AT~2018bwo} and the presence of
very blue sources in its vicinity suggest a negligible contribution of the host galaxy to the total line-of-sight reddening. For this reason, hereafter we assume
that the total reddening towards \object{AT~2018bwo} is entirely due to the Milky Way contribution \citep[$E(B-V)_{\rm MW} = 0.02$~mag;][]{sch11}. 

\object{AT~2021afy}\footnote{The object is also known as ZTF21aaekeqd.} was discovered by the Zwicky Transient Facility \citep[ZTF;][]{bel19,gra19} survey on 2021 January 10.52, 
at a magnitude of $r = 20.48$ \citep{mun21}. 
Alert of the discovery was released by the ALeRCE broker\footnote{\url{http://alerce.online/object/ZTF21aaekeqd}.} \citep{car20}. The coordinates 
of the transient are $\alpha = 15^{h}48^{m}43\fs172$ and $\delta = +21\degr51\arcmin09\farcs62$ (J2000). 
The object lies above the disk plane of the edge-on spiral (Sbc-type) galaxy \object{UGC~10043} (Fig. \ref{Fig:21afy_21blu_FC}, middle panel). 
For the host galaxy, a Tully-Fisher distance of about 49.2~Mpc was inferred by \citet[][with H$_0 = 73$~km~s$^{-1}$~Mpc$^{-1}$, and
assuming $\Omega_{\rm matter} = 0.27$ and $\Omega_{\rm vacuum} = 0.73$]{tul16}. Hence, the adopted distance modulus is 
$\mu = 33.46 \pm 0.45$~mag\footnote{The distance to \object{UGC~10043} is debated, as Tully-Fisher values reported in the literature range
from about 40 to almost 60 Mpc, but are still within the (large) error bars adopted in \protect\citet{tul16} estimate.}.

%
%

While the Milky Way reddening towards \object{AT~2021afy} is modest, $E(B-V)_{\rm MW} = 0.05$~mag, the detection of prominent absorption of the interstellar
Na~I doublet (Na~I\,D) $\lambda\lambda$5890,~5896 in the transient's spectra at the redshift of the host galaxy (see Sect. \ref{sect:spec}) suggests significant 
reddening, which is unexpected given the peripheral location of \object{AT~2021afy} from the nucleus of \object{UGC~10043}.
For this reason, we speculate that the gas and dust cloud is circumstellar, or located in the proximity of the object. 
Accounting for the contribution of the host-galaxy reddening (see Sect. \ref{sect:2021afy_spec} for details), we infer a total
line-of-sight colour excess of $E(B-V)_{\rm tot} = 0.43 \pm 0.11$~mag. 

\object{AT~2021blu}\footnote{Alternative survey names are ATLAS21dic, ZTF21aagppzg, PS21akb, and Gaia21cwl.} was discovered by the Asteroid 
Terrestrial-impact Last Alert System \citep[ATLAS;][]{ton18,smi20} survey on 2021 February 1.47, at an ATLAS-orange magnitude of $o = 18.486$ \citep{ton21}.
The coordinates of the transient are $\alpha = 10^{h}42^{m}34\fs340$ and $\delta = +34\degr26\arcmin14\farcs60$ (J2000). The object lies in a remote location of
the irregular (Im type) galaxy \object{UGC~5829}. While a distance of 8~Mpc (with Hubble constant H$_0 = 75$~km~s$^{-1}$~Mpc$^{-1}$)
was estimated by \citet{tf88}, the kinematic distance corrected for Local Group infall into Virgo gives $d = 8.64 \pm 0.61$~Mpc \citep{mou00}
(computed adopting H$_0 = 73$~km~s$^{-1}$~Mpc$^{-1}$), and a distance modulus of $\mu = 29.68 \pm 0.15$~mag. 
 The site of \object{AT~2021blu} is shown in Fig. \ref{Fig:21afy_21blu_FC} (bottom panel).

The remote location of the transient in the host galaxy and the nondetection of the Na~I\,D narrow interstellar feature at the redshift of \object{UGC~5829}
suggest no reddening due to host galaxy dust. For this reason, we assume that extinction is only due to the Galactic contribution, $E(B-V)_{\rm MW} = 0.02$~mag \citep{sch11}.

We remark that \object{AT~2021blu} was initially classified as a luminous blue variable outburst by \citet{uno21}. However, as we detail in the next sections,
follow-up data indicate that both this object and \object{AT~2021afy} are LRNe.

\section{Photometric data} \label{sect:photo}

Basic information on the instrumental configurations used for the photometric campaigns of the three LRNe is provided in Appendix \ref{Appendix:A}.
The reduction of the optical photometric data collected with our facilities was carried out with the 
{\sc SNOoPY}\footnote{{\sc SNOoPY} is a package for supernova photometry using point-spread-function (PSF) fitting and/or template subtraction developed by E. Cappellaro. 
A package description can be found at \url{http://sngroup.oapd.inaf.it/ecsnoopy.html}.} package. 
Science frames were first bias-subtracted and flatfield-corrected. {\sc SNOoPY} allows us to carry out the astrometric calibration of the images, and 
PSF-fitting photometry of the target after template subtraction, if required.  Owing to their remote locations in the host galaxies, simple 
PSF-fitting photometry was used to obtain the photometric data for \object{AT~2021afy} and \object{AT~2021blu}, while
template subtraction was necessary for \object{AT~2018bwo}. Deep template images of the \object{AT~2018bwo} explosion site (with Johnson $U$, $B$, $V$; 
and Sloan $g$, $r$, $i$ filters) were obtained on 2021 July 7 with one of the 1~m telescopes of the Las Cumbres Observatory global telescope network.

The instrumental magnitudes in the Sloan filters were then calibrated using zero points and colour-term corrections with reference to 
the Sloan Digital Sky Survey (SDSS) catalogue. As the field of \object{AT~2018bwo} was not
sampled by SDSS, the Sloan-filter photometry of this LRN was calibrated using reference stars taken from the Pan-STARRS catalogue.
A catalogue of comparison stars to calibrate photometry in the Johnson-Cousins  filters was obtained 
by converting Sloan and Pan-STARRS magnitudes to Johnson-Cousins magnitudes using the transformation relations of \citet{chr08}. 
Finally, for the $o$- and $c$-band ATLAS data, we directly used the template-subtracted forced photometry \protect\citep{ton18,smi20} released through 
the ATLAS data-release interface\footnote{\url{https://fallingstar-data.com/forcedphot/queue/}.}.

{\it Swift} optical and ultraviolet (UV) magnitudes (see  Appendix \ref{Appendix:A}) were measured
with the task {\sc uvotsource} included in the {\sc UVOT} software package {\sc HEAsoft}\footnote{\url{https://heasarc.gsfc.nasa.gov/docs/software/heasoft/}.} distribution v. 6.28.
We performed aperture photometry using a $3''$ radius, while the sky contribution was computed in a ring placed between $5''$ and $10''$ from the source.

NIR images required some preliminary processing steps. We first constructed sky images for each filter by median-combining several dithered science frames. 
The contribution of the bright NIR sky was hence subtracted from individual science frames. To improve the signal-to-noise ratio (S/N), we finally combined 
the sky-subtracted frames. For the NOTCam data (see Appendix \ref{Appendix:A}), the above steps were performed using a version of the NOTCam Quicklook v2.5 reduction 
package\footnote{\url{http://www.not.iac.es/instruments/notcam/guide/observe.html}.} with a few functional modifications (e.g., to increase the field of view of the reduced image).
The following steps (astrometric calibrations, PSF-fitting photometry, and zero-point corrections) were made using  {\sc SNOoPY} and the same prescriptions as for the optical images. 
Reference stars from the Two Micron All-Sky Survey (2MASS) catalogue \citep{skr06} were used for the photometric calibration.

\begin{figure}
   \centering
   \includegraphics[angle=0,width=9.1cm]{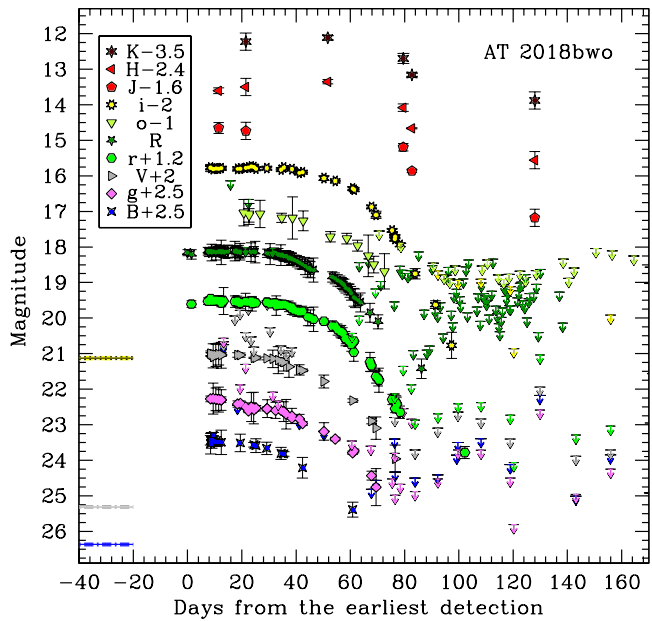}
      \caption{\object{AT~2018bwo} optical and NIR photometry.
The light curves include selected data from the public surveys and from \citet{bla21} to fill our observational gaps. The very few data points in the ATLAS-$c$ band
are not shown. The solid lines on the left represent the magnitudes of the quiescent progenitor
from \protect\citet{bla21}, converted to Johnson-Cousins $B$, $V$, and $I$ following the prescriptions of \protect\citet{har18}.
To facilitate the comparison with the $i$-band light curve of \object{AT~2018bwo}, the $I$-band magnitude of the progenitor is reported 
in the AB magnitude system. The dot-dashed lines represent the uncertainties in the progenitor detections, which are of $\sim 0.04$~mag in all filters. 
The phases are in days from the earliest LRN detection (MJD = 59252.9).
}
         \label{Fig:18bwo_lc}
   \end{figure}

\begin{figure}
   \centering
   \includegraphics[angle=0,width=9.1cm]{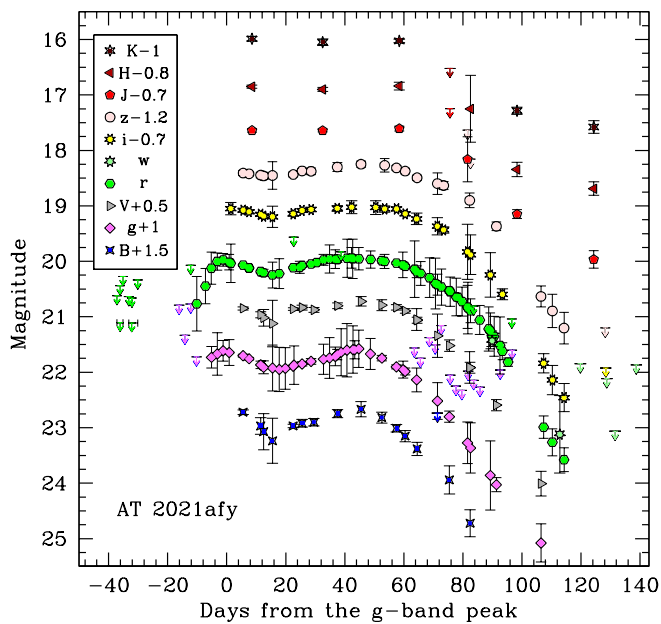}
      \caption{AT~2021afy optical and NIR photometry.
The light curves also include data from public surveys. The phases are from the Sloan $g$-band maximum (MJD = 59231.7).}
         \label{Fig:21afy_lc}
   \end{figure}

The final magnitudes of \object{AT~2018bwo}, \object{AT~2021afy}, and \object{AT~2021blu} in the optical bands are given in Tables A1, A2, and A3, 
respectively\footnote{The tables are only available in electronic form at the CDS via anonymous ftp to cdsarc.u-strasbg.fr (130.79.128.5)
or via \url{http://cdsweb.u-strasbg.fr/cgi-bin/qcat?J/A+A/}.}. The  light curves of \object{AT~2018bwo} and \object{AT~2021afy}
are shown in Fig. \ref{Fig:18bwo_lc} and \ref{Fig:21afy_lc}, respectively. The long-term light curves of \object{AT~2021blu} from the UV to the NIR are shown in  
Fig. \ref{Fig:21blu_lc} (top panel). The bottom-left panel of Fig. \ref{Fig:21blu_lc} displays in detail the UV light curves of the \object{AT~2021blu}
outburst during the first peak, while the bottom-right panel shows the evolution of the optical and NIR light curves during the LRN outburst, before the 
seasonal gap. 

\subsection{AT~2018bwo} \label{sect:2018bwo_lc}

Although \object{AT~2018bwo} was discovered on 2018 May 22, the object was also visible in DLT40 images taken eight and six days prior,
at a comparable brightness. Earlier images are not available as the object was in  solar conjunction.
These early DLT40 images are unfiltered, but were calibrated to match the Johnson-Cousins $R$ band. In all these frames, the brightness 
remains nearly constant at $R \approx 18.1$--18.2~mag. The lack of earlier images prevents us from setting a stringent limit on the LRN onset. 
Monthly unfiltered DLT40 stacked images obtained from June to August 2017 do not show signs of the LRN down to a limit of $R \approx 21.8$~mag.
A closer nondetection is provided by the {\it Gaia} Alert team\footnote{\url{http://http://gsaweb.ast.cam.ac.uk/alerts/home}.}, 
which reports that no source is visible at the location of the object on 2018 January 15,
hence about four months before the discovery. Therefore, we can only estimate a lower 
limit for the LRN outburst duration. The last positive detection of the LRN is  $\sim 2.5$ months after the discovery, while observations at later epochs 
only provide upper detection limits.

We find some differences between our Sloan-band light curves and those presented by \citet{bla21}. 
Our data have smaller scatter, and they appear to be $\sim 0.15$~mag brighter in the $g$ band and 
0.2~mag fainter in the $r$ band, while there is a fair agreement in the $i$ band. As both datasets were obtained
after template subtraction, this mismatch is puzzling. We note, however, that we used the Pan-STARRS reference
catalogue for the calibration. Other possible explanations are the low S/N of the source in individual images taken with 
1~m-class telescopes, or inaccurate colour-term corrections.

%

Our optical data reveal that the LRN remained in a sort of plateau for over three weeks after the discovery, at average magnitudes of $g = 19.78 \pm 0.27$ and $V = 19.02 \pm 0.17$~mag, 
which provide absolute magnitudes of $M_g = -9.45 \pm 0.45$ and  $M_V = -10.14 \pm 0.45$~mag. We also obtain the
intrinsic colours in this phase, $g-r = 1.44 \pm 0.28$~mag and  $B-V = 1.90 \pm 0.24$~mag. The plateau is
followed by a luminosity drop in all bands. In its initial phase, the light-curve decline is relatively slow, but it becomes very steep $\sim 50$ days 
after the discovery. As for most of LRNe, the object leaves the plateau earlier in the bluer bands than in the redder bands. 

The overall shape of the light curve of \object{AT~2018bwo} is reminiscent of those of LRNe during the late plateau phase (or soon after the broad, red 
light-curve maximum). This similarity is corroborated by spectroscopic clues, as the observed spectra of \object{AT~2018bwo} resemble the late-time spectra of canonical LRNe 
(see Sect. \ref{sect:spec}).  \citet{bla21} suggested that  the merger's photosphere was initially at a much lower temperature and with a larger radius than typical LRNe. 
However, this statement is not supported by stringent observational constraints. In particular, from the available data, we can presume 
that the outburst onset occurred a few months before the LRN discovery, and we cannot rule out that the intrinsic colour was initially much bluer than that observed. 
Consequently, while we agree that the red colour of \object{AT~2018bwo} is an indication of a more expanded and cooler photosphere, this is
possibly due to the late discovery of the transient \citep{pasto19a,pasto19b,cai19,pasto21a,pasto21b}.

%
%

\subsection{AT~2021afy} \label{sect:2021afy_lc}

The light curve of the \object{AT~2021afy} outbursts is well sampled in the optical and NIR bands (Fig. \ref{Fig:21afy_lc}). In contrast, only limited information is available for the pre-outburst phases.

To better constrain the epoch of the LRN outburst onset, we analysed the ZTF DR3 images, finding a weak ($\sim 2.7\sigma$) detection of the transient at $r = 20.77\pm0.49$~mag
on 2021 January 7.555, three days before the discovery announced by \citet{mun21}. However, nondetections are registered on the same day in the $g$ band ($> 20.72$~mag) or at earlier epochs. 
To increase the S/N, we also stacked\footnote{Information on the pre-outburst staked images of the \object{AT~2021afy} field is given in Appendix \ref{Appendix:A.1} (Table \ref{Table:A1.4}).} 
the highest-quality images obtained in December 2020, and no source was detected down to $r > 21.1$~mag. 
No activity was revealed in earlier images provided by ZTF. In particular, we stacked images in the $g$, $r$, and $i$ bands obtained 
over several months in mid-2018, and no source was detected at the LRN location to the following limits: $g \gtrsim 20.95$, $r \gtrsim 22.05$, and $i \gtrsim 21.33$~mag.

The available data allow us to constrain a first light-curve rise, which lasts at least 10 days. The $g$-band maximum, 
derived through a low-order polynomial fit to the light curve, is reached on MJD = $59231.7 \pm 1.6$, at $g = 20.63 \pm 0.03$~mag. 
Hereafter, this epoch will be used as a reference for \object{AT~2021afy}.
The $r$-band peak is reached 0.7~days later. Accounting for the total line-of-sight extinction, the intrinsic colour at the first maximum is
$g-r \approx 0.16$~mag. 
After the first peak, the light curves decline in all bands for about two weeks, reaching a relative minimum 0.3--0.4 mag fainter, followed by a second, 
broader maximum about one month later. At the second peak, we measure $g = 20.59 \pm 0.02$~mag, while $g-r$ is similar to the colour 
of the first peak. This broad peak is then followed by a rapid decline in all bands, and the colours become rapidly much redder ($g-r \approx 1.1$~mag, $\sim 90$~days after the first peak).

We note that the minimum between the two light-curve peaks is more pronounced in the blue bands than in the red bands, while the NIR 
light curves show sort of a long-lasting plateau after the first maximum, although the NIR light curves are not well sampled. 
Regardless of the filter, and in contrast with the behaviour of other LRNe, the luminosity 
of  \object{AT~2021afy} at the time of the first maximum is very similar to that of the second peak in all bands. Accounting for the reddening and the distance 
adopted in Sect. \ref{Sect:hosts}, we obtain the 
following $g$-band absolute magnitudes at the two peaks: $M_{g,{\rm max1}} = -14.46 \pm 0.63$ and $M_{g,{\rm max2}} = -14.48 \pm 0.63$~mag.

\subsection{AT~2021blu} \label{sect:2021blu_lc}

\begin{table*}
\caption{\label{tab:photo_pre} Archival data, obtained from February 2006 to May 2017, of the source at the \object{AT~2021blu} location. }
\centering
\begin{tabular}{ccccc}
\hline\hline
UT Date & MJD & Filter & Magnitude & Instrumental configuration\\
\hline 
2006-02-25 & 54094.15 & $B$ & 23.50 (0.14) & INT + WFC      \\
2006-02-25 & 54094.13 & $V$ & 23.03 (0.26) & INT + WFC      \\
2006-12-25 & 54094.12 & $r$ & 22.98 (0.12) & INT + WFC      \\
2010-03-21 to 2013-02-12 & 55801.07$^\ast$ & $g$ & 23.27 (0.21) & PS1 (stack) \\
2011-03-14 to 2014-02-09 & 56246.25$^\ast$ & $r$ & 22.98 (0.18) & PS1 (slack) \\ 
2011-05-16 to 2015-01-12 & 56246.96$^\ast$ & $z$ & 22.83 (0.27) & PS1 (slack) \\
2010-12-31 to 2015-01-22 & 56271.52$^\ast$ & $y$ & $>$21.86     & PS1 (slack) \\ 
2011-03-14 to 2015-01-12 & 56582.35$^\ast$ & $i$ & 22.86 (0.20) & PS1 (slack) \\ 
2016-03-09 & 57456.33 & $g$ & 23.03 (0.34) & Bok + 90prime  \\
2016-02-03 & 57421.37 & $r$ & $>$22.73     & Bok + 90prime  \\
2016-02-04 & 57422.40 & $r$ & 22.79 (0.44) & Bok + 90prime  \\
2016-02-04 & 57422.40 & $z$ & 22.51 (0.35) & KPNO4m + Mosaic3 \\
2016-02-06 & 57424.38 & $z$ & 22.50 (0.33) & KPNO4m + Mosaic3 \\
2016-02-15 & 57433.38 & $r$ & $>$22.70     & Bok + 90prime  \\ 
2017-03-22 & 57834.34 & $z$ & 22.03 (0.18) & KPNO4m + Mosaic3 \\
2017-03-25 & 57837.31 & $z$ & 22.06 (0.16) & KPNO4m + Mosaic3 \\
2017-05-17 & 57890.21 & $g$ & $>$22.60     & Bok + 90prime  \\
\hline
\end{tabular}
\\
\tablefoot{Johnson-Bessell $B$ and $V$ data are in the Vega magnitude scale, while Sloan $g$, $r$, $i$, $z$ and Pan-STARRS $y$ data are in the AB magnitude scale.
The Pan-STARRS data were obtained after stacking individal images collected from March 2010 to January 2015.\\
$(\ast)$ Average MJD of the stacked image.}
\end{table*}

%
   \begin{figure*}
   \centering
   {\includegraphics[angle=270,width=18.2cm]{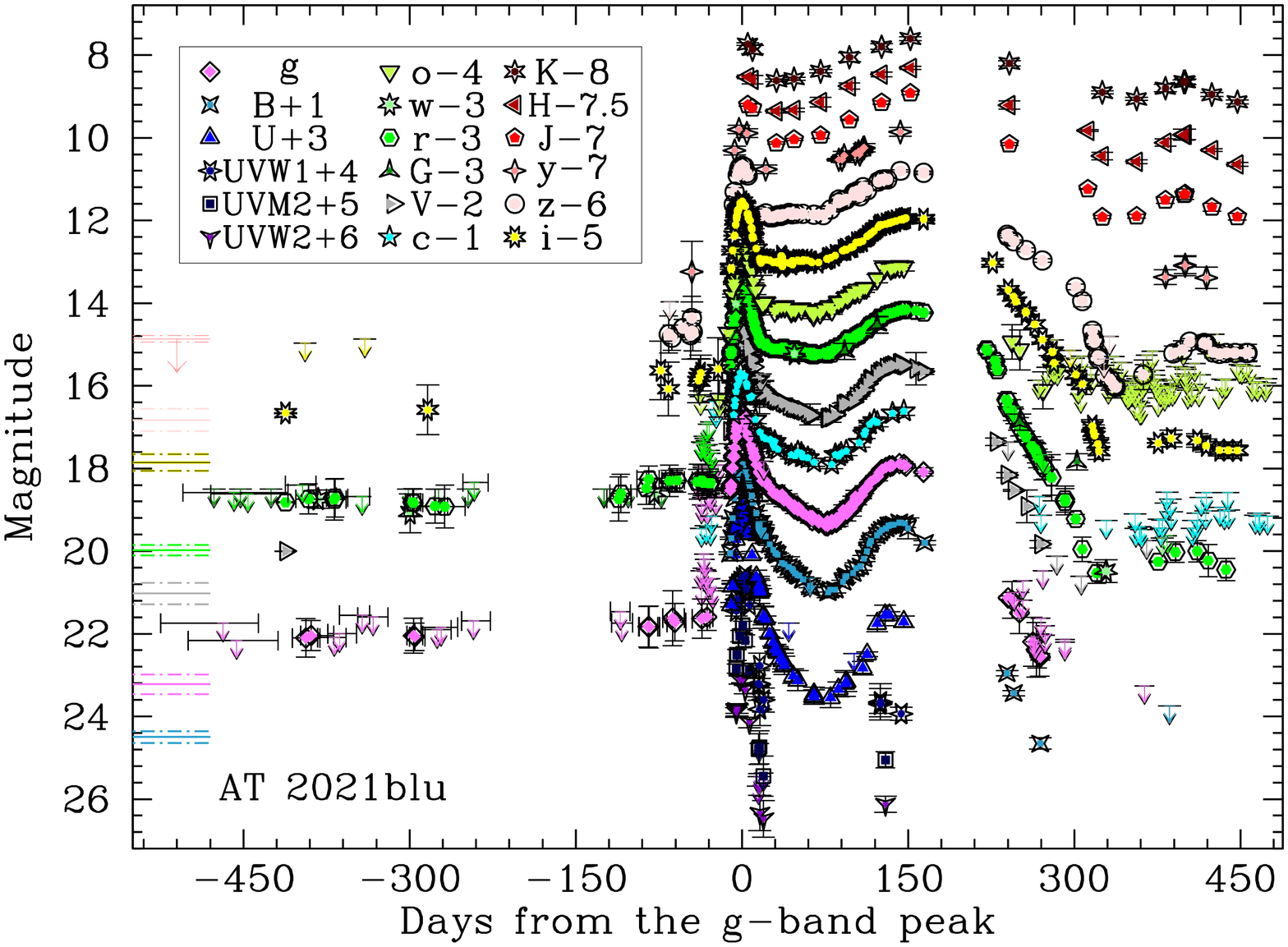}
\includegraphics[angle=0,width=9.0cm]{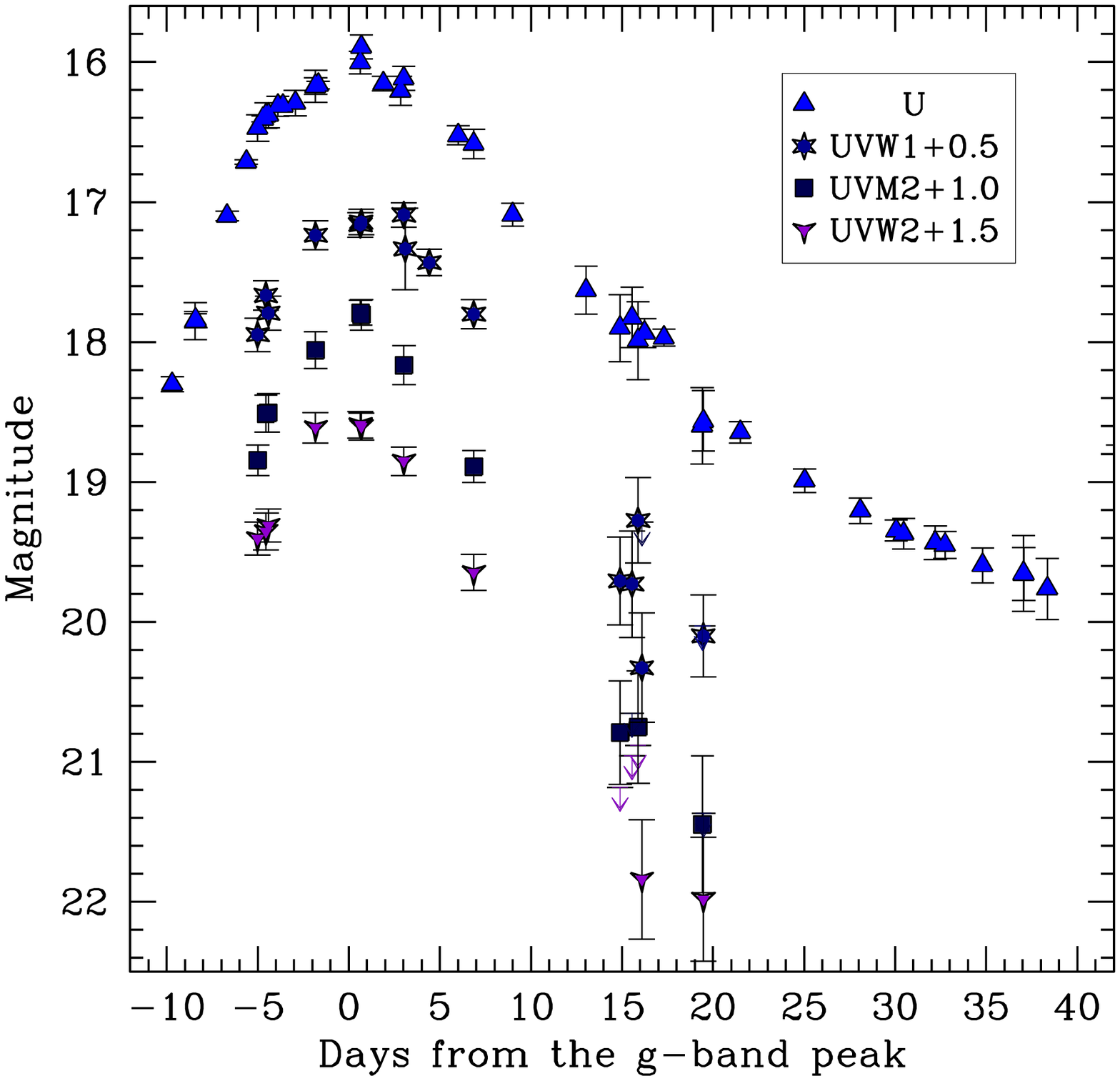}
\includegraphics[angle=0,width=9.0cm]{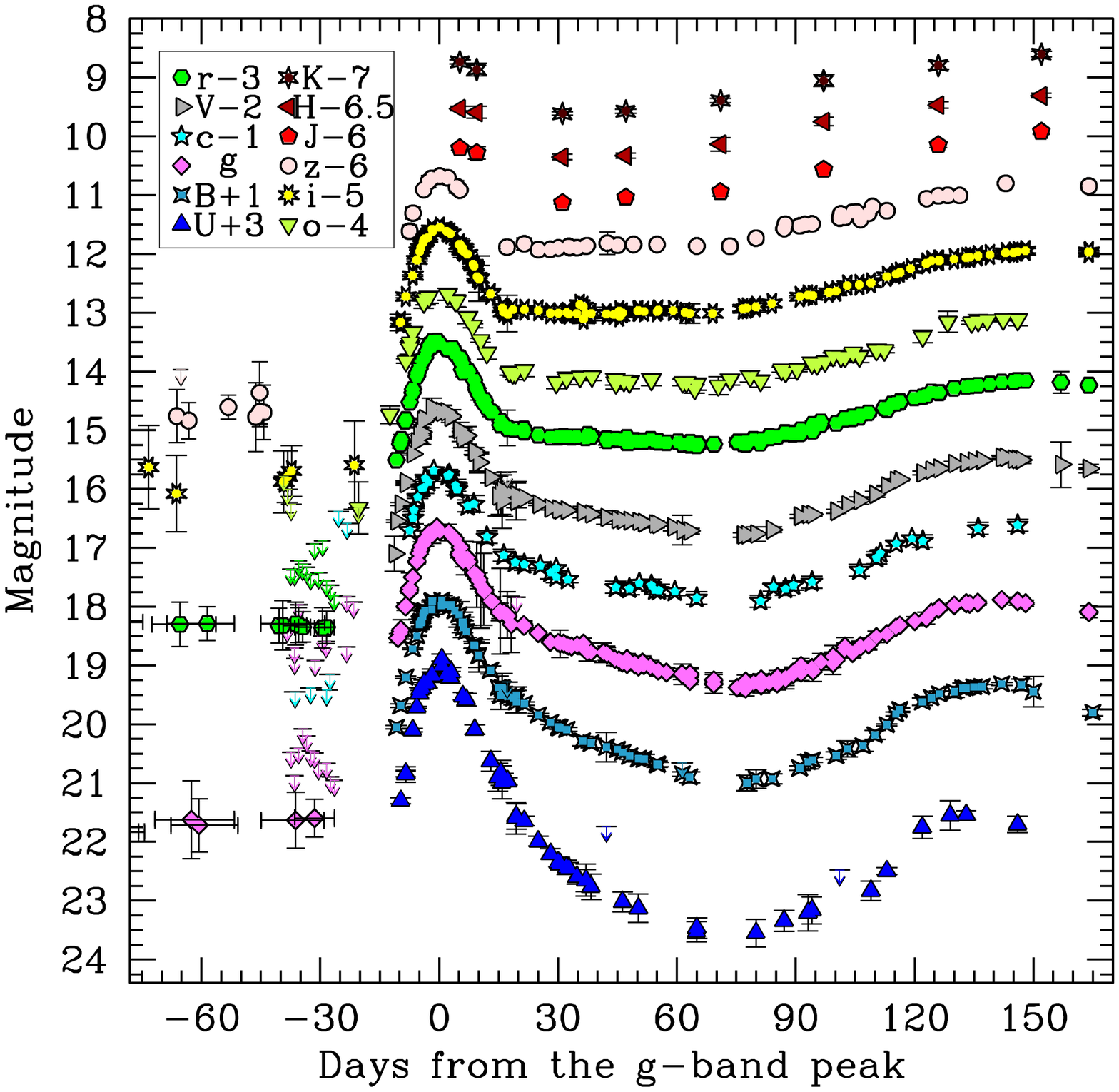}}
      \caption{Photometric evolution of  \object{AT~2021blu}. {\sl Top panel:} Long-term light curves in all filters. Transformation relations from \protect\citet{jes05}
are used to convert Sloan $u$-band photometry to Johnson-Bessell $U$.
The solid lines on the left represent
the magnitudes of the quiescent progenitor of \object{AT~2021blu}.
The down arrows represent upper detection limits.
The dot-dashed lines represent the uncertainties of the progenitor detections in the different bands (see Table \ref{tab:photo_pre}).
{\sl Bottom-left panel}: Close-up view of the 
peak of the outburst in the UV bands.  {\sl Bottom-right panel}: Detail of the LRN light curves in the optical and NIR bands from $-75$~d to $+170$~d
from the $g$-band peak (MJD = $59258.89$). Data from the public surveys are also included.
              }
         \label{Fig:21blu_lc}
   \end{figure*}
%

   
   \begin{figure*}
   \centering
   \includegraphics[angle=0,width=17.3cm]{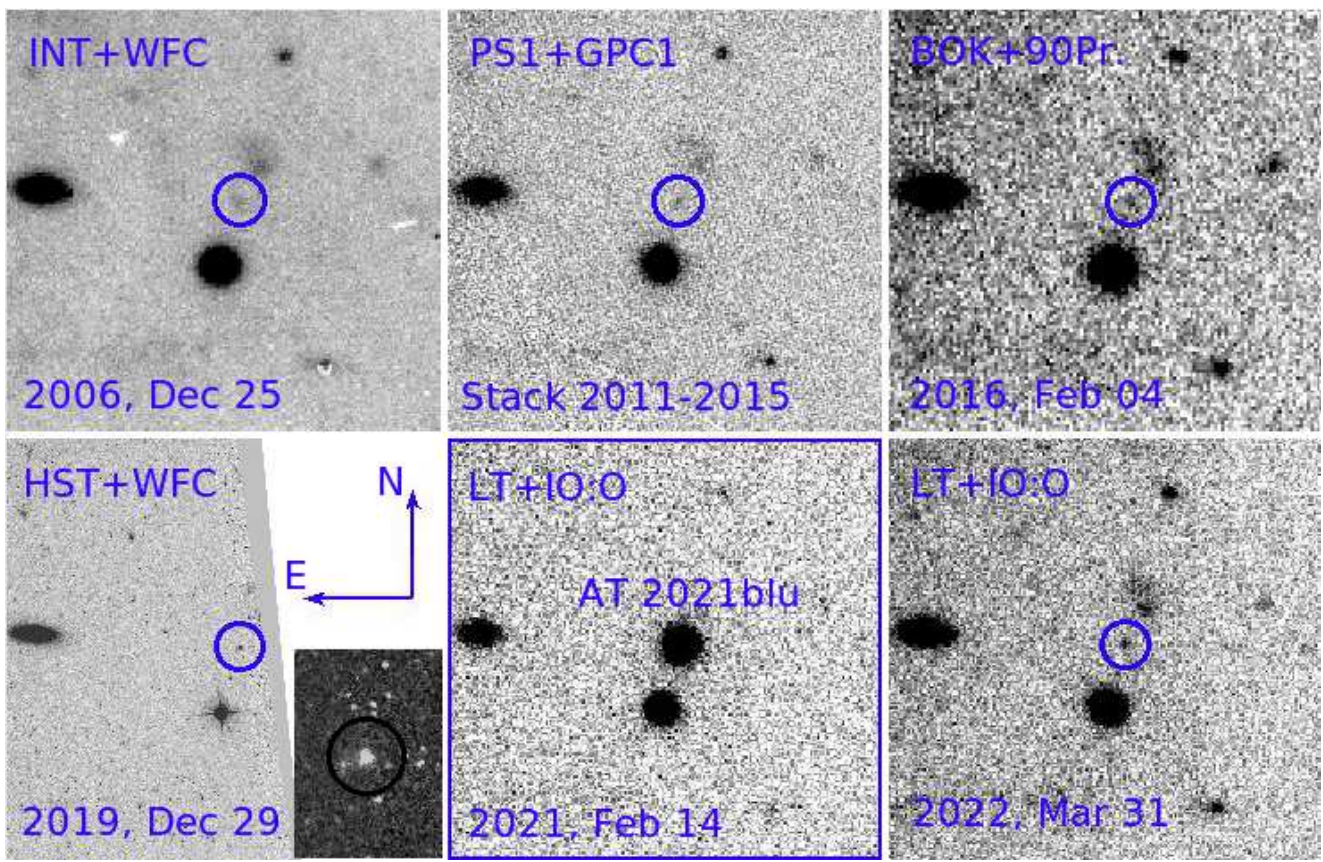}
      \caption{Evolution with time of the source coincident with the position of \object{AT~2021blu} in the Sloan $r$ band.
From 2006 to 2016 (top panels) the source is barely detected, with negligible magnitude changes. In late 2019 (bottom-left panel), 
the source is imaged by {\it HST} in the $F606W$ filter, and it is about 1~mag brighter than in the decade before. LT images show \object{AT~2021blu}
in outburst (in February 2021, approximately at maximum; bottom-middle panel), and at very late phases (late March 2022; bottom-right panel).
              }
         \label{Fig:HST}
   \end{figure*}


\begin{table*}
\caption{\label{tab:photo_maxima} Epochs (MJDs) and apparent magnitudes of the two light-curve peaks of \object{AT~2021blu} in the different filters.}
\centering
\begin{tabular}{cccccc}
\hline\hline
Filter & $\lambda_{mean}$~(\AA) & MJD~(peak~1) & Magnitude~(peak~1) &  MJD~(peak~2) & Magnitude~(peak~2) \\
\hline
$UVW2$   & 2140 & $59258.81 \pm 0.27$ &   $17.07 \pm 0.02$  &   --   &  -- \\
$UVM2$   & 2273 & $59258.86 \pm 0.21$ &   $16.86 \pm 0.04$  &   --   &  -- \\
$UVW1$   & 2688 & $59258.85 \pm 0.32$ &   $16.63 \pm 0.05$  &   --   &  -- \\
$U$      & 3416 & $59258.86 \pm 0.22$ &   $16.04 \pm 0.04$  &   $59393.2 \pm 1.8 $ &  $18.51 \pm 0.05$ \\
$B$      & 4313 & $59258.88 \pm 0.08$ &   $16.90 \pm 0.02$  &   $59399.3 \pm 2.7 $ &  $18.34 \pm 0.03$ \\
$g$      & 4751 & $59258.89 \pm 0.10$ &   $16.69 \pm 0.02$  &   $59401.8 \pm 5.3 $ &  $17.89 \pm 0.02$ \\
$cyan$   & 5409 & $59258.89 \pm 0.18$ &   $16.69 \pm 0.04$  &   $59403.9 \pm 15.4$ &  $17.62 \pm 0.11$ \\
$V$      & 5446 & $59258.91 \pm 0.07$ &   $16.69 \pm 0.02$  &   $59404.2 \pm 5.4 $ &  $17.48 \pm 0.03$ \\
$r$      & 6204 & $59258.96 \pm 0.16$ &   $16.51 \pm 0.02$  &   $59410.0 \pm 8.9 $ &  $17.18 \pm 0.06$ \\ 
$orange$ & 6866 & $59258.97 \pm 0.13$ &   $16.61 \pm 0.02$  &   $59412.8 \pm 10.6$ &  $17.08 \pm 0.07$ \\    
$i$      & 7519 & $59258.99 \pm 0.07$ &   $16.56 \pm 0.02$  &   $59412.3 \pm 4.0 $ &  $16.93 \pm 0.02$ \\
$z$      & 8992 & $59259.16 \pm 0.17$ &   $16.66 \pm 0.02$  &   $59415.6 \pm 7.0 $ &  $16.82 \pm 0.03$ \\
$J$      & 12350&$>59264.11$&   $<16.20$           &   $59419.7 \pm 10.8$ &  $15.91 \pm 0.08$ \\
$H$      & 16620&$>59264.12$&   $<16.03$           &   $59421.6 \pm 10.3$ &  $15.81 \pm 0.08$ \\    
$K$      & 21590&$>59264.12$&   $<15.74$           &   $59425.3 \pm 8.7 $ &  $15.59 \pm 0.09$ \\
\hline 
\end{tabular}
\\
\tablefoot{Johnson-Bessell $U$, $B$ and $V$, UV and NIR magnitudes are the Vega system, while Sloan $g$, $r$, $i$, $z$ data are in the AB magnitude system. }
\end{table*}

\subsubsection{Pre-outburst data} \label{sect:pre2021blu}

The field of \object{AT~2021blu} was extensively observed in the last few years. We inspected images released by the main surveys through public archives. 
To increase the S/N, we created periodic stacks\footnote{Information on the pre-outburst ZTF stacked images of the \object{AT~2021blu} field is provided 
in Appendix \ref{Appendix:A.1} (Table \ref{Table:A1.5}).} using good-quality ZTF images, and a source was detected at the location of the
LRN already in 2018. In addition, very deep images taken with ground-based, mid-sized telescopes in 2006 and early 2016 show a 
source of $\sim 23$~mag at the LRN location (Fig. \ref{Fig:HST}). In particular, Johnson-Bessell $B$ and $V$, and Sloan $r$ images taken in February 2006 with the 
Isaac Newton Telescope (INT) equipped with the wide-field camera (WFC) reveal a faint source at the LRN position, with
$B = 23.50 \pm 0.14$,  $V = 23.03 \pm 0.26$, and $r = 22.98 \pm 0.12$~mag. The source is also detected in deep PS1 reference 
images determined by stacking frames obtained from March 2010 to January 2015. Specifically, the stack PS1 frame in the $r$ band shows the source
being at the same magnitude as in 2006, with an intrinsic colour of $g-r \approx 0.27$~mag. The magnitudes of the source at the position of  \object{AT~2021blu}
in the 2006--2017 period are reported in Table \ref{tab:photo_pre}. We further discuss these archival data 
in Sect. \ref{Sect:progenitors_archive}, as they likely provide us with the most stringent information on the quiescent progenitor of \object{AT~2021blu}.
We note, however, that the low spatial resolution and the relatively low S/N of these images do not allow us to rule out the presence of contaminating 
sources in the proximity of the LRN location. 

Furthermore, archive frames in the Sloan $g$, $r$ and $z$ filters obtained in 2016 with the 2.3~m Bok and the 4~m Mayall telescopes (both hosted at the Kitt Peak Observatory)
equipped with mosaic cameras still show the putative progenitor of \object{AT~2021blu}. Over the decade, this source experienced modest magnitude evolution, and in February 2016 
it had marginally brightened by $\sim 0.15$--0.2~mag in the $g$ and $r$ bands (see Fig. \ref{Fig:HST}, and Table \ref{tab:photo_pre}).

More-recent images show this source becoming progressively more luminous: in one year (in March 2017) it has brightened by $\sim 0.5$~mag in the $z$ band, and the object has been repeatedly 
detected at later epochs. The $r$-, $w$- and $i$-band light curves from December 2019 to January 2021 (approximately from $-420$~d to $-40$~d before $g$-band maximum) show some luminosity 
fluctuations superposed on a global, slow luminosity rise (Fig. \ref{Fig:21blu_lc}, top panel), very similar to those observed in other LRNe \citep{bla17,bla20,pasto19b,pasto21a,pasto21b}. 
The $r-i$ colour remains at about 0.15--0.2~mag during that period. As for other members of this family, this slow luminosity rise follows the ejection of the common envelope, and it is possibly 
powered by collisions between circumbinary shells. 

\subsubsection{Photometric evolution of the outburst} \label{sect:outburstlc2021blu}

 The object is later observed in outburst (in early February 2021) by ATLAS on MJD = 59246.49 (at an $o$-band magnitude of $18.73 \pm 0.15$).
The object experiences a fast rise, reaching the first (blue) maximum light in a bit less than two weeks. The epoch of the $g$-band maximum is MJD = $59258.89  \pm 0.10$, which
is used hereafter as a reference for \object{AT~2021blu}. From the apparent magnitudes at the first peak, $g = 16.69 \pm 0.02$~mag ($V = 16.69 \pm0.02$~mag),
we estimate the following absolute magnitudes and intrinsic colours: $M_{g,{\rm pk1}} = -13.07 \pm 0.15$ and $M_{V,{\rm pk1}} = -13.06 \pm $ 0.15~mag, with 
$g-r = 0.16 \pm 0.03$, $B-V = 0.19 \pm 0.03$~mag.
The UV light curves obtained with {\it Swift} rapidly reach maximum brightness at nearly the same time as the $g$-band peak, at magnitudes 
between 16.5 and 17 (depending on the {\it Swift} UV filters; Fig. \ref{Fig:21blu_lc}, bottom-left panel).

The first peak is followed by a luminosity decline which lasts about 75~days, during which \object{AT~2021blu} fades by $\sim 4.5$~mag in the $U$ band, 3.1~mag in the $B$ band,
2~mag in the $V$ band, 2.7~mag in the $g$ band, 1.6~mag in the $r$ band, and 1.5~mag in the $i$ band (see Fig. \ref{Fig:21blu_lc}, bottom-right). A decline similar to that of the red 
optical bands is also observed in the NIR domain, although this phase was not well sampled. The UV light curves exhibit a very rapid post-peak decline, more rapid than the one 
observed in the $U$ band, with the LRN fading below the detection threshold of {\it Swift}/UVOT about three weeks after maximum.

Later, the luminosity rises again in all bands, reaching a second, broader peak, earlier in the blue filters. In particular, the light curve reaches the second  $g$-band maximum 
on MJD~=~$59401.8 \pm 5.3$, which is about 143~days after the first peak. The apparent magnitude at the second maximum  is $g=17.89 \pm 0.02$~mag, which provides an absolute magnitude 
of $M_{g,{\rm pk2}} = -11.87 \pm 0.16$~mag, while the reddening-corrected colour at this epoch is  $g-r =  0.71 \pm 0.08$~mag.
The second peak is reached slightly later (on MJD~=~$59404.2 \pm 5.4 $) in the $V$-band, at a magnitude of $V = 17.48 \pm 0.03$~mag ($M_{V,{\rm pk2}} = -12.26 \pm 0.15$~mag).
At this epoch, we determine a reddening-corrected colour of $B-V = 0.84 \pm 0.04$~mag.
The times and the apparent magnitudes of the two light-curve peaks were estimated through a low-order polynomial fit to the photometric data, 
and the resulting values for the different filters are reported in Table \ref{tab:photo_maxima}.
While the first light-curve maximum is observed nearly at the same time in the different bands, the second maximum is reached earlier in the blue filters than in 
the red and NIR ones, as expected from a cooling photosphere.

%
   \begin{figure}
   \centering
   \includegraphics[angle=0,width=9.0cm]{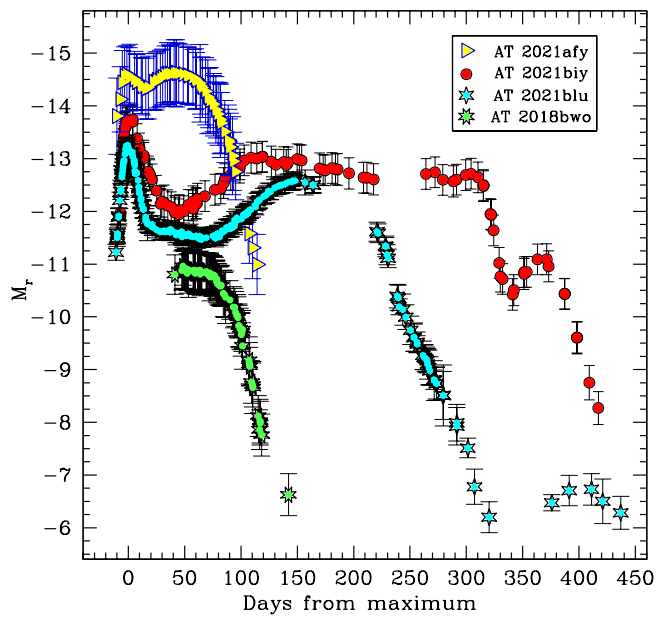}
      \caption{Comparison of the Sloan $r$-band absolute light curves of the three LRNe discussed in this paper with that of \object{AT~2021biy} \protect\citep{cai22}. For phasing \object{AT~2018bwo}, 
we assume that the early light curve maximum occurred 40~d before the earliest DLT40 detection of the transient (Sect. \ref{sect:2018bwo_lc}, and Table A.1).}
         \label{Fig:LRN_absolute}
   \end{figure}
%

Then, the object disappeared behind the Sun soon after the second maximum, and it was recovered
two months later, showing a very fast decline in all the bands lasting about 100~days, with a slower decline rate in the NIR bands.
After a faint minimum at $i = 23.06 \pm 0.31$~mag ($M_i = -6.66 \pm 0.35$~mag), the luminosity shows a short-duration hump lasting about 30--40~days in the red-optical and NIR bands, 
which is $\sim 0.5$~mag brighter than the minimum. Finally, the light curves settle to nearly constant magnitude in all bands ($i = 22.80 \pm 0.18$~mag, hence $M_i = -6.92 \pm 0.23$~mag). 
We note that a similar red hump was observed at late phases in other LRNe, including \object{AT~2021jfs} \citep{pasto19b} and \object{AT~2021biy} \citep{cai22}. 

A comparison of the Sloan $r$ absolute light curve for the three LRNe presented in this paper with that of \object{AT~2021biy} \citep{cai22} is shown in Fig. \ref{Fig:LRN_absolute}. 
While the late-time red hump is evident in \object{AT~2021biy}, it is a lower-contrast but more~persistent feature in the light curve of \object{AT~2021blu}. Although the nature of these 
bumps has not been convincingly expained so far, extra energy radiated by ejecta collisions with circumstellar shells is a plausible explanation. We note that the two LRNe with short-lasting 
outbursts in Fig. \ref{Fig:LRN_absolute}, \object{AT~2021afy} and  \object{AT~2018bwo}, display a fast-declining light curve without evident late brightenings.

\subsubsection{{\it Hubble Space Telescope} imaging of AT~2021blu} \label{sect:HSTima}

We used a deep ($15 \times 60$~s) Liverpool Telescope (LT) plus IO:O $r$-band image of \object{AT~2021blu} as a reference to search for a possible progenitor in archival {\it Hubble Space Telescope}
({\it HST}) ACS-WFC data\footnote{Program GO-15922, PI R. B. Tully.} taken on 2019 December 29, and available through the Mikulski Archive for Space Telescopes\footnote{\url{https://archive.stsci.edu/}.}. 
A second epoch\footnote{Program GO-16691, PI R. J. Foley.} of {\it HST} imaging of the \object{AT~2021blu} location was obtained on 2022 February 24, $\sim 1$~yr after the LRN outburst.

Unfortunately, AT~2021blu lies at the edge of the available ACS image obtained in December 2019. In order to match the pre- and post-explosion images,
we had to align the two frames using sources in the field that were situated east of \object{AT~2021blu}. Furthermore, only very few point 
sources were detected in both the LT and {\it HST} data. We hence used a collection of sources in the field to align the images, including 
compact clusters (that were unresolved by LT) and background galaxies.
Using 11 such sources, the position of \object{AT~2021blu} on the {\it HST}+ACS $F606W$ image was localised with a root-mean-square uncertainty of 67~mas (see Fig. \ref{Fig:HST}, bottom-left panel). 
Within this region, we find a single, bright source which we suggest to be the progenitor. 

The {\sc DOLPHOT} package \citep{Dolphin00} was used to perform PSF-fitting photometry on the progenitor candidate. While $2 \times 380$~s exposures were 
taken with ACS in each of the $F606W$ and $F814W$ filters, these images were dithered and the location of AT~2021blu lies outside the 
field of view of one of the dither positions. We hence are left with only a single 380~s image in each of the $F606W$ and $F814W$ filters. We carefully 
examined this image for cosmic rays, but found that our photometry is unaffected by them.
The following magnitudes are measured for the progenitor candidate: $F606W = 21.826 \pm 0.008$ and $F814W = 21.226 \pm 0.009$~mag (in the Vega magnitude system). 
All other sources within $1''$ from this candidate are much fainter, and their  integrated flux is about $6\%$ and $10\%$ of that of the \object{AT~2021blu} 
precursor in the  $F606W$ and $F814W$ filters, respectively.
Given the distance and extinction values adopted in Sect. \ref{Sect:hosts}, we obtain the following absolute magnitudes for the precursor of  \object{AT~2021blu}:
$M_{F606W} = -7.91 \pm 0.15$ and  $M_{F814W} = -8.49 \pm 0.15$~mag.

Assuming a 5800~K blackbody consistent with the observed colour, we used the {\sc IRAF} task {\sc SYNPHOT} to calculate a conversion to Sloan filters,
which provides $r = 21.82$ and $i = 21.66$~mag (AB system). These magnitudes are significantly brighter than earlier detections from ground-based telescopes,
suggesting that the system was already in a pre-eruptive phase. For this reason, these
{\it HST} data do not provide striking information on the nature of the quiescent progenitor system.

The second epoch of {\it HST} observations of the \object{AT~2021blu} field was obtained about 26 months later, when the LRN was very faint after the long luminosity 
decay following the second peak, and before the short-duration hump discussed at the end of Sect. \ref{sect:outburstlc2021blu}. The source at the LRN location was much 
redder than at the first {\it HST} epoch, with $F606W = 23.392 \pm 0.016$ and $F814W = 21.700 \pm 0.012$~mag (in the Vega system). 
At this epoch, the integrated flux contribution of all faint sources within a radius of $1''$ from the transient is $33\%$ in $F606W$ and $11\%$ in $F814W$ 
of the \object{AT~2021blu} flux. This may help to guess the contamination of background sources to late-time photometry of the LRN obtained with low spatial resolution 
ground-based facilities. Finally, applying the same strategy as above to convert magnitudes from the {\it HST} to the Sloan systems, we infer
$r = 23.26$ and $i = 22.38$~mag (in the AB system).

\section{Luminosity, radius, and temperature evolution} \label{sect:TLR}
   
%
   \begin{figure}
   \centering
   \includegraphics[angle=0,width=8.3cm]{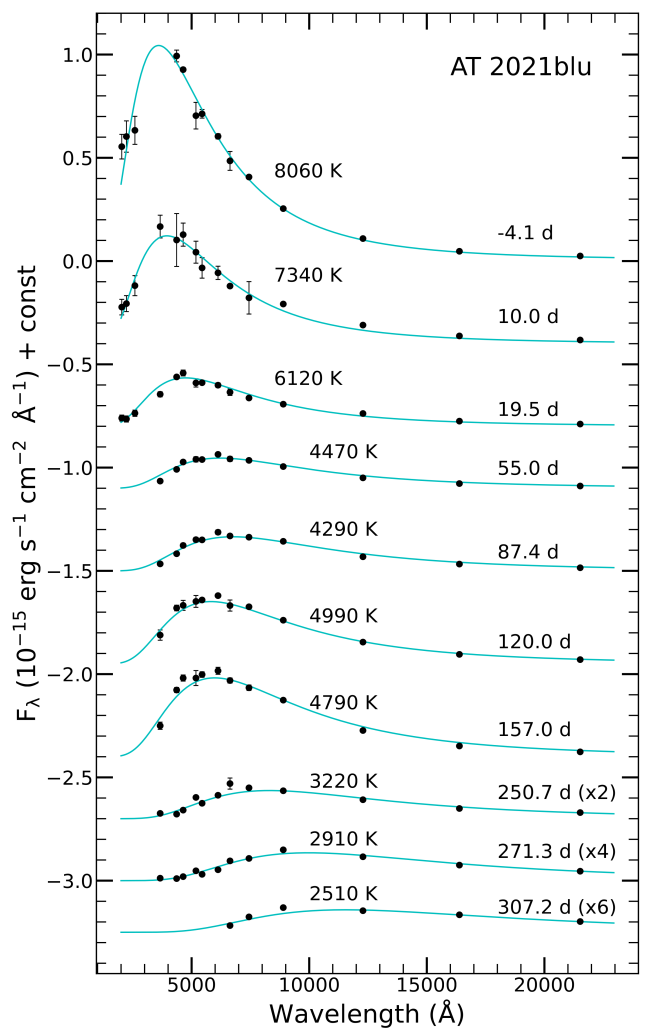}
      \caption{Blackbody fits to the multi-band observed data for \object{AT~2021blu}, showing the evolution of the SED at ten representative epochs. A scaling factor has been applied to the 
late-time flux data to improve the visibility of the fits.}
         \label{Fig:BBfits}
   \end{figure}
%

%
   \begin{figure*}
   \centering
   \includegraphics[angle=270,width=18.2cm]{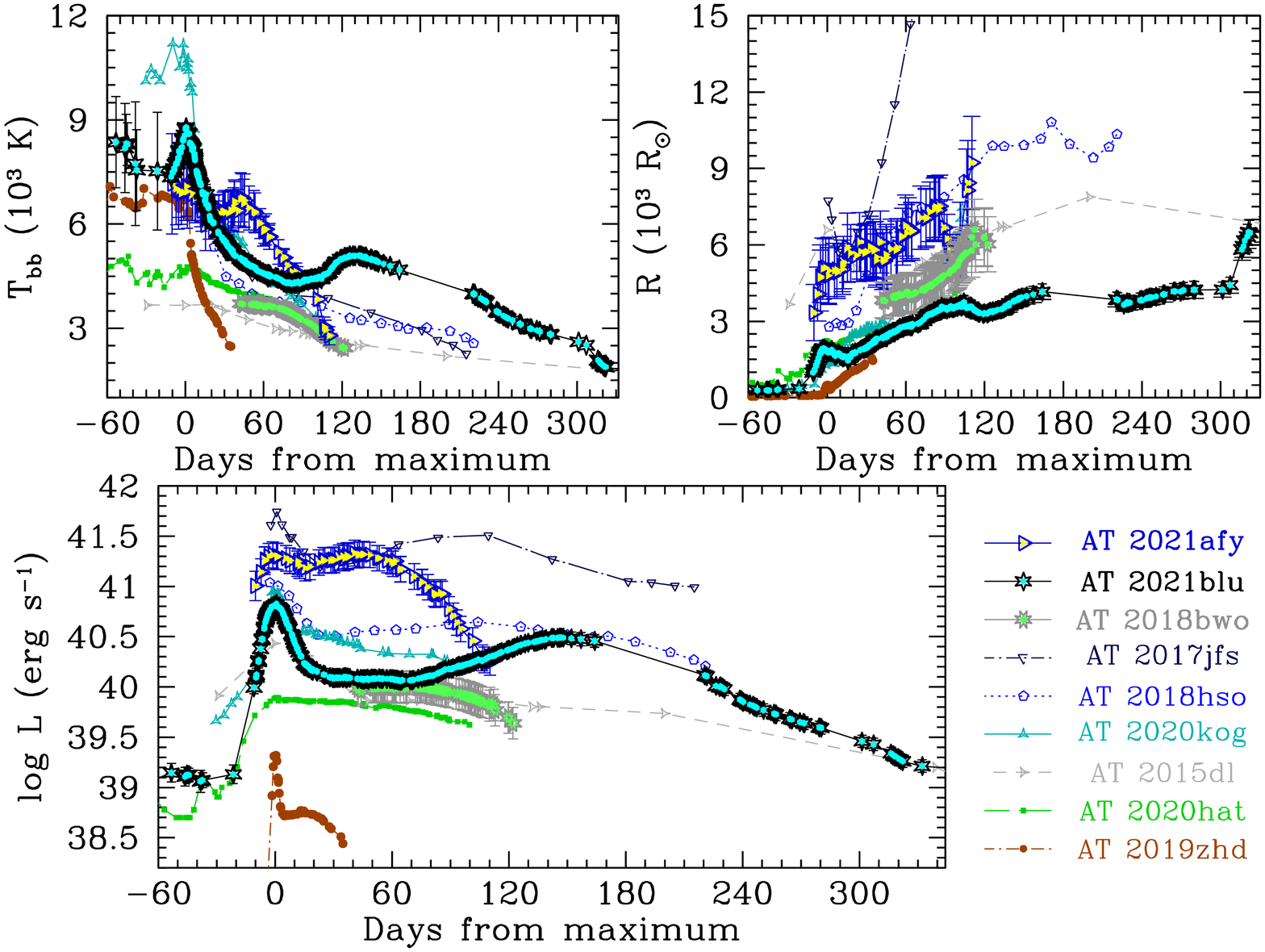}
      \caption{Evolution with time of $T_{\rm eff}$ (top left), $R_{\rm ph}$ (top right), and the bolometric light curve (bottom) for \object{AT~2018bwo}, \object{AT~2021afy}, and
\object{AT~2021blu}, along with the comparison objects \object{AT~2015dl} \protect\citep{bla17}, \object{AT~2017jfs} \protect\citep{pasto19b},  \object{AT~2018hso} \protect\citep{cai19}, 
\object{AT~2019zhd} \protect\citep{pasto21a}, \object{AT~2020hat}, and \object{AT~2020kog} \protect\citep{pasto21b}. }
         \label{Fig:TLR}
   \end{figure*}
%

Adopting a similar approach as for other LRN studies \citep[see, e.g.][]{cai19,bla20,bla21,pasto21b}, we now estimate the bolometric light curves
and the evolution of the temperature and the radius for the three objects.
The broad-band light curves illustrated in Sect. \ref{sect:photo} were used to infer the bolometric ones for the three LRNe.
To obtain the bolometric luminosity at a selected epoch, we fit the reddening-corrected SED 
of the object at that epoch with a blackbody function. If the observation in one band is not available at that epoch, its flux contribution is
estimated through an interpolation of available photometric data in adjacent epochs. Blackbody fits to the data of \object{AT~2021blu}
at some selected epochs are shown for illustrative purposes in Fig. \ref{Fig:BBfits}. The bolometric flux and the blackbody temperature ($T_{\rm bb}$),
along with their uncertainties, are determined through Monte Carlo simulations, as detailed by \citet{val22}. 
The procedure is repeated for all epochs with multi-band observations. The resulting bolometric curves of the three LRNe are 
shown in Fig. \ref{Fig:TLR} (bottom panel) and are compared with those of six well-studied LRNe.

\object{AT~2021afy} is one of the brightest objects in our sample. The two bolometric peaks have a very similar luminosity
 $L_{\rm bol} \approx 2.1 (\pm 0.6) \times 10^{41}$~erg~s$^{-1}$ (which accounts for the errors in the host galaxy distance and the reddening estimate; see Sect. \ref{Sect:hosts}).
Only \object{AT~2017jfs} is more luminous than \object{AT~2021afy}.
In contrast with the expectation for a bright LRN, \object{AT~2021afy}  remains luminous for a relatively short time ($\sim 3$ months).

\object{AT~2021blu} has a quite luminous first peak, with a $L_{\rm bol} \approx 6.5 \times 10^{40}$~erg~s$^{-1}$, followed five months later 
by a fainter, second, broad maximum at $L_{\rm bol} \approx 3.1 \times 10^{40}$~erg~s$^{-1}$. The overall bolometric evolution
resembles that of  \object{AT~2018hso} \citep{cai19}, which is only marginally brighter than \object{AT~2021blu}.

As already mentioned in Sect.\ref{sect:2018bwo_lc}, we could not follow the early-time evolution of \object{AT~2018bwo}. Consequently, we cannot 
precisely constrain the time of the early maximum,
along with the duration of the LRN outburst. However, we argue that the outburst onset occurred several weeks before the discovery.
We arbitrarily fixed the epoch of the early maximum at 40~days before our earliest detection. The object already appears to be on the plateau (or on
a low-contrast second broad peak), with an average bolometric luminosity slightly exceeding $L_{\rm bol} \approx 10^{40}$~erg~s$^{-1}$. In this phase, it is
marginally brighter than \object{AT~2020hat}, an object that did not show a high-contrast early peak, and one order of magnitude brighter 
than  \object{AT~2019zhd}, the lowest-luminosity object of the sample.

\begin{table*}
\caption{\label{tab:speclog}Log of spectroscopic observations of the three LRNe discussed in this paper. }
\centering
\begin{tabular}{lcccccc}
\hline\hline
UT Date & MJD & Phase &Instrumental configuration & Exp. time (s) & Range (\AA) & Res. (\AA) \\
\hline
\multicolumn{7}{c}{AT~2018bwo} \\
\hline
2018-05-23 &    58261.18 &    +8.3  &   $11.1\times9.8$~m~SALT + RSS + PG0900  & 900  & 3640-7260 & 6 \\
2018-05-26 &    58263.42 &   +10.5  & 8.1~m Gemini-South + Flamingos-2 + JHG5801  & 2400  & 8920-18000 & $\cdots$\\
2018-06-05 &    58274.41 &   +21.5  & 6.5~m Magellan/Baade + FIRE &  2029 & 8200-24680 & $\cdots$ \\
2018-06-18 &    58287.30 &   +34.4  & 4.1~m SOAR + Goodman + grt.400 & 1200 & 3700-7120 & 6.4 \\
2018-07-11 &    58310.59 &   +57.7  & 10~m~Keck-I +LRIS +600/4000   &      1108 &     5600-10200   &     6    \\    
2018-08-25 &    58355.15 &  +102.2  & 10.4~m~GTC+OSIRIS + R1000B + R1000R     &   2400+2400 &     3630-9800    &  7,8    \\
2018-09-12 &    58373.26 &  +120.4  & 6.5~m Magellan/Baade + FIRE &  1522 & 8200-24680 & $\cdots$ \\
2018-09-17 &    58380.90 &  +128.0  &   10~m~Keck-I +LRIS +600/4000   &      3600 &     5700-10200   &     6    \\    
\hline 
\multicolumn{7}{c}{AT~2021afy} \\
\hline
2021-01-25 &    59239.26 &   +7.6  &   10.4~m GTC + OSIRIS + R1000B   &      3000 &     3630-7870   &     7    \\    
2021-02-17 &    59262.25 &  +30.6  &   10.4~m GTC + OSIRIS + R1000B   &      3600 &     3640-7870   &     7    \\    
2021-02-24 &    59269.19 &  +37.5  &   10.4~m GTC + OSIRIS + R1000R   &      3600 &     5080-10200  &     8    \\    
2021-04-06 &    59310.10 &  +78.4  &   10.4~m GTC + OSIRIS + R1000R   &      3600 &     5100-10400  &     8    \\    
2021-04-23 &    59327.02 &  +95.3  &   10.4~m GTC + OSIRIS + R1000R   &	     2700 &     5100-10400  &     8    \\    
\hline
\multicolumn{7}{c}{AT~2021blu} \\
\hline
2021-02-06 &    59251.45 &  $-$7.4 & 2.0~m FNT + FLOYDS                  &   3600      &   3500-10000 &    15     \\
2021-02-07 &    59252.30 &  $-$6.7 & 3.05~m Shane +  Kast + 600/4310+300/7500   & 2460+2400   &   3620-10700 &   5,9   \\ 
2021-02-10 &    59255.02 &  $-$3.9 & 2.56~m NOT + ALFOSC + gm4           &   1800      &   3400-9650  &    14     \\ 
2021-02-10 &    59255.46 &  $-$3.4 & 2.0~m FNT + FLOYDS                  &   2700      &   3500-10000 &    15     \\ 
2021-02-11 &    59256.24 &  $-$2.6 & 3.05~m Shane +  Kast + 452/3306+300/7500   & 1230+1200   &   3300-10300 &   5,9    \\
2021-02-16 &    59261.05 &  +2.2 & 1.82~m Copernico + AFOSC + VPH7     &   2700      &   3250-7270  &    14     \\ 
2021-02-16 &    59261.05 &  +2.2 & 3.6~m DOT + ADFOSC + 676R           &    900      &   3550-8850  &    12     \\ 
2021-02-18 &    59263.02 &  +4.1 & 2.56~m NOT + ALFOSC + gm4           &   2440      &   3400-9700  &    14     \\ 
2021-02-18 &    59263.29 &  +4.4 & 3.05~m Shane + Kast + 600/4310+300/7500 & 2160+2100 &   3630-10740 &   5,10  \\ 
2021-02-21 &    59266.12 &  +7.2 & 3.6~m DOT + ADFOSC + 676R           &    1200     &   3800-8880  &   11.5    \\ 
2021-02-23 &    59268.37 &  +9.5 & 10.0~m Keck-II + NIRES                &             &   9640-24660 &  $\cdots$ \\
2021-02-25 &    59270.96 & +12.1 & 10.4~m GTC + OSIRIS + R1000B        &    540      &   3630-7880  &     7     \\ 
2021-03-02 &    59275.18 & +16.3 & 3.6~m DOT + ADFOSC + 676R           &   1800      &   3700-8870  &   11.5    \\ 
2021-03-05 &    59278.04 & +19.2 & 3.6~m DOT + ADFOSC + 676R           &   1800      &   3900-8890  &   11.5    \\ 
2021-03-07 &    59280.39 & +21.5 & 3.05~m Shane + Kast + 600/4310+300/7500    & 3060+3000   &   3620-10730 &   5,9    \\ 
2021-03-14 &    59287.02 & +28.1 & 1.82~m Copernico + AFOSC + VPH7     &    3600     &   3350-7270  &    15     \\ 
2021-03-15 &    59288.45 & +29.6 & 2.0~m FNT + FLOYDS                  &    2700     &   3500-10000 &    15     \\ 
2021-03-18 &    59291.09 & +32.2 & 3.58~m TNG + LRS + LRB/LRR          & 1800+1800   &   3350-9700  &   10,10   \\ 
2021-03-30 &    59303.47 & +44.6 & 2.0~m FNT + FLOYDS                  &    2700     &   4000-10000 &    15     \\ 
2021-04-02 &    59306.92 & +48.0 & 2.56~m NOT + ALFOSC + gm4           &    3600     &   3400-9680  &    14     \\  
2021-04-08 &    59312.46 & +53.6 & 2.0~m FNT + FLOYDS                  &    3600     &   3500-10000 &    15     \\ 
2021-04-19 &    59323.89 & +65.0 & 10.4~m GTC + OSIRIS + R1000B + R1000R & 1500+1500   &   3630-10400 &    7,8    \\ 
2021-05-05 &    59339.99 & +81.1 & 3.58~m + TNG+ LRS + LRB/LRR         & 5400+3600   &   3400-9600  &   10,10   \\ 
2021-05-12 &    59346.32 & +87.4 & 10.0~m Keck-I + LRIS + 600/400+400/8500 & 900+900  &   3150-10150 &   5,6     \\
2021-05-18 &    59352.98 & +94.1 & 2.56~m NOT + ALFOSC + gm4           &    3800     &   3400-9600  &    18     \\ 
2021-05-30 &   	59364.97 & +106.1 & 2.56~m NOT + ALFOSC + gm4           &    3000     &   3400-9650  &    14     \\ 
2021-06-04 &    59369.28 & +110.4 & 3.05~m Shane +  Kast + 452/3306+300/7500   & 1230+1200   &   3400-10000 &   5,9    \\
2021-06-15 &    59380.92 & +122.0 & 10.4~m GTC + OSIRIS + R2000B/R2500R &  1200+900   &   3850-7680  &  3.1,3.4  \\ 
2021-06-29 &    59394.91 & +136.0 & 10.4~m GTC + OSIRIS + R1000B        &     900     &   3630-7870  &     7     \\  
2021-07-08 &    59404.91 & +146.0 & 10.4~m GTC + OSIRIS + R1000B        &    1200     &   3630-7870  &     7     \\ 
\hline
\end{tabular}
\tablefoot{For \object{AT~2018bwo}, the phases 
are computed from the first LRN detection (2018 May 14; MJD = 58252.905). 
The phases for the other two objects (\object{AT~2021afy} and \object{AT~2021blu}) are computed with respect to their $g$-band light-curve peaks,
that occurred on MJD = $59231.7\pm1.6$ and MJD = $59258.89\pm0.10$, respectively. The resolution reported here are the FWHM of the night-sky lines. 
 For further information on the instruments, and identification
of the acronyms, see Appendix \ref{Appendix:B}.}
\end{table*}
 
The evolution of $T_{\rm bb}$ is shown in the top-left panel of Fig.~\ref{Fig:TLR}. \object{AT~2021blu} is one of the hottest objects 
in the sample. The lack of simultaneous observations in multiple bands before the LRN outburst makes the $T_{\rm bb}$ estimates very 
uncertain. However, during the slow luminosity rise of the  pre-outburst phase, $T_{\rm bb}$ remains between 7000 and 8000~K. Then, 
the temperature rises while the LRN reaches the first maximum. At peak, $T_{\rm bb} \approx 8800$~K, then declines to a relative minimum ($T_{\rm bb} \approx 4300$~K)
three months after the bolometric maximum. During the photometric rise to the second broad peak, the temperature grows again and reaches a 
maximum of $T_{\rm bb} \approx 5000$~K. Finally, it declines monotonically to $T_{\rm bb} \approx 2600$~K at $\sim 300$~d, and more rapidly later,
reaching $\sim 1800$~K one month later, when the bolometric light curve reaches a local minimum before the red hump discussed in Sect. \ref{sect:outburstlc2021blu}.
One may wonder if the assumption of a thermal continuum at such late phases is appropriate in the case of \object{AT~2021blu}. 
However, although  \object{AT~2021blu} was not observed in spectroscopy after $\sim 5$ months past maximum (see Sect. \ref{sect:2021blu_spec} and Table \ref{tab:speclog}), 
the SED is still consistent with a blackbody (Fig. \ref{Fig:BBfits}). Furthermore, LRN \object{AT~2021biy} \citep{cai22} showed a similar behaviour in the late-time
light curve and in the temperature evolution, while its spectra resembled those of intermediate M-type stars. All of this makes the assumption of thermal radiation
at very late epochs plausible also for  \object{AT~2021blu}.

\object{AT~2021afy} has a smoother temperature evolution. From the first days after the outburst onset and up 
to maximum light, $T_{\rm bb}$ remains nearly constant, at $\sim 7000$~K. Then,  two weeks after maximum,
it slowly declines to a minimum of $T_{\rm bb} \approx 6000$~K, and rises again up to $T_{\rm bb} \approx 6700$~K  at the time of the 
second light-curve maximum. The late phases are characterised by a linear temperature decline, which
fades to $T_{\rm bb} \approx 2800$--2900~K at phase 110~d.

Finally, the blackbody temperature of \object{AT~2018bwo}
slowly declines from $T_{\rm bb} \approx 3700$~K to $\sim 2500$~K over the observed follow-up campaign, similar to \object{AT~2020hat} 
during the plateau \citep{pasto21a} and \object{AT~2015dl} at the time of the second 
light-curve peak \citep{bla17}. Regardless of the exact explosion epoch, \object{AT~2018bwo} appears to have a cooler photosphere
than other comparison objects, in agreement with \citet{bla21}.

We can then estimate the evolution of the photospheric radius ($R_{\rm ph}$) for the three LRNe (Fig. \ref{Fig:TLR}, top-right panel).
The $R_{\rm ph}$ value for  \object{AT~2021afy} initially rises from 3300~R$_\odot$ to 
5000~R$_\odot$ at the first maximum. After maximum, it increases more slowly, reaching $R_{\rm ph} \approx 8000$--9000~R$_\odot$ 
over three months later. The radius evolution of  \object{AT~2018bwo} is somewhat similar, with
$R_{\rm ph}$ ranging from about 3800~R$_\odot$ to 6500~R$_\odot$ over the two months following the discovery.

We note that  both \object{AT~2021afy} and \object{AT~2018bwo} were observed in the optical bands at phases later than 110--120~d, but these
observations mostly provide detection limits. In contrast, the two LRNe are clearly detected in the NIR bands, indicating a predominant
emission in the IR domain \citep{bla21}. This incomplete SED information leads us to give uncertain bolometric luminosity estimates 
inferred from single-blackbody fits and, consequently, unreliable values of the temperature and the radius at very late phases.

%
\begin{figure*}
\centering
{\includegraphics[angle=270,width=15.0cm]{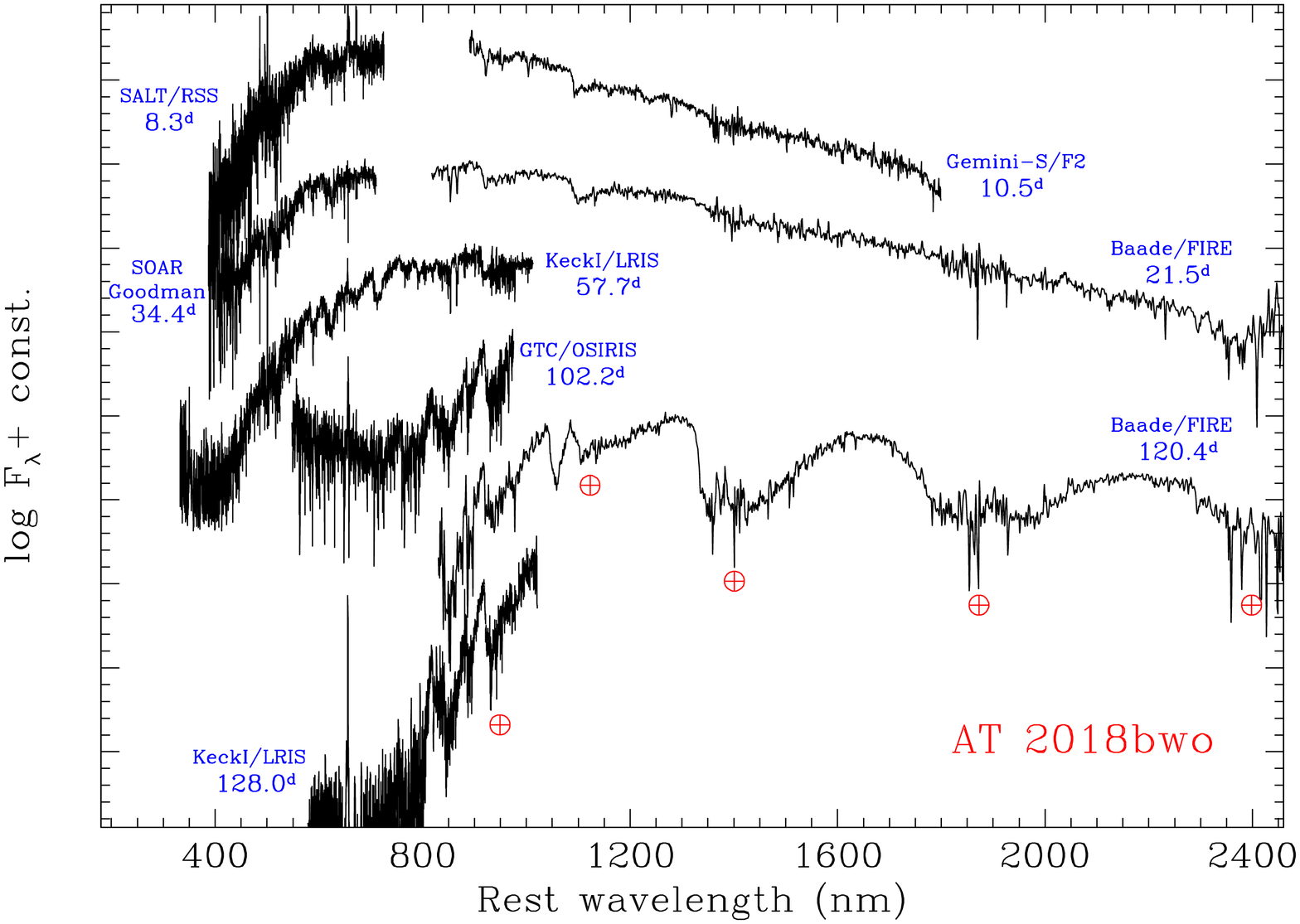}
\includegraphics[angle=270,width=15.0cm]{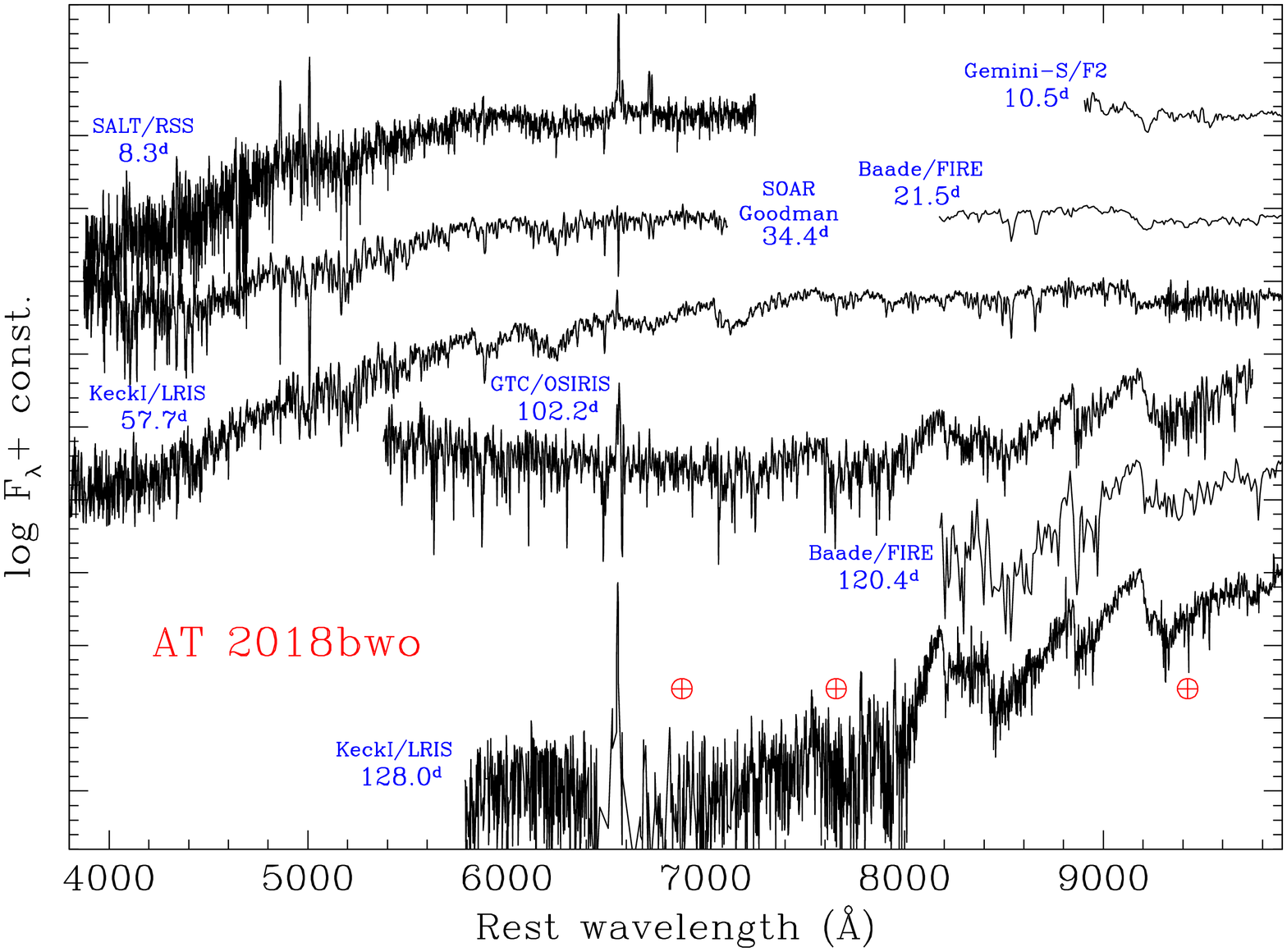}}
\caption{ Optical and NIR spectra of AT~2018bwo, corrected for the
  redshift ($z=0.001558$) and for the reddening ($E(B-V)_{\rm MW}
  = 0.02$~mag). {\sl Top panel:} Full wavelength range.  {\sl Bottom panel:} Close-up view of the spectral set in the optical domain. 
The instrumental configurations and the phases from the earliest detection are also reported. We remark that the epoch
  of the outburst onset adopted in Sect. \ref{sect:TLR} was 40~days earlier.}
         \label{Fig:specseq18bwo}
   \end{figure*}
%

The well-sampled panchromatic light curve of \object{AT~2021blu} allows us to study in detail how its $R_{\rm ph}$ evolves with time.
In the pre-outburst phase, $R_{\rm ph}$ remains in the range 260--350~R$_\odot$. At this phase, we expect that the photosphere is located
in the common envelope. From phase about $-11$~d to maximum light, $R_{\rm ph}$ rises from 1000~R$_\odot$ to 1900~R$_\odot$.
After the peak, the photospheric radius initially declines to a local minimum observed two weeks after maximum ($R_{\rm ph} \approx 1500$~R$_\odot$)
and then rises again until $\sim 105$~d, reaching a value of $R_{\rm ph} \approx 3750$~R$_\odot$. 
This is followed by a shallow dip ($R_{\rm ph} \approx 3300$~R$_\odot$ at nearly 120~d) and a further increase.
The radius, in fact, reaches a new maximum ($R_{\rm ph} \approx 4200$~R$_\odot$) soon after the broad light-curve peak,
and then the photosphere recedes again by a few hundred solar radii when the object was re-observed after the seasonal gap.
This phase is then followed by a new increase of the photospheric radius, which is initially slow, but later (at $\sim 310$~d)
rapidly rises to a value of $R_{\rm ph} \approx 6500$~R$_\odot$ at $\sim 330$~d, when the light curve reaches a minimum before 
the very late red and NIR hump. This feature, noticed also in \object{AT~2021biy} \citep{cai22} at a similar phase, can result from
an additional source of energy, such as the CSM interaction.

The comparisons in Fig. \ref{Fig:TLR} suggest not only that LRNe span a wide range of luminosities, but that there is also an evident heterogeneity 
in the bolometric light-curve shapes, with some objects showing a luminous early peak, while others (including \object{AT~2020hat} and, to a lesser extent, 
\object{AT~2021afy}) have a low-contrast first peak. The same heterogeneity is observed in the evolution of the temperature and radius at the
photosphere; if LRNe are produced by the coalescence of the stellar cores in a binary system, this diversity can be considered as an indication that 
the two stellar components span a wide range of parameters.

\section{Spectroscopic data} \label{sect:spec}

\citet{bla21} presented some optical and NIR spectra of \object{AT~2018bwo}. We complement their observations with an additional set of spectra obtained from a few days 
after the LRN discovery to $\sim 5$ months later.  The spectra cover three 
phases of the LRN evolution: soon after the discovery, at the end of the plateau, and at very late phases when most of the LRN flux is emitted in the IR domain. 
We obtained five epochs of spectroscopy for \object{AT~2021afy}. All observations were performed after the first light-curve peak, until $\sim 95$~d. Given the faint apparent 
magnitude of the object, all spectra were obtained using the 10.4\,m Gran Canarias Telescopio (GTC) with the Optical System for Imaging and low-Intermediate-Resolution 
Integrated Spectroscopy (OSIRIS).
Finally, \object{AT~2021blu} has a more extensive spectroscopy, ranging from one week before maximum light to  $\sim 146$~d. The instruments used for
the spectroscopic observations of the three objects are listed in Appendix \ref{Appendix:B}, and basic information on the spectra is provided in Table~\ref{tab:speclog}.

All spectra were taken at the parallactic angle \citep{fil82}, except those obtained at Keck-I, where an atmospheric dispersion corrector is employed.
The spectra were reduced using tasks in {\sc IRAF}\footnote{{\sc IRAF} was distributed by the National Optical Astronomy Observatory,
which was operated by the Association of Universities for Research in Astronomy (AURA), Inc., under a cooperative agreement with the National
Science Foundation (NSF).} or with dedicated pipelines such as {\sc FOSCGUI}\footnote{{\sc FOSCGUI} is a {\sc Python}-based graphic user interface (GUI) developed by E. Cappellaro, and
aimed at extracting supernova spectroscopy and photometry obtained  with FOSC-like instruments. A package description can be found at \url{http://sngroup.oapd.inaf.it/foscgui.html}.}
tool. The different tools perform the usual preliminary reduction steps, including  bias subtraction and flatfield corrections of the two-dimensional images. Then, the spectra are
calibrated in wavelength using comparison-lamp spectra and the night-sky lines, and 1-D spectra are optimally extracted. The spectra are flux-calibrated using spectra 
of standard stars taken during the night, and the calibration is
finally checked versus the available photometry. Finally, the broad absorption bands of O$_2$ and H$_2$O due to Earth's atmosphere are removed using the spectra of 
early-type stars, which are characterised by a nearly featureless continuum at the wavelengths of the telluric bands.

\subsection{AT~2018bwo} \label{sect:2018bwo_spec}

The spectra of \object{AT~2018bwo}, shown in Fig. \ref{Fig:specseq18bwo}, have a red continuum with a number of 
molecular bands (primarily TiO), prominent both in the optical and the NIR regions.
While our spectra complement those available in the literature, for a detailed line identification we direct the reader to
 \citet{bla21}.

%
\begin{figure*}
\centering
\includegraphics[angle=270,width=17.6cm]{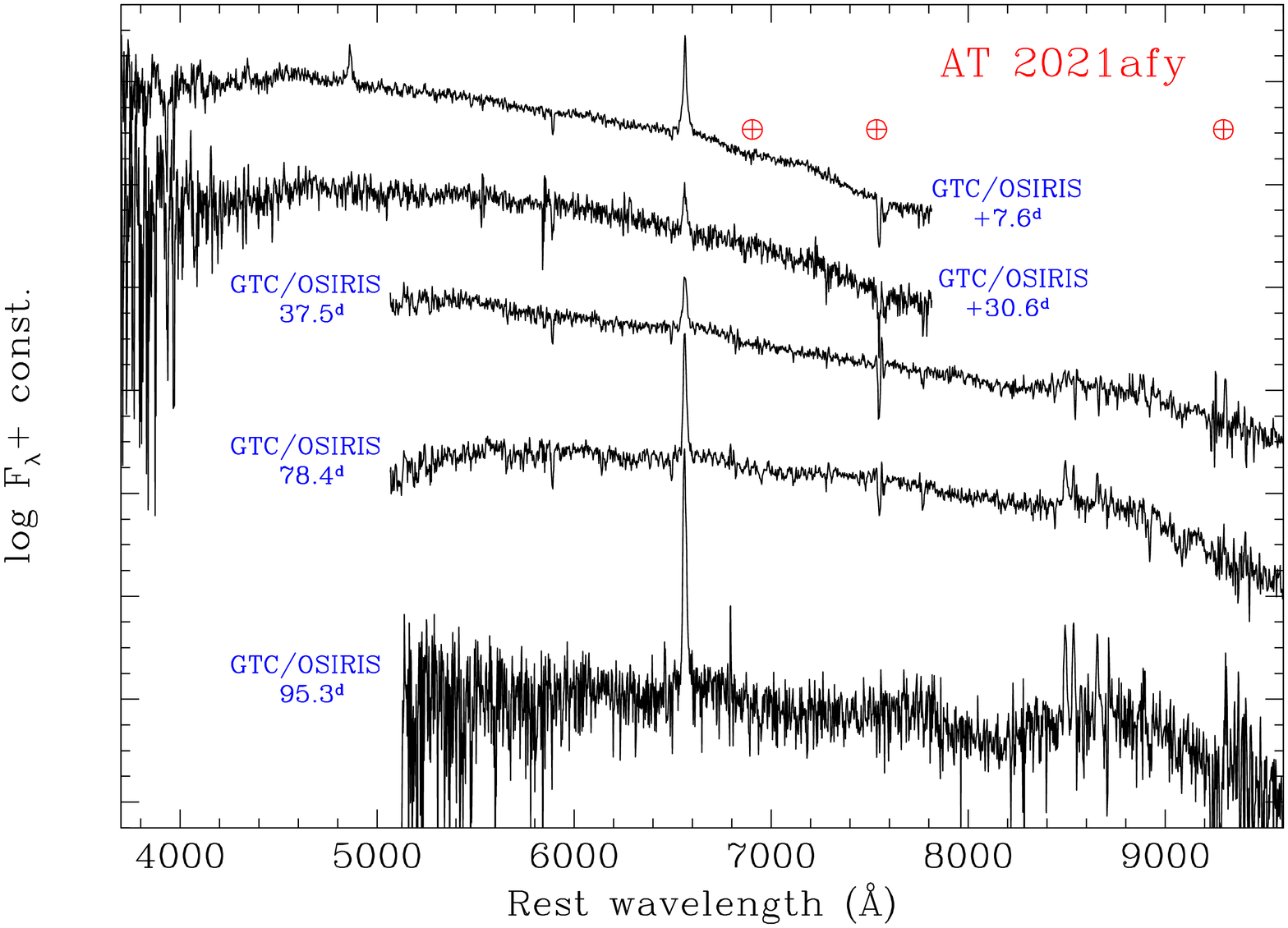}
\caption{Optical spectra of AT~2021afy, corrected for
  $z=0.007522$ and a total colour excess of $E(B-V)_{\rm tot} = 0.43$~mag.
  The instrumental configurations and the phases from the $g$-band
  maximum are also marked.}
         \label{Fig:specseq21afy}
   \end{figure*}
%

Our spectra are corrected only for Milky Way reddening, as specified in Sect. \ref{Sect:hosts}.
Hereafter, the phases will be with reference to the time of the earliest LRN detection (MJD = 59252.9). 
Our first optical spectrum, at phase $+8.3$~d, is noisy; hence, the narrow metal lines in absorption 
typical of LRNe in this phase cannot be discriminated from noise patterns. We detect narrow emission lines
(H, [O~II], [O~III], [N~II], [S~II]) caused by contamination from host-galaxy H~II regions, along with some bumps in the continuum
which are possibly due to the emerging TiO features (in particular at 5200--5400~\AA). The second spectrum was obtained
two days later, and covers only the NIR domain (Fig. \ref{Fig:specseq18bwo}, top panel). It is characterised by a strong red continuum, but a few broad
absorption features are observed at $\sim 10,900$--11,300~\AA\ (a known combination of CN and TiO features), and at $\sim 12,250$--12,650~\AA\ due 
to AlO and TiO, as proposed by \citet{bla21}.
Combining the $+8.3$~d optical spectrum with the NIR spectrum at $+10.5$~d, we measure the continuum temperature
with a blackbody fit and find it to be $T_{\rm bb} = 3750 \pm 250$~K.

A NIR spectrum was also obtained at  $+21.5$~d (Fig. \ref{Fig:specseq18bwo}, top panel); it shows most of the features detected before, along with a prominent absorption 
band of TiO at 9100--9850~\AA\ \citep{val98}. The AlO plus TiO blend at $\sim 12,250$--12,650~\AA\ is now less evident.
The temperature of the continuum, $T_{\rm bb} = 3850 \pm 300$~K, is similar to that observed 
11~days earlier, and is also consistent with those reported in Fig. \ref{Fig:TLR} at a similar phase.

A second optical spectrum of \object{AT~2018bwo}, with higher S/N, was obtained at $+34.4$~days. In this case, we see a red continuum ($T_{\rm bb} = 3600 \pm 400$~K),
a forest of narrow unresolved metal lines \citep[also detected by][]{bla21}, along with some TiO bands, with the strongest being at 6100--6400~\AA.
The clear detection of narrow absorption lines of Ba~II and Fe~II allows us to estimate the photospheric velocity at this phase, $\sim 220$~km~s$^{-1}$. The H$\alpha$ feature due to the LRN is barely visible, and cannot be disentangled from the narrow H$\alpha$ of a nearby H~II region.

%
   \begin{figure*}
   \centering
   {\includegraphics[angle=270,width=15.0cm]{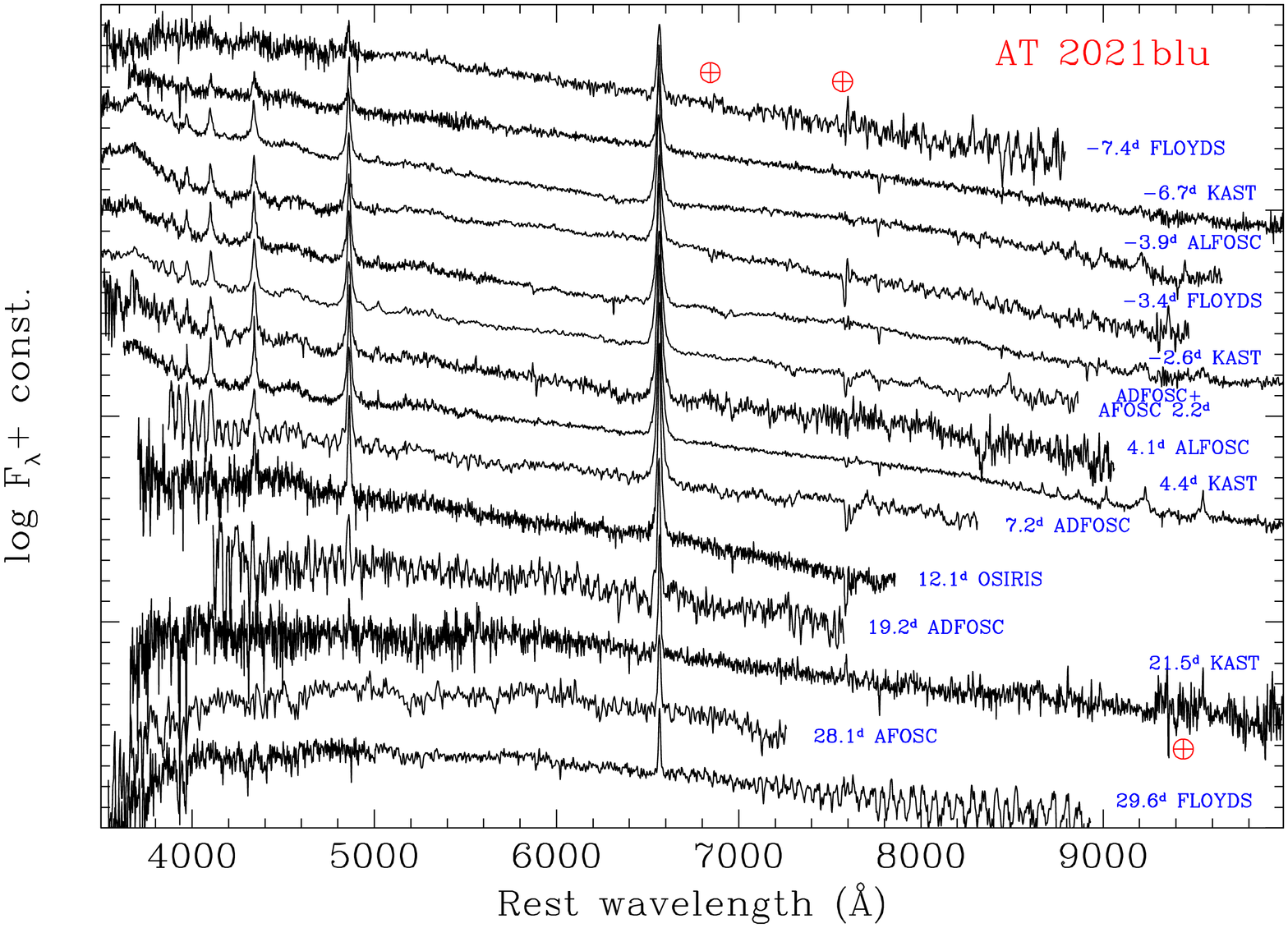}
\includegraphics[angle=270,width=15.0cm]{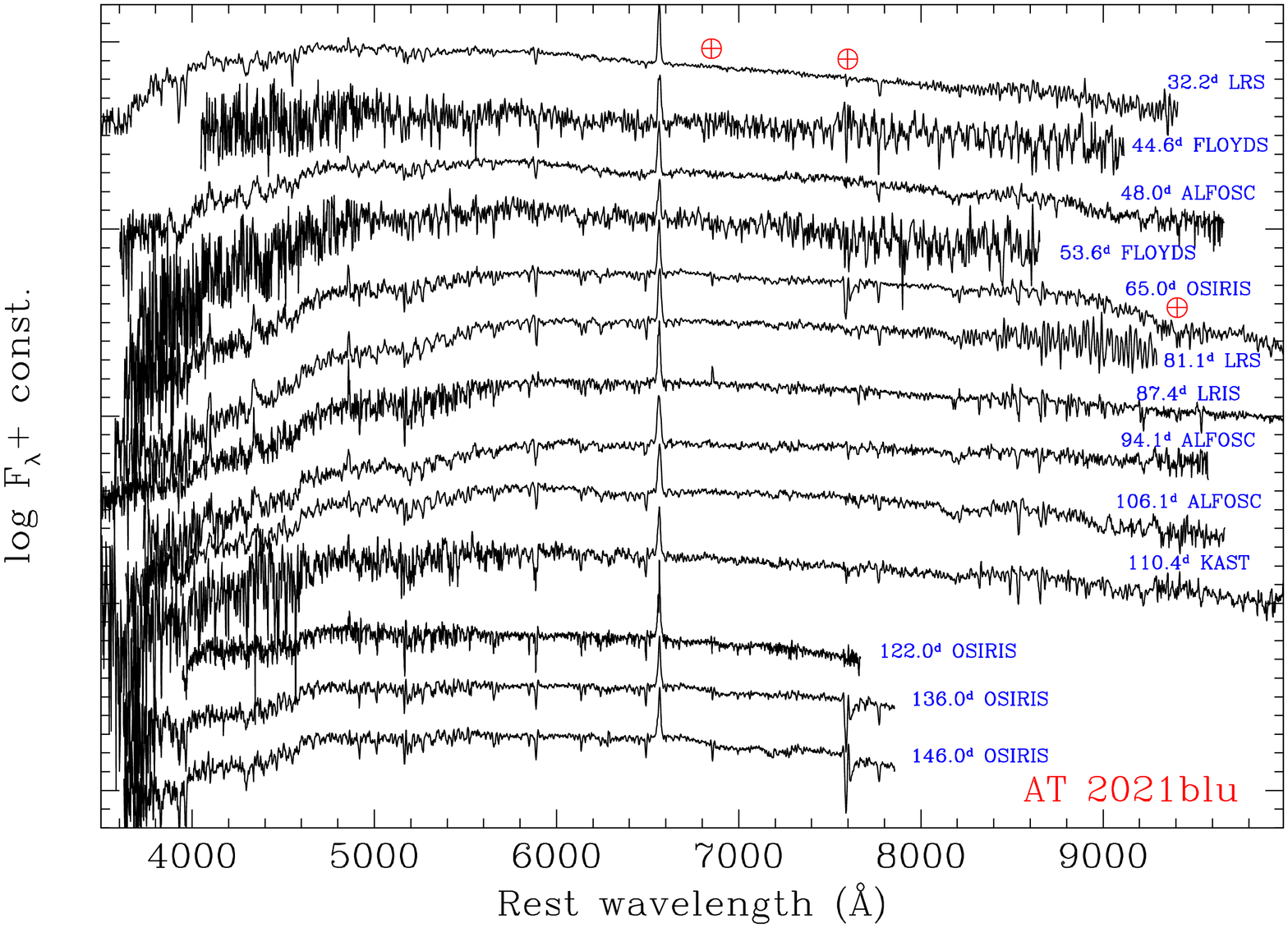}
}
      \caption{Optical spectra of AT~2021blu, corrected for 
        the redshift ($z=0.002098$) and the reddening ($E(B-V)_{\rm MW} = 0.02$~mag). {\sl Top panel:} Spectra from $-7$~d to $+30$~d. {\sl Bottom panel:} Spectra from $+32$~d to $+146$~d. A few spectra are quite noisy, and a poor S/N spectrum at $+16.3$~d is not shown. Residual fringing is visible in the FLOYDS spectra. The instruments and the phases from the $g$-band maximum are also indicated.}
         \label{Fig:specseq21blu}
   \end{figure*}
%

The third optical spectrum (phase $+57.7$~d) taken with the 10~m Keck-I telescope has good S/N. It shows a remarkably red continuum ($T_{\rm bb} = 2750 \pm 200$~K) dominated 
by broad TiO absorption features. Metal lines (with Ba~II being particularly strong) are still well visible. From the position of the minimum of the absorption
metal lines, we infer a photospheric velocity of $\sim 220$~km~s$^{-1}$, still constant, and consistent (marginally higher) with that reported by  \citet{bla21} 
for an almost coeval spectrum. H$\alpha$ has a P~Cygni profile, with an unresolved emission component, and an absorption which is blueshifted by $\sim 500$~km~s$^{-1}$.

Very-late-time optical spectra (at $+102.2$ and $+128.0$~d; see Fig. \ref{Fig:specseq18bwo}, bottom panel) show a continuum flux only above 7300~\AA, along with very pronounced absorption bands
at 7600--8000~\AA, 8200--8750~\AA, 8850--9050~\AA, and above 9200~\AA\ due to TiO, VO, and CN, usually visible at these phases in LRNe \citep[e.g.][]{mar99}.
We also obtained a third NIR spectrum at $+120.4$~d, which is very similar to the spectrum obtained 110.6~d after the first LRN 
detection\footnote{This phase corresponds to $+103.1$~d  adopting their reference epoch.} shown by \citet{bla21}. We confirm the detection of a number 
of molecular bands (TiO, VO, CN, and AlO), along with that of the CO band heads. All of these features are in common with late M-type to early L-type 
cool stars, as reported by \citet{bla21}. However, while we confirm the detection of the broad molecular bands, our spectrum of \object{AT~2018bwo} does 
not convincingly support the detection of the numerous narrow metal lines identified in the late-time NIR spectrum of \citet[][see their Fig. 7]{bla21}.

%
   \begin{figure}
   \centering
   \includegraphics[angle=0,width=8.6cm]{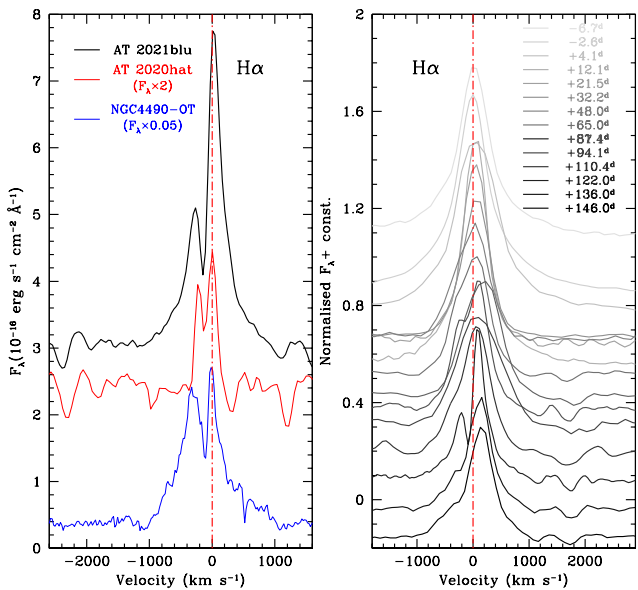}
      \caption{The profile of the H$\alpha$ line \object{AT~2021blu}. {\it Left panel:} Comparison between the H$\alpha$ profile in the highest resolution spectra of \object{AT~2021blu} (at phase $\sim +122$~d),  \object{AT~2020hat} 
 \protect\citep[at phase $\sim +33$~d;][]{pasto21a},
and  \object{NGC4490-2009OT1} obtained almost 200~days after maximum brightness \protect\citep{smi16}.
{\it Right panel:} Evolution of the H$\alpha$ profile in the spectra of  \object{AT~2021blu}.}
         \label{Fig:Halpha}
   \end{figure}
%
   
%
   \begin{figure}
   \centering
   \includegraphics[angle=0,width=8.6cm]{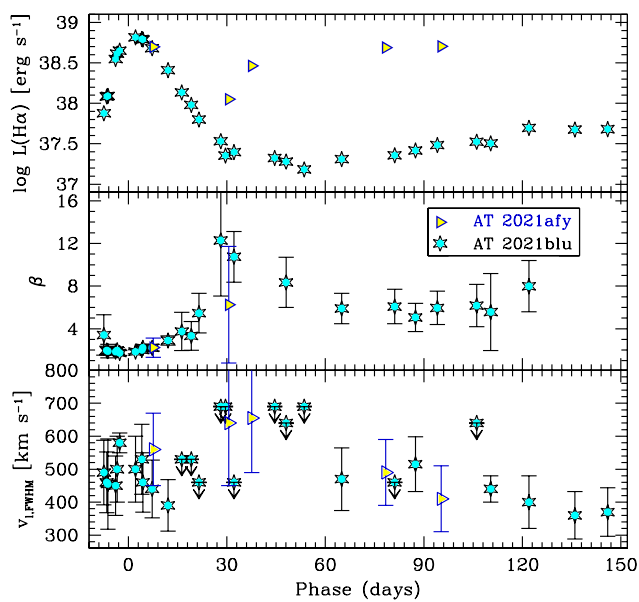}
      \caption{Evolution of the luminosity of H$\alpha$ (top panel), the Balmer decrement $\beta$ (the intensity ratio between H$\alpha$ and H$\beta$ lines;  middle panel), and the FWHM velocity obtained from Lorentzian fits to the H$\alpha$ profile  (bottom panel) in the spectra of \object{AT~2021blu} and  \object{AT~2021afy}. The H line parameters inferred from the classification spectrum of \protect\citet{uno21}, available through the Weizmann Interactive Supernova Data Repository \protect\citep[WISeREP;][]{yar12}, are also included.}
         \label{Fig:Halpha_flux}
   \end{figure}
%

\subsection{AT~2021afy} \label{sect:2021afy_spec}
 
We obtained five GTC+OSIRIS spectra of \object{AT~2021afy}, spanning a period from a week to over 3 months after maximum brightness. The spectral sequence is shown in Fig. \ref{Fig:specseq21afy}.
Deep interstellar absorption of Na~I\,D is present at the host-galaxy redshift, which is attributed  to material along 
the LRN line of sight. Assuming a standard gas-to-dust ratio, we expect a significant extinction of the transient's light 
in the host galaxy.
We measure this Na~I\,D absorption in the three higher-S/N spectra and find an equivalent width (EW) of $2.4\pm0.7$~\AA. Following \citet{tur03}, we 
obtain the amount of host-galaxy extinction from the relation between EW of the Na~I\,D and colour excess, $E_{\rm host}(B-V) = 0.38 \pm 0.11$~mag. 
Accounting for the Milky Way reddening component, we obtain a total colour excess of $E_{\rm tot}(B-V) = 0.43$~mag (see Sect. \ref{Sect:hosts}).

The five spectra, after the correction for the total reddening estimated above, show the typical evolution of LRNe \citep[see, e.g.][]{pasto19a}. 
The spectrum at $+7.6$~d shows a moderately blue continuum with $T_{\rm bb} = 8100 \pm 700$~K, prominent lines of the Balmer series in emission
with a Lorentzian profile and a full width at half-maximum (FWHM) velocity ($v_{\rm FWHM}$) of $\sim 560 \pm 100$~km~s$^{-1}$ (after 
correction for instrumental resolution). Some line
blanketing of metal lines (mostly Fe~II) is likely responsible for the flux drop at wavelengths shorter than $\sim 4500$~\AA. 

The second spectrum, at phase $+30.6$~d, is more noisy. It appears to be slightly redder ($T_{\rm bb} = 7300 \pm 700$~K), and H$\alpha$ is significantly weaker but marginally broader, with 
$v_{\rm FWHM} \approx 640 \pm 190$~km~s$^{-1}$.
A higher-S/N spectrum was taken at $+37.5$~d, and now shows a number of absorption metal lines (Fe~II, Ba~II, Sc~II),
as observed in other LRNe during the second photometric peak \citep{pasto19a,pasto21a}.
 The continuum temperature is $T_{\rm bb} = 6900 \pm 600$~K and H$\alpha$ is still visible in emission, with $v_{\rm FWHM} \approx 655 \pm 165$~km~s$^{-1}$.

The narrow metal lines in absorption become more prominent at  $+78.4$~d, the spectral continuum indicates a much lower temperature ($T_{\rm bb} = 5400 \pm 600$~K), and H$\alpha$
becomes more pronounced, although narrower (with $v_{\rm FWHM} \approx 490 \pm 100$~km~s$^{-1}$). Its profile cannot be well fitted by a Gaussian function, so its FWHM has been
obtained through a Lorentzian fit.
In this phase, the Ca~II NIR triplet is in emission, becoming the second strongest spectral feature.
The last spectrum (at phase $\sim +95.3$~d; $T_{\rm bb} = 5200 \pm 900$~K) has lower S/N, and shows prominent  NIR Ca~II lines and H$\alpha$, the latter 
with  $v_{\rm FWHM} \approx 410 \pm 100$~km~s$^{-1}$.

\subsection{AT~2021blu} \label{sect:2021blu_spec}

   \begin{figure*}
   \centering
  {\includegraphics[angle=270,width=15.8cm]{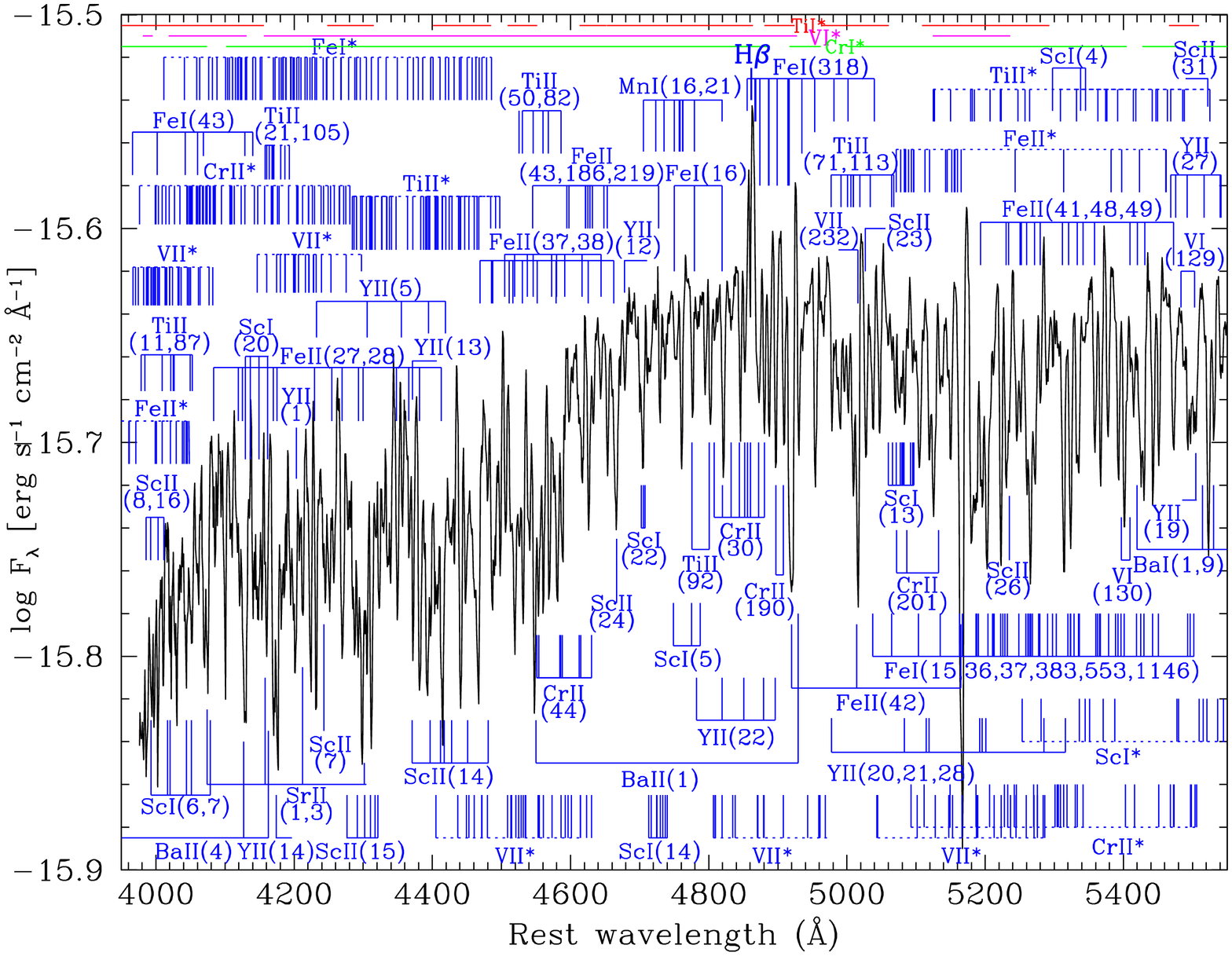}
  \includegraphics[angle=270,width=15.8cm]{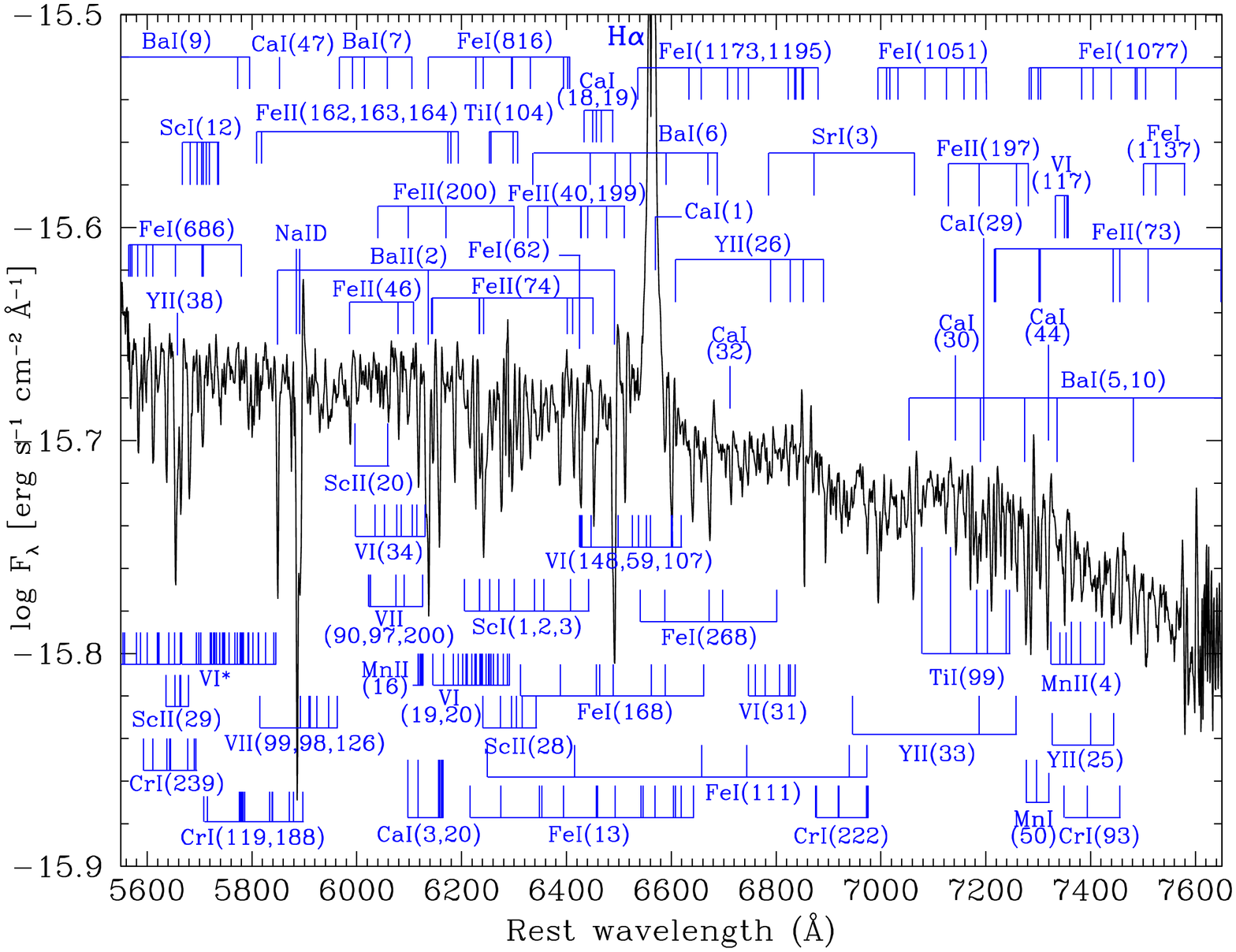}}
      \caption{Line identification on the best-resolution optical spectrum of \object{AT~2021blu} taken at phase +122~d. The spectrum was corrected for the redshift ($z = 0.002098$) and the reddening ($E(B-V)_{\rm MW} = 0.02$~mag). The markers identify the minimum line wavelengths, blueshifted by 250 km s$^{-1}$ from the rest wavelength. For ions marked with the ``$\ast$'' symbol,
individual multiplets are not discriminated in the figure.
}
         \label{Fig:specid21blu}
   \end{figure*}
%

Optical spectra of \object{AT~2021blu} were obtained from $\sim 1$~week before the first blue peak to $\sim 5$~months later, corresponding approximately 
to the time of the second (red) maximum. We collected almost thirty spectra, although not all of them have good S/N. The sequence with the best-quality spectra
is shown in Fig. \ref{Fig:specseq21blu}.

All spectra obtained during the first peak (from $-7.4$~d to $+12.1$~d) are very similar, being characterised by a blue continuum (with $T_{\rm bb}$ in the range between 7500~K and 8000~K)
and Balmer emission lines having Lorentian profiles and a typical $v_{\rm FWHM} \approx 400$--500~km~s$^{-1}$. Paschen lines are also detected in the good-quality 
$+4.4$~d spectrum, along with numerous multiplets of Fe~II in emission. The H lines are marginally resolved, with $v_{\rm FWHM} = 460 \pm 90$~km~s$^{-1}$.
The continuum temperature remains between 7500 and 8000~K over the entire period.

From about $+19.2$~d to $+32.2$~d, the spectra become progressively redder, the  Fe~II emission lines are replaced by absorption features, and H$\alpha$ becomes fainter,
although its profile always remains in pure emission. A residual Lorentzian profile still seems to persist, but the highest-resolution spectra in this phase are only
marginally resolved (with $v_{\rm FWHM} < 460$~km~s$^{-1}$).
In the two spectra at $+21.5$~d and $+22.2$~d, the continuum temperature declines to $T_{\rm bb} = 6500 \pm 600$~K and $T_{\rm bb} = 5950 \pm 250$~K, respectively. Other metal lines are now visible in absorption,
including Fe~II, Sc~II, Ba~II, Na~I\,D, Ca~II (H\&K and the NIR triplet), and O~I.

The subsequent spectra show even more pronounced metal lines (in particular, the Ba~II multiplets), while the continuum temperature continues its decline from
$T_{\rm bb} = 5500 \pm 350$~K at $+48.0$~d
to $T_{\rm bb} = 4350 \pm 350$~K at $+94.1$~d (see also Fig. \ref{Fig:TLR}, top-left panel). In this phase, the profile of the
H$\alpha$ emission line becomes more asymmetric, with a redshifted emission peak. The FWHM velocity at  $+65.0$~d obtained through a Lorentzian fit is $470 \pm 95$~km~s$^{-1}$.

Hereafter, the continuum temperature rises again, reaching $T_{\rm bb} = 5300 \pm 450$~K at $+146.0$~d.
At this phase (starting $\sim 100$~d after the blue peak), the light curve reaches the broader and redder second maximum. 
The spectra are dominated by a forest of metal lines, while H$\beta$, visible until now, disappears in the last available spectra (at phases above $\sim +130$~d). At the same time, the H$\alpha$ profile becomes more markedly 
asymmetric with time (see Fig. \ref{Fig:Halpha}, right panel).

%
\begin{figure}
\centering
{\includegraphics[angle=270,width=9.4cm]{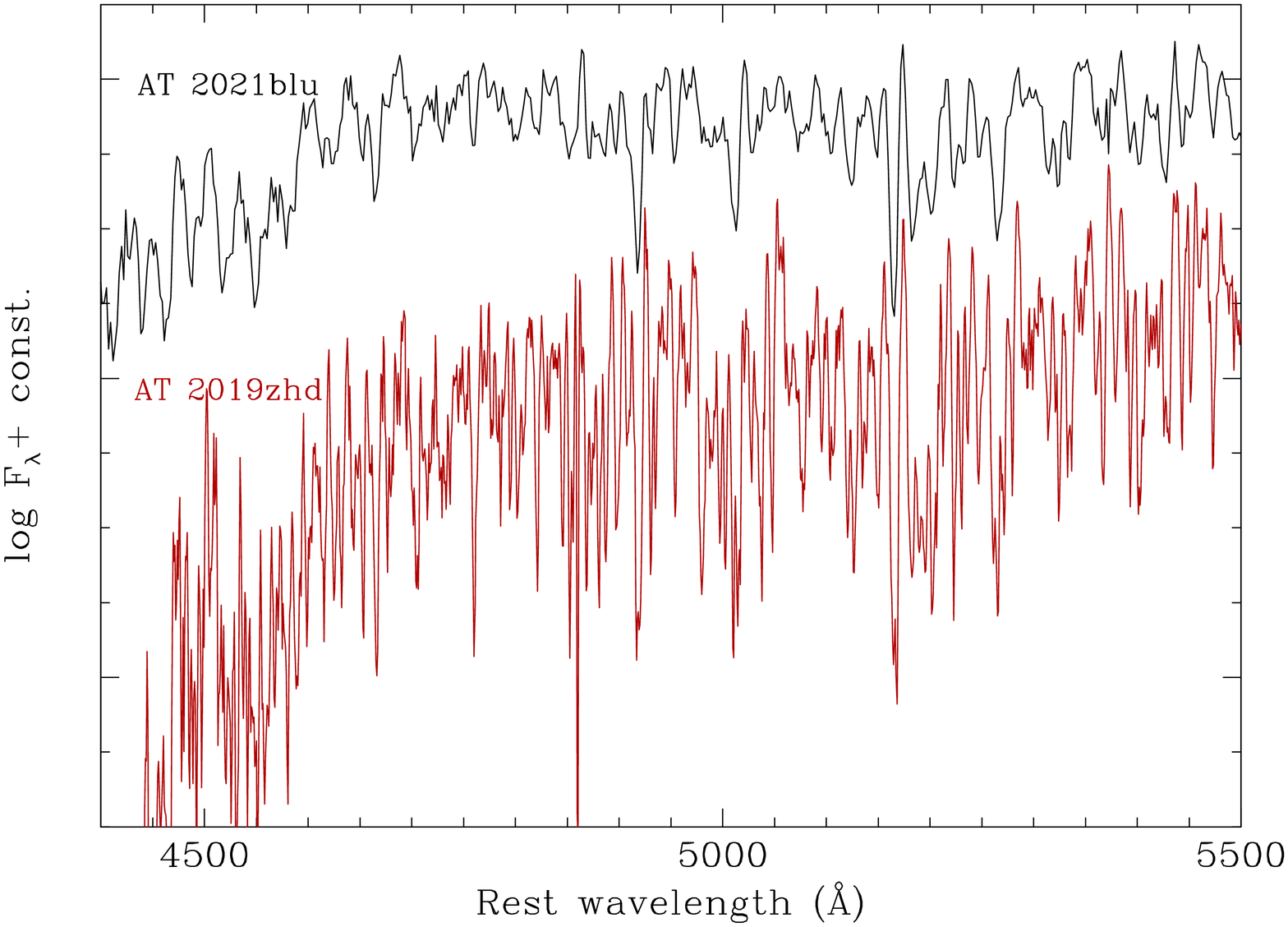}
\includegraphics[angle=270,width=9.4cm]{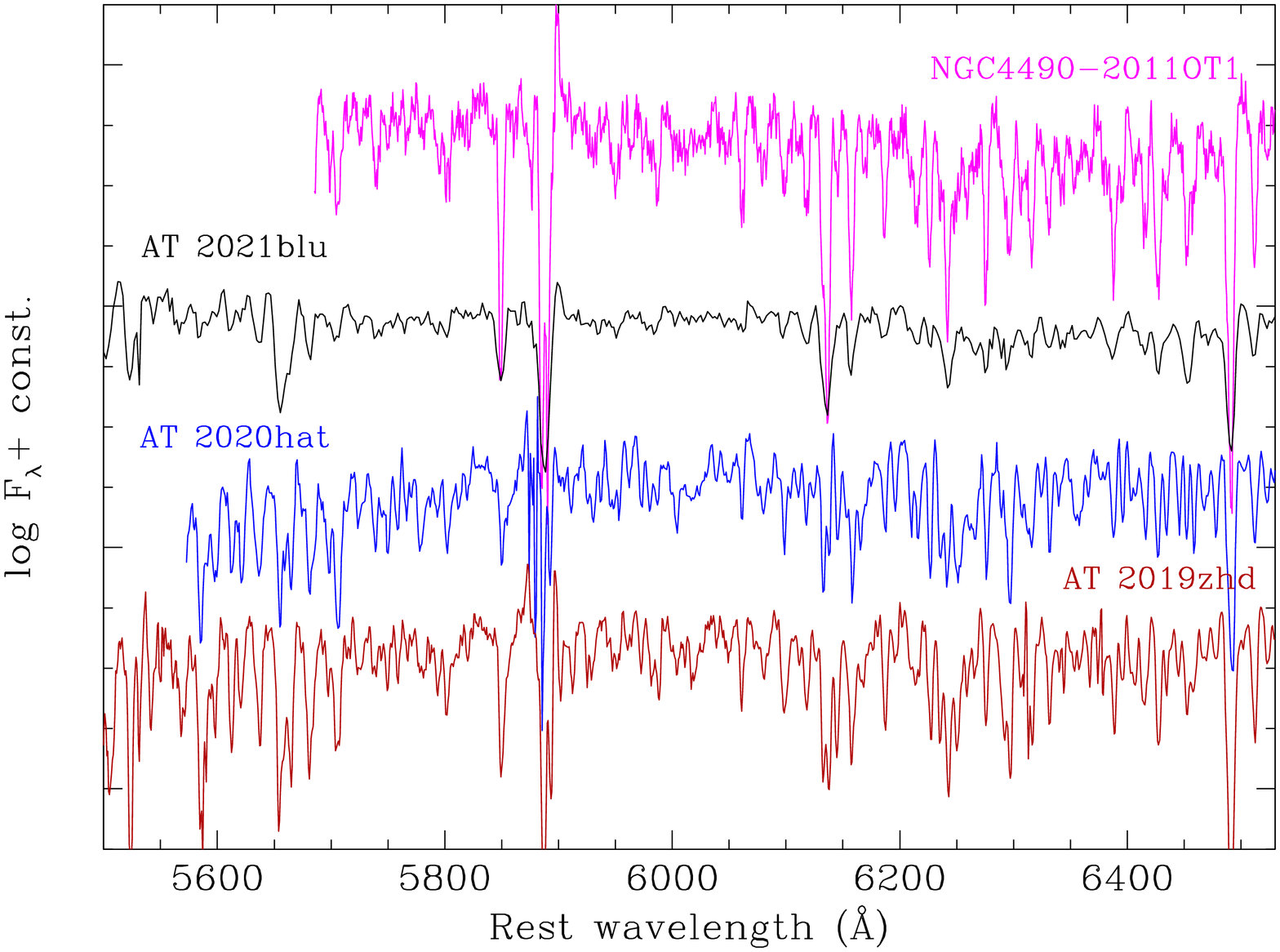}
\includegraphics[angle=270,width=9.4cm]{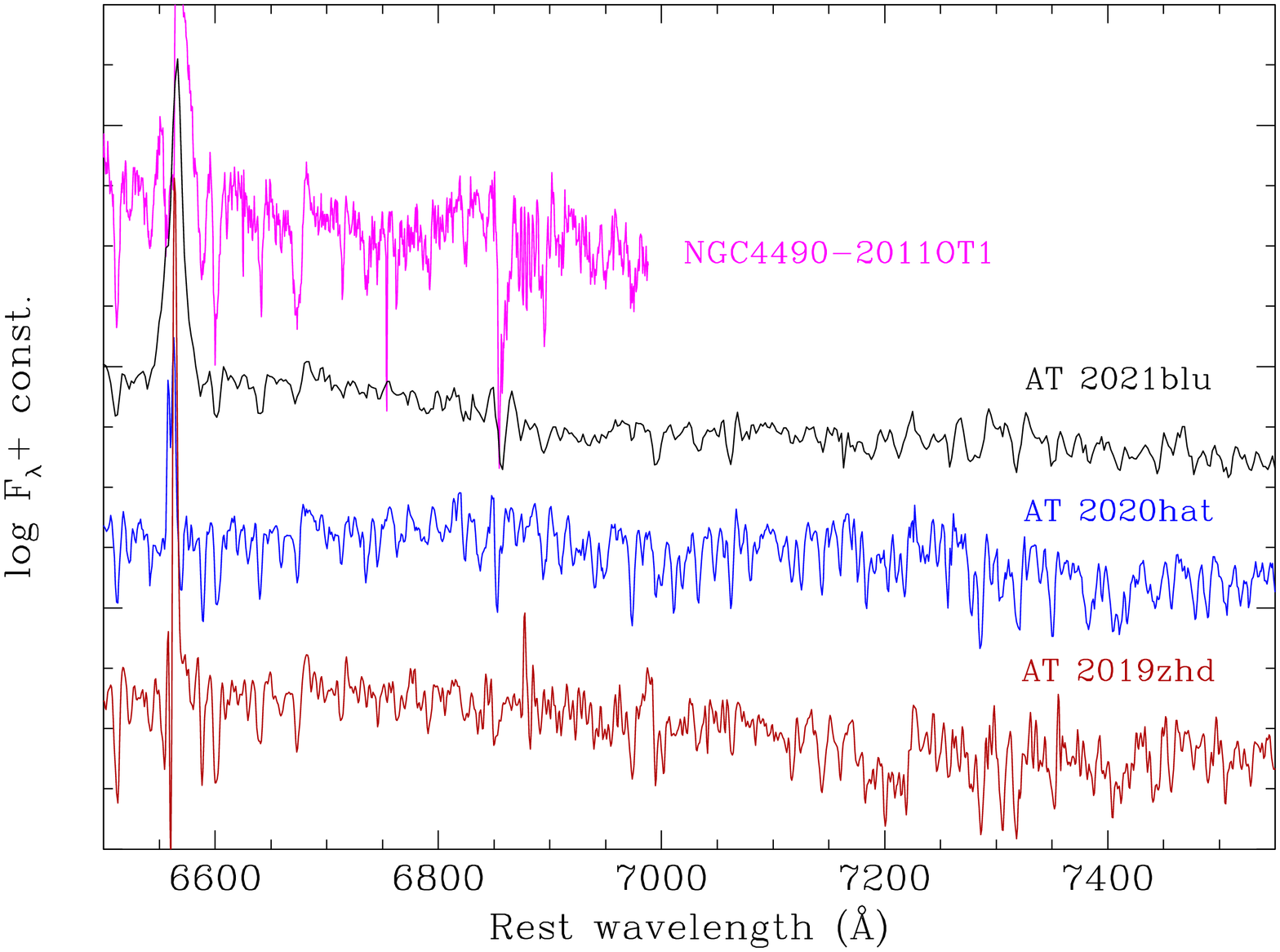}}
\caption{Comparison between medium resolution spectra of \object{AT~2021blu}, \object{NGC4490-2011OT1} \protect\citep{smi16}, \object{AT~2020hat} \protect\citep{pasto21a}, and \object{AT~2019zhd} \protect\citep{pasto21b} obtained during the plateau or the red peak phase. {\sl Top panel:} Close-up view of the region between 4400 and 5500~\AA.  {\sl Middle
 panel:} View of the region from 5500 to 6500~\AA.  {\sl Bottom panel:} View of the region from 6500 to 7570~\AA. Despite the objects are different, most of narrow metal lines are observed in all the spectra. }
         \label{Fig:cfr_highres}
   \end{figure}
%

While our spectroscopic monitoring campaign of \object{AT~2021blu} stopped $\sim 5$~months after maximum brightness, an optical spectrum was obtained 
by \citet{sora22} $\sim 8$~months after maximum, showing the typical TiO bands observed in LRN spectra at late epochs.

The H$\alpha$ luminosity evolution of \object{AT~2021blu} is shown in the top panel of Fig. \ref{Fig:Halpha_flux}, while the evolution of the Balmer decrement (the H$\alpha$/H$\beta$ flux ratio) is reported in the middle panel. 
The values inferred for  \object{AT~2021blu} are compared with those of  \object{AT~2021afy}, while no measure was performed on the  \object{AT~2018bwo} spectra because of the strong contamination of the narrow lines owing to nearby H~II regions.
We note that the evolution of the H$\alpha$ luminosity of both \object{AT~2021blu} and \object{AT~2021afy} roughly follows the global trend of the bolometric light curves.
Except for the very early phases, when the  H$\alpha$ luminosity of the two objects is comparable,
it is systematically fainter in \object{AT~2021blu}.

The evolution of $v_{\rm FWHM}$ for \object{AT~2021blu} and  \object{AT~2021afy}, obtained by fitting the H$\alpha$ line with a Lorentzian function, 
is shown in the bottom panel  of Fig. \ref{Fig:Halpha_flux}. We note that $v_{\rm FWHM}$ has a very slow evolution in both objects, and tends to decrease with time.
The Balmer decrement ($\beta$) of  \object{AT~2021blu} (Fig. \ref{Fig:Halpha_flux}, middle panel) has a minimum value of $\beta \approx 2$ at around the time of the early light-curve peak, and it is similar to that observed in the $+7.2$~d spectrum of  \object{AT~2021afy}. These values are only slightly smaller than those expected from Case B recombination.
The Balmer decrement of \object{AT~2021blu} increases up to $\beta \approx 11$--12 
one month later, then declines to  $\beta \approx 6$ about two months past maximum light, and finally remains nearly constant during the long-lasting light-curve minimum.

%
   \begin{figure}
   \centering
{\includegraphics[angle=270,width=9cm]{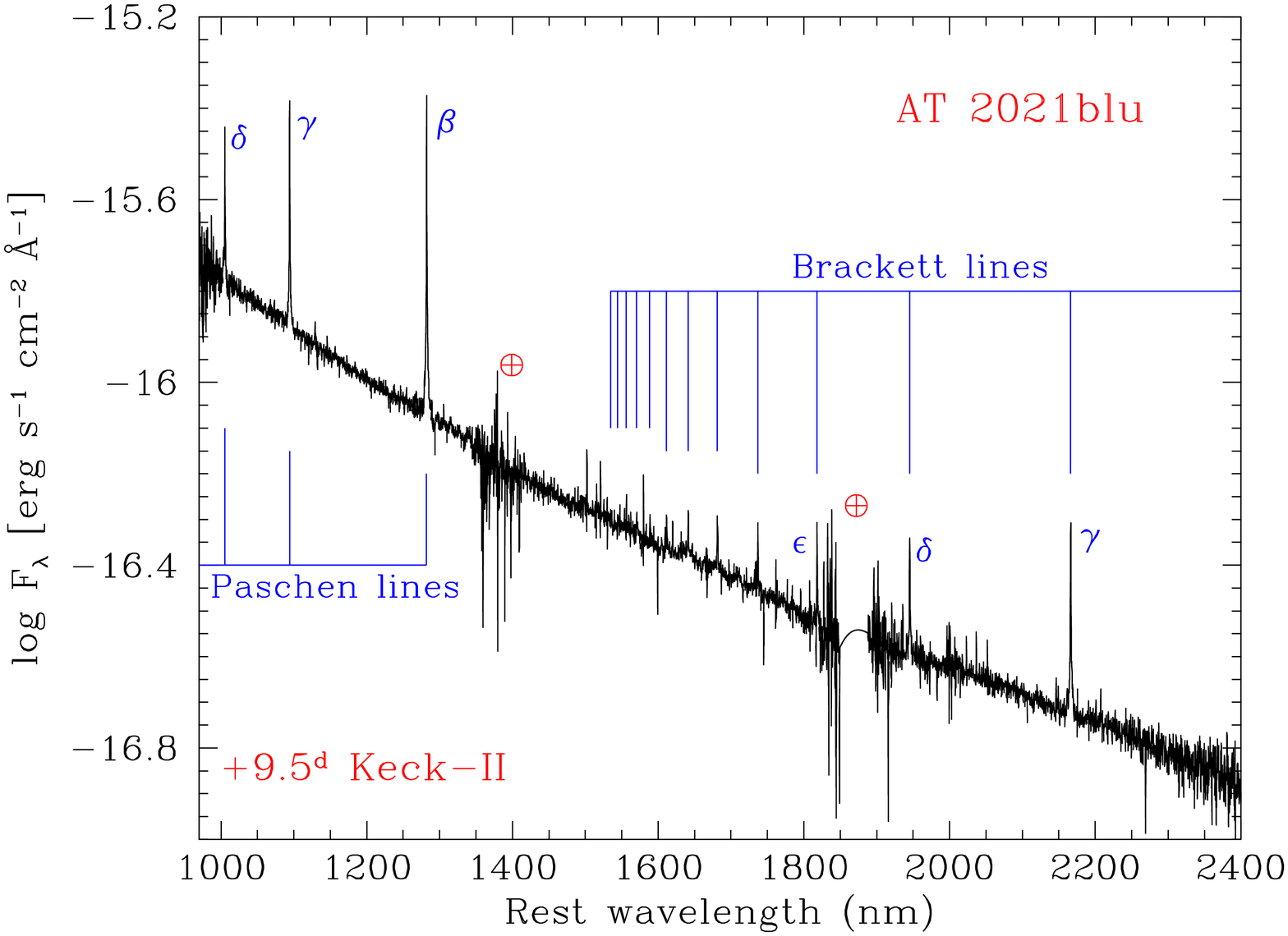}
\includegraphics[angle=270,width=9cm]{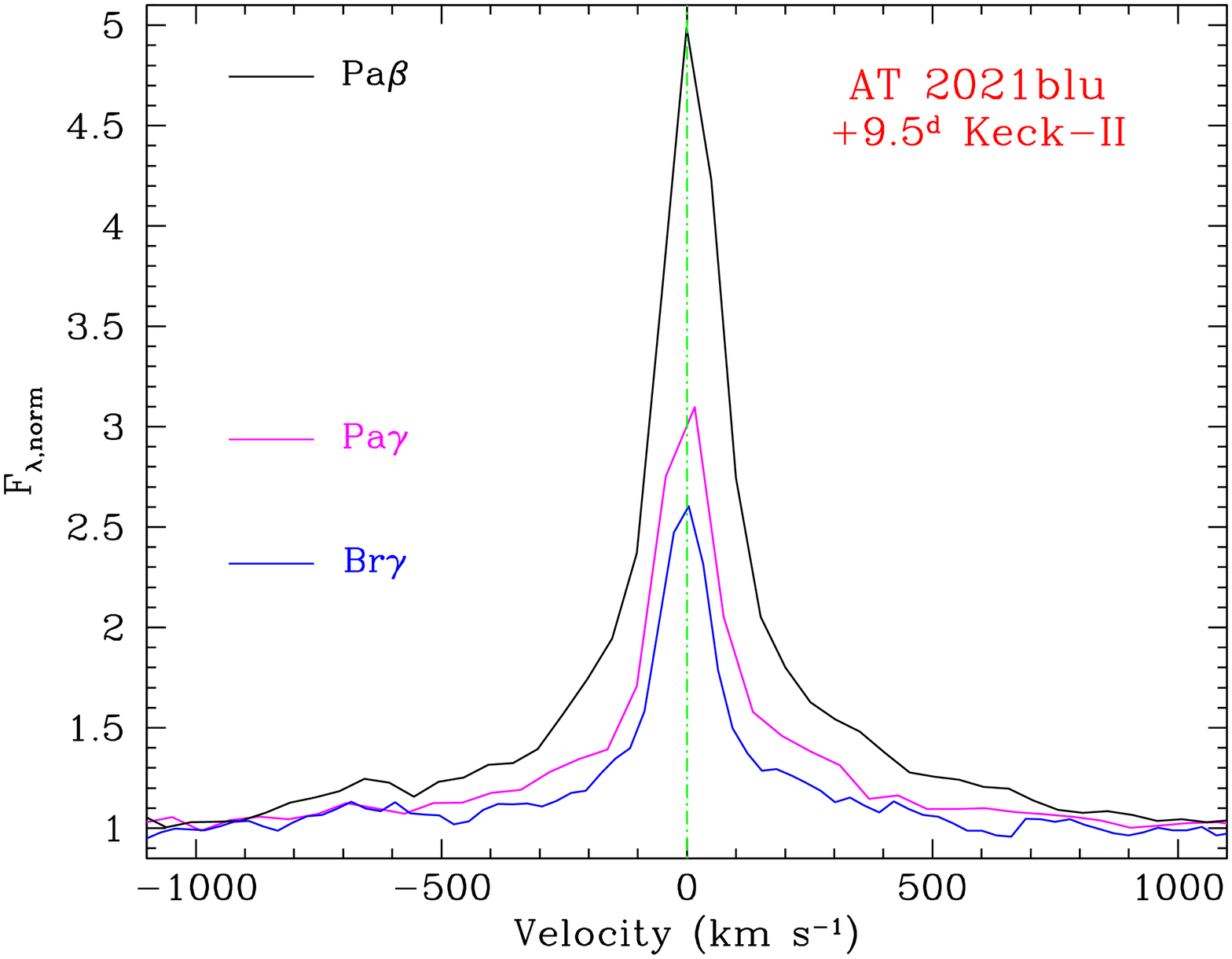}}
      \caption{NIR spectroscopy of \object{AT~2021blu}. {\it Top panel}: Line identification in the NIR spectrum of \object{AT~2021blu} taken about 10~d after maximum brightness with the Keck-II telescope equipped with NIRES. {\it Bottom panel}: Profile of the main H lines in the NIR domain.  The spectrum has been normalised to the flux level of the continuum.}
         \label{Fig:specNIR21blu}
   \end{figure}
%

The high-quality, moderate-resolution GTC spectrum obtained at phase $+122.0$~d reveals the nature of the asymmetry of H$\alpha$. The line is mostly in emission, but
a narrow absorption component is observed, blueshifted by $110 \pm 20$~km~s$^{-1}$ (Fig. \ref{Fig:Halpha}, left panel), similar to that observed in good-resolution spectra of 
LRNe \object{AT~2020hat}  \citep{pasto21a} and \object{NGC4490-2009OT1} \citep{smi16}.
This spectrum of \object{AT~2021blu} allows us to identify the forest of lines visible at the time of the second photometric peak (Fig. \ref{Fig:specid21blu}). 
For the 5600--7600~\AA\ region, we follow the identification  performed in the  \object{AT~2020hat} spectrum presented by \citet{pasto21a}, given the excellent 
match of the lines observed in the two spectra, while for the bluest spectral region, we use the transitions listed by \citet{moo45}. 
The forest of narrow features identified in the \object{AT~2021blu} spectrum in Fig. \ref{Fig:specid21blu} are real metal lines and not noise patterns, as they are also 
detected in the best-resolution spectra of other LRNe (see Fig. \ref{Fig:cfr_highres}) at a similar evolutionary stage. In particular, we find evidence for the  presence of neutral 
and singly ionised Fe, Ti, Cr, Sc, V, Sr, Ba, and Y, along with Mn~II. While the detection of Ca~I lines is only tentative, the main Ca~II lines are outside the range of the $+122.0$~d 
spectrum. However, the H\&K and the NIR triplet of Ca~II are unequivocally detected in the low-resolution spectra at earlier and later epochs. The very strong absorption lines of Ba~II 
allow us to precisely estimate the photospheric velocity as $250 \pm 20$~km~s$^{-1}$.

A NIR spectrum of \object{AT~2021blu} was obtained with the Keck-II telescope equipped with  the Near-Infrared Echellette Spectrometer (NIRES; see Fig. \ref{Fig:specNIR21blu}, top panel)
about 10~d after the first peak. The continuum matches that of a blackbody with $T_{\rm bb} = 6600 \pm 70$~K. Searching for individual features,
we identify only H lines in emission of the Paschen and Brackett series, with a profile which is approximately Lorentzian, with a FWHM velocity of $160 \pm 20$~km~s$^{-1}$  
(Fig. \ref{Fig:specNIR21blu}, bottom panel).

\section{Discussion} \label{Sect:discussion}
From the data for the three transients discussed in this paper, it is evident that LRNe span a very wide range of observational parameters, as 
reported in previous studies \citep[see, e.g.][]{pasto19a,bla21}.
In particular, the light curve of \object{AT~2021afy} exhibits a very small luminosity difference between the two peaks, 
while in \object{AT~2021blu} the luminosity of the early peak is largely predominant over that of the second peak.
Apparently, \object{AT~2018bwo} does not show an early blue peak, although the observations suggest
that the object was discovered in a late stage of its evolution. In this small sample, \object{AT~2021blu} is the object with
the best observational coverage: we constrained its long-lasting phase with a slowly rising light curve prior to the LRN outburst, 
the classical double-peaked light curve of the outburst, and finally a late-time hump in the red optical and NIR light curves. All of this 
makes  \object{AT~2021blu} one of the rare LRNe with comprehensive information on the main photometric parameters along
its entire evolution.

For \object{AT~2018bwo} and \object{AT~2021blu}, we can also constrain the properties of the progenitor system through the
inspection of archival images, when the stars were likely in quiescence. As we subsequently see in Sect. \ref{Sect:progenitors_archive},
this photometric information is crucial for inferring the progenitor mass. Other parameters of the progenitor can be estimated 
through simple models available in the literature (see Sect. \ref{Sect:merging}).
Finally, correlations among observational parameters of LRNe are systematically investigated in Sect. \ref{sect:corr}.

\subsection{Progenitors} \label{Sect:progenitors_archive}

\citet{bla21} performed a detailed analysis of the nature of the stellar system that produced \object{AT~2018bwo}. In particular, they found a yellow source 
at the location of  \object{AT~2018bwo} in the {\it HST} images obtained in 2004, 14~yr before the LRN outburst. At that epoch, the progenitor system was 
assumed to be in a quiescent stage. The progenitor's photometry reported by \citet{bla21}, with our assumptions regarding the host-galaxy distance and reddening, provides 
$M_{F555W} = -5.85$~mag, and colours of  $F435W-F555W = 0.49$~mag and $F555W-F814W = 0.67$~mag. Adopting the standard transformations between 
magnitudes in the natural {\it HST} photometric system and the Johnson-Bessell system (for an F6 star), we obtain $M_V = -5.92 \pm 0.36$~mag. With this absolute 
magnitude, the binary system producing \object{AT~2018bwo} belongs to the intermediate-luminosity population of LRN progenitors. 

As discussed by \citet{bla21}, the absolute magnitude of the quiescent progenitor and the luminosity of the LRN outburst are tightly correlated with the mass of the 
progenitor system. \citet{bla21} compared the photometric parameters of the progenitor of \object{AT~2018bwo} (adopting slightly different reddening assumptions) with both single and 
binary stellar evolution models, and found that the best-matching progenitor was a binary with a massive yellow supergiant primary, whose mass ranged from 
11 to 16~M$_\odot$. The binary interaction then led to the ejection of a common envelope as massive as 0.15--0.5~M$_\odot$ \citep{bla21}. Unfortunately, the 
photometric evolution of the system after the ejection of the common envelope is poorly constrained, as only a shallow upper limit to the total 
magnitude of the system is available at that phase ($M_V \gtrsim -7.5$~mag, using the stacked unfiltered images obtained in mid-2017 by DLT40, and scaled to 
Johnson-Bessell $V$-band photometry). Furthermore, \object{AT~2018bwo} was not observed at early phases because of the gap caused by solar 
conjunction. \citet{bla21} suggested that if the object was not very old when discovered, a very expanded photosphere at the time of the coalescence was 
necessary to explain its initial red colour. However, we  cannot rule out that the object was discovered when it was already at the red peak (or the plateau) 
phase. Our interpretation is supported by the detection of \object{AT~2018bwo} about 1~week before  the discovery epoch (Sect. \ref{sect:2018bwo_lc}), at 
a similar magnitude. In this respect, \object{AT~2018bwo} was likely discovered at a similar evolutionary stage as LRN \object{UGC12307-2013OT1} presented 
by \citet{pasto19a}, where the early blue peak was missed owing to the seasonal gap.

Given the relatively large distance of the host galaxy ($\sim 49.2$~Mpc), we  have limited information about the \object{AT~2021afy} progenitor. {\it HST} imaged the 
LRN field\footnote{Program GO-8645, PI R. Windhorst.} on 2000 September 7. From an inspection of the available $F300W$ and $F814W$ images, no source is visible 
at the LRN location down to $\sim 23.6$~mag and $\sim 23.4$~mag, respectively. Furthermore, public stacked images obtained by Pan-STARRS several years before the 
outburst do not show sources at the location of \object{AT~2021afy}, with upper limits of $g = 23.05$, $r = 23.20$, $i = 23.40$, and $z = 22.81$~mag (Table \ref{Table:A1.4}).
Adopting the Johnson-to-Sloan band transformation relations of \citet{jes05} for normal stars, we obtain an upper detection limit of $M_V > -11.66$~mag for 
the quiescent system. With the ZTF stacked images obtained in mid-2018, shallow upper limits for the slow pre-LRN rise are also derived ($g = 20.95$, $r = 22.05$, 
$i = 21.33$~mag).
Again, using \citet{jes05} conversions, we infer a limit of $M_V > -13.20$~mag for the pre-LRN brightening. This phase is then followed by the classical LRN 
luminosity evolution, characterised by two peaks with almost the same luminosity, separated by a shallow minimum (see Sect. \ref{sect:2021afy_lc}).

%
   \begin{figure}
   \centering
  \includegraphics[angle=0,width=8.2cm]{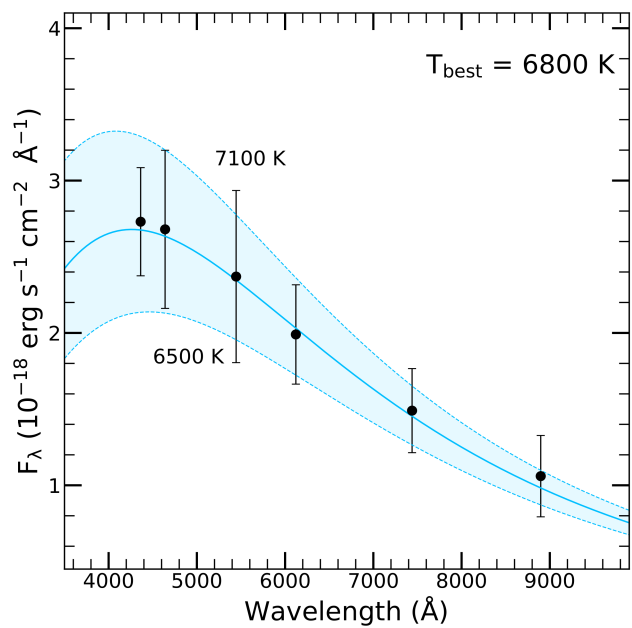}
      \caption{Blackbody fit to the SED of the  \object{AT~2021blu} progenitor candidate.}
         \label{Fig:SEDprogenitor}
   \end{figure}
%

The information available for the quiescent progenitor of \object{AT~2021blu} is less robust than that of \object{AT~2018bwo}. The only pre-outburst {\it HST} 
observation was taken in December 2019 (Sect. \ref{sect:HSTima}), when the object was already in the slow brightening phase. For this reason, we inspected 
earlier images taken with ground-based telescopes and found a source with minor variability across the 2006 to 2016 decade (see Sect. \ref{sect:pre2021blu}). In 2006, we infer the following 
absolute magnitudes and intrinsic colours for the precursor of \object{AT~2021blu}: $M_V = -6.72 \pm 0.30$~mag, $B-V = 0.45 \pm 0.29$~mag, and 
$M_r = -6.76 \pm 0.19$~mag. We also inspected PS1 template images obtained by stacking numerous frames taken before early 2015, and we inferred the 
following reddening-corrected absolute magnitudes and colours: 
$M_g = -6.49 \pm 0.26$~mag, $g-r = 0.27 \pm 0.28$~mag, $r-i = 0.11 \pm 0.27$~mag, $i-z = 0.02 \pm 0.34$~mag, and $M_y > -7.84$~mag.

%
\begin{sidewaysfigure*}   
   \centering
   {\includegraphics[angle=270,width=10.2cm]{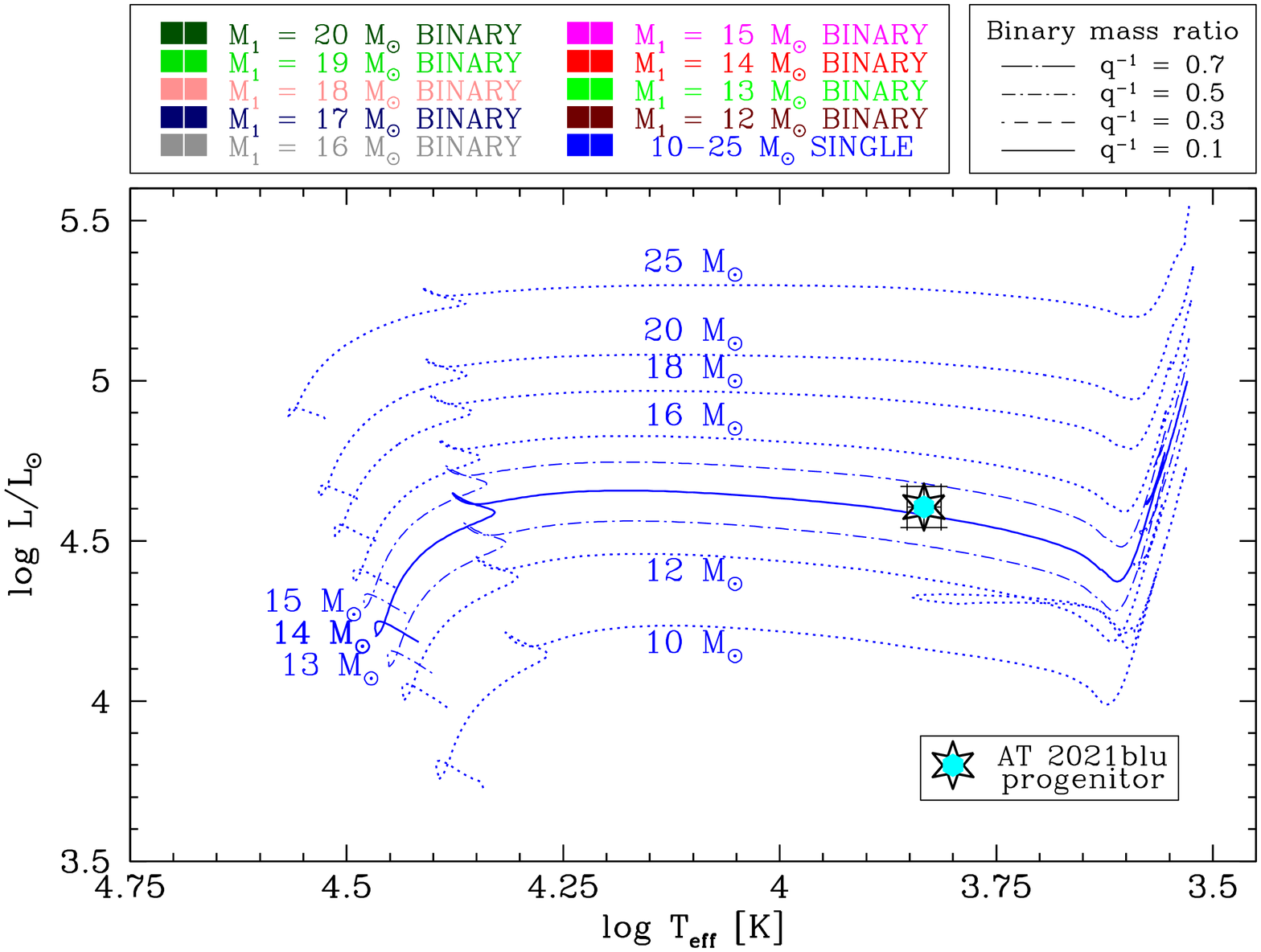}
\includegraphics[angle=270,width=10.2cm]{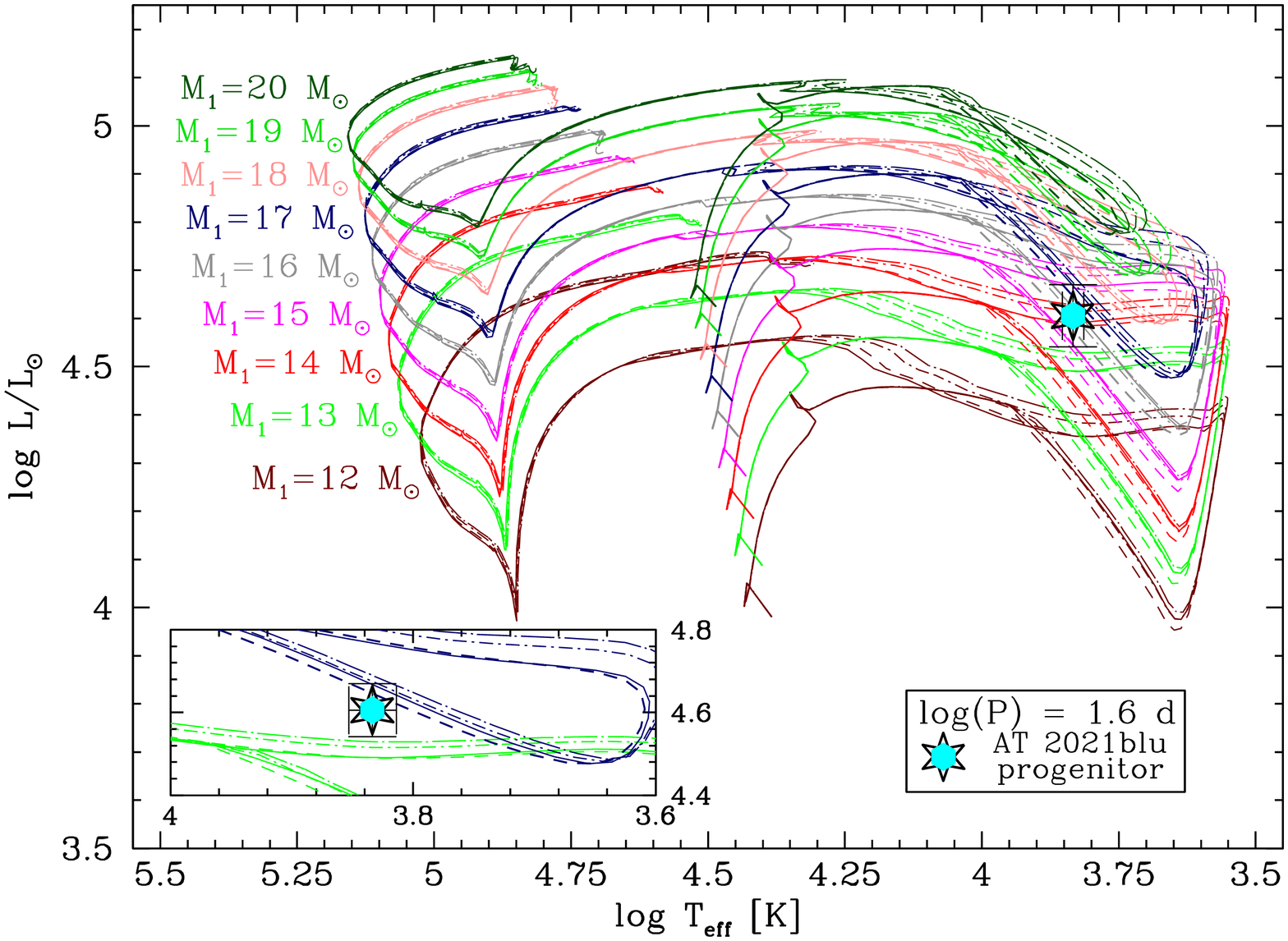}
\includegraphics[angle=270,width=10.2cm]{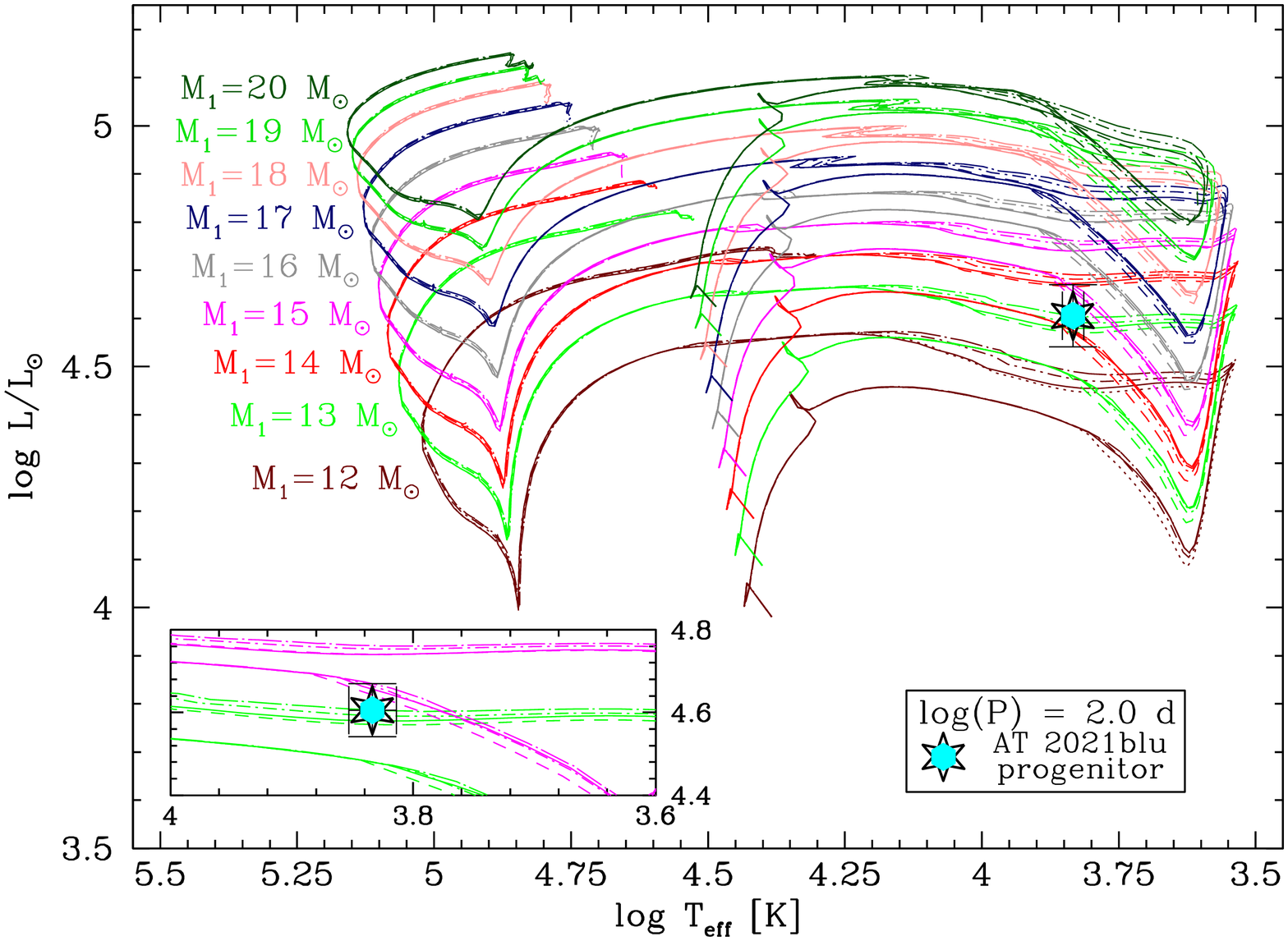}
\includegraphics[angle=270,width=10.2cm]{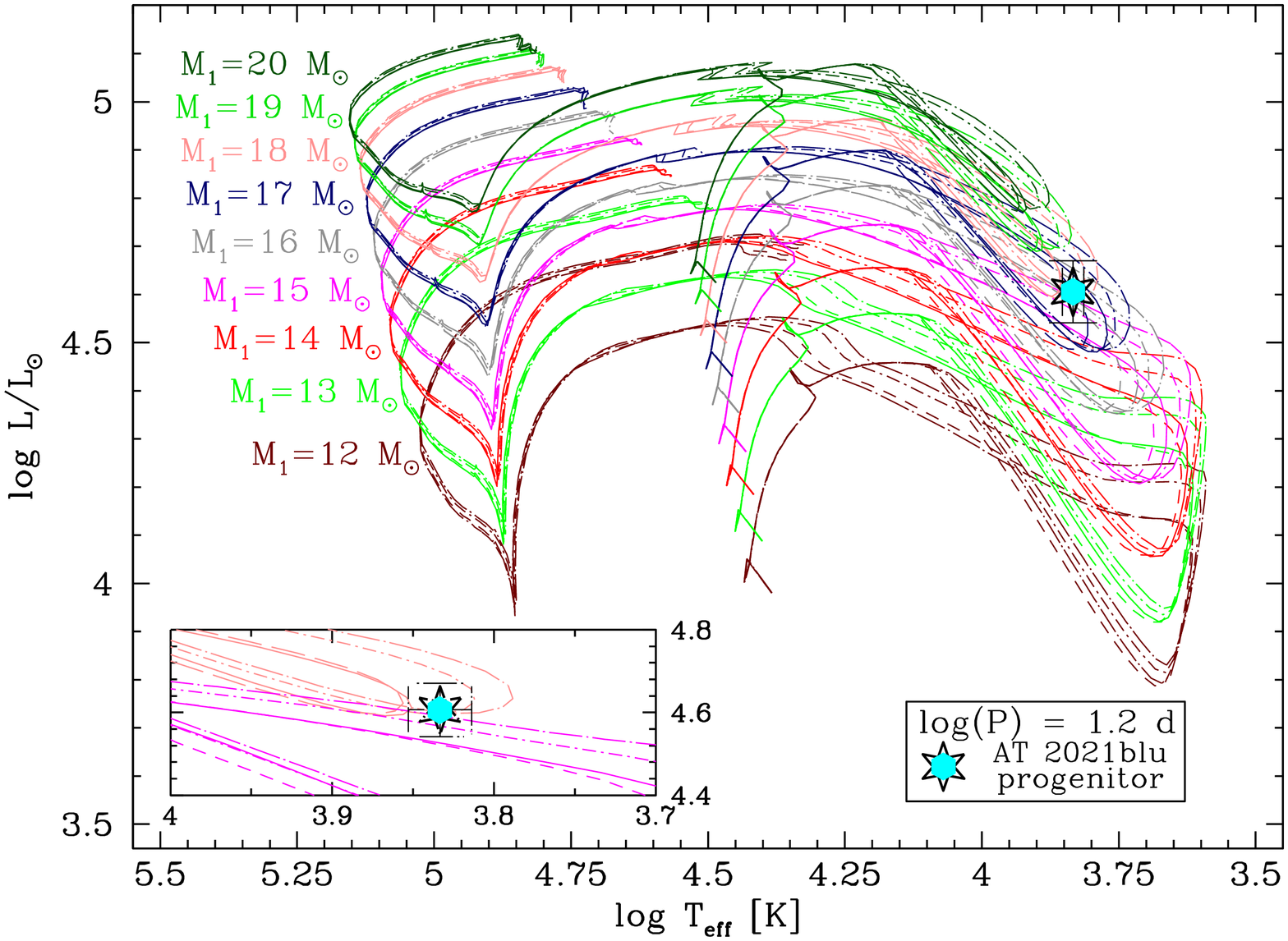}}
      \caption{Location of the \object{AT~2021blu} progenitor in the HRD (cyan starred symbol), and comparison with different solar-metallicity evolutionary 
tracks.
{\it Top-left panel:} Comparison with evolutionary tracks for single stars. The track  of a 14~M$_\odot$ star is indicated with a blue solid line, those of 13 and 15~M$_\odot$ with dot-dashed lines,
all other tracks with blue dotted lines. {\it Bottom-left panel:} Comparison with tracks for binary systems with initial orbital periods $P \approx 100$~days
($log(P) = 2$). {\it Top-right panel:}  Comparison with tracks for binary systems with initial orbital periods $P \approx 40$~days ($log(P) = 1.6$).
{\it Bottom-right panel:} Comparison with tracks for binary systems with initial orbital periods $P \approx 15$~days ($log(P) = 1.2$).
Binary tracks with primaries of 12 to 20~M$_\odot$ are shown as lines with different colours. Binary tracks for different mass ratios 
($q = M_1/M_2$, where $M_1$ indicates the mass of the primary star and $M_2$ that of the secondary star) are also shown. 
All tracks are taken from the BPASS database \protect\citep[][]{eld17,sta18}. The insets show a close-up view of the location of the \object{AT~2021blu} progenitor 
in the HRD, along with the tracks of different configurations of binaries of 13--18~M$_\odot$ that intersect
 the error bars of the \object{AT~2021blu} progenitor.
}
         \label{Fig:HRDprogenitor}
\end{sidewaysfigure*}

First, we assume that the measured source is the progenitor star and that the flux contamination of nearby stars is negligible (see Sect. \ref{sect:HSTima}). 
The main parameters of this source can be estimated by fitting the SED with a blackbody function, as detailed in Sect. \ref{sect:TLR}. The best fit to the SED 
is obtained with a blackbody of $T_{\rm eff} = 6800 \pm 300$~K (Fig. \ref{Fig:SEDprogenitor}). Given the above colours and accounting for the error bars, the 
source at the progenitor's location corresponds to a star of F3--F4 spectral type. We also infer  $L_{\rm bol} = (1.55 \pm 0.23) \times 10^{38}$~erg~s$^{-1}$ and 
$R_{0}= 144 \pm 14$~R$_\odot$ for the putative progenitor of \object{AT~2021blu}. 

We now discuss the issue of the flux contamination from nearby sources in the photometry of the \object{AT~2021blu} progenitor. In Sect \ref{sect:HSTima}, we estimated that on 2019 
December 29 the flux of the contaminating sources within a radius of $1''$ from the transient was $\sim6\%$ in $F606W$ and $\sim10\%$ in $F814W$ of the
 LRN precursor flux. If we consider the flux of the quiescent progenitor in the Sloan $r$ and $i$ bands during the 2006--2016 period, the total flux of other 
stars within $1''$ is estimated to be about $18\%$ and $32\%$ (respectively) of the progenitor flux. Although {\it HST} did not observe the field of \object{AT~2021blu} in 
blue filters, we note that the contaminating sources are significantly redder than the \object{AT~2021blu} progenitor. For this reason, the contamination 
is expected to be modest in the blue bands. Removing the contribution of the contaminating source would probably make the progenitor slightly bluer, changing 
the intrinsic colour to $r-i \approx -0.04$~mag and thus shifting its classification towards an early-F star \citep[see, e.g.][]{fin07,fuk11}. However, 
given that precise information on contaminating sources is only available for two {\it HST} filters, hereafter we assume that the flux of the source observed
from 2005 to 2016 is largely dominated by the progenitor's contribution, with the caveat that the progenitor is possibly slightly hotter ($T \approx 7200$~K) 
and fainter.

To constrain the mass of the \object{AT~2021blu} progenitor, we made use of a grid of BPASS evolutionary-track models \footnote{The tracks are taken from 
\url{https://bpass.auckland.ac.nz/index.html}.} for single stars and binary systems at solar metallicity \citep{eld17,sta18}, and plotted them in the 
Hertzsprung-Russell diagram (HRD). Single-star models from 10 to 25~M$_\odot$ are shown in Fig.~\ref{Fig:HRDprogenitor} as blue dotted lines, along with 
binary models with primary stars having ZAMS mass ($M_1$) ranging from 12 to 20~M$_\odot$ (the tracks for the different stellar masses are shown with 
different colours). For each value 
of $M_1$, we report tracks computed for different mass ratios of the two members of the binary ($q^{-1} = M_2/M_1 = 0.1$, 0.3, 0.5, and 0.7) and for three 
indicative initial orbital periods ($P \approx 15$, 40, and 100~days). The cyan starred symbol in Fig.~\ref{Fig:HRDprogenitor} represents the photometric point 
of the \object{AT~2021blu} progenitor obtained without removing the contribution of stars within $1''$. Single-star models for $M_1 = 14 \pm 1$~M$_\odot$ 
provide an excellent match to the observed photometry of the candidate progenitor of \object{AT~2021blu}. Evolutionary tracks for binary systems also well 
match the position of the observed progenitor in the HRD, in particular binaries whose primary has a mass ranging between 13 and 16~M$_\odot$. We note, 
however, that even systems with more massive primaries are consistent with the progenitor's photometry when the initial orbital period decreases, as we can 
expect for the binary progenitor of \object{AT~2021blu}. Consequently, if we include systems with $log(P) = 1.2$ (nearly 15~days), the range of possible 
masses for the primary star widens to 13--18~M$_\odot$ (see the insets in Fig.\ref{Fig:HRDprogenitor}).

Unfortunately, the mass of the secondary companion is poorly constrained, as the evolutionary 
tracks are less sensitive to its mass; hence, we can only provide crude limits to the total binary mass, which lie in the 
$13\leq M/M_\odot<36$ range\footnote{The lower binary mass limit is computed assuming $M_1 = 13$~M$_\odot$ and $q^{-1} \ll 0.1$, while the upper limit is computed assuming 
$M_1 = 18$~M$_\odot$ and $q^{-1} \lesssim 1$.}.
We remark that the above mass estimates assume that the observed progenitor in quiescence is not affected by significant circumstellar reddening. 
Additional reddening would make the progenitor more luminous and hotter, hence leading to a larger mass. Furthermore, removing the flux of the contaminants 
within $1''$ would shift the location of the progenitor in the HRD to a slightly higher effective temperature and a marginally lower bolometric luminosity, 
without significantly changing the mass estimates.

\citet{koc14} noted the existence of a possible correlation between the LRN absolute magnitude at maximum brightness and the progenitor mass, which was later 
confirmed by \citet{bla21} based on a wider compilation of data from the literature. The analysis of  \citet{bla21} has been recently revised by 
\citet{cai22}, with different assumptions about the distance and the reddening, and after adopting Johnson-Bessell $V$ as the reference band. Finally, 
\citet{cai22} considered the magnitude of the second peak (or the plateau) instead of the magnitude of the first peak, as the former is likely dependent on 
the mass of the recombining hydrogen, while the latter is probably more sensitive to the parameters (mass and velocity) of the high-velocity gas ejected in the
polar direction during the merging process \citep{met17}. 

The empirical relation between the absolute magnitudes in the $V$ band at the second peak (or the plateau) and the mass (weighted by the uncertainties) obtained 
by \citet{cai22} is

\begin{equation} \label{eq1}
$$log\,\biggl(\frac{M_{\rm prog}}{{\rm M}_\odot}\biggl) = (-0.162\pm0.020)\,M_V -(0.701\pm0.048) $$.
\end{equation}
\noindent
This relation can be used to infer an independent estimate of the mass of the LRN progenitors when the early-time light curve is not available. The masses of progenitors 
of LRNe with known photometric information during the second peak (or the plateau) inferred using Eq. \ref{eq1} are reported in Table \ref{tab:param_LRNe} 
(Column 12). For \object{AT~2021blu}, we obtain $M_{\rm prog} = 19.3_{-9.5}^{+18.6}$~M$_\odot$, consistent (within the large uncertainties) with the mass 
derived through the comparison of the archival progenitor imaging with the evolutionary tracks discussed above. This makes more plausible our suggestion 
that the faint source imaged in archival frames at the \object{AT~2021blu} location was dominated by the flux of the LRN  progenitor. As a consistency check, 
we note that \object{AT~2021blu} is marginally brighter than \object{AT~2015dl}  \citep[whose progenitor system was estimated to have a primary
of $18\pm1$~M$_\odot$;][]{bla17}; thus, a total mass of $\sim 19$~M$_\odot$ is a realistic estimate for the primary progenitor of  \object{AT~2021blu}.

For  \object{AT~2021afy} we infer a much larger progenitor mass, as expected from the high luminosity of its second light-curve peak: 
$M_{\rm prog} = 46_{-25}^{+54}$~M$_\odot$. We note, however, that the large error in the $V$-band absolute magnitude at maximum (Sect. \ref{Sect:hosts}, and Table \ref{tab:param_LRNe})
makes this mass estimate quite uncertain. At the adopted distance, \object{AT~2021afy} is slightly less luminous than  \object{SNhunt248} \citep{kan15,mau15,mau18} at the 
first brightness maximum, and is somewhat similar to \object{NGC4490-2011OT1} \citep{smi16,pasto19a} (see Table \ref{tab:param_LRNe}). 
For this reason, we expect that its binary progenitor system belongs to the massive edge of the LRN distribution.

\subsection{The merger scenario} \label{Sect:merging}

While there is a consensus that the LRN phenomenon is an outcome of common-envelope binary evolution 
\citep[][and references therein]{iva17,bar17,jon20}, whether the two stars merged or rearranged into a new stable binary configuration is more  debated \citep{how20}. 
Convincing arguments favouring the final merging scenario were provided by the detailed study of  the Galactic LRN \object{V1309~Sco} \citep{mas10,tyl11,mas22}.
Specifically, as mentioned in Sect. \ref{sect:intro}, the decade-long follow-up observations of \object{V1309~Sco} performed by the OGLE survey, and the thorough
observational studies by \citet{tyl11}, revealed the multiple stages leading to the LRN eruption. \object{V1309~Sco} showed a long-lasting phase of slow 
luminosity rise with a superposed binary variability, with a period of $\sim 1.44$~days. Then, the photometric period started to shorten as a consequence of the 
loss of systemic angular momentum, finally leading to the orbital decay. In 2007, the light curve showed a sudden decline, and the signatures of binary 
modulation disappeared. This was interpreted as the consequence of an outflow of material from the primary that generated an optically thick, expanding 
envelope. The common envelope engulfed the binary system, hiding the binary modulated variability.  As a consequence of a continuous optically thick outflow 
\citep{pej14}, the photometric minimum was followed by a gradual luminosity rise which lasted about half a year. In that phase, \object{V1309~Sco} brightened by 
$\sim 4$~mag. A steep brightening by a further $\sim 4$~mag in about 10 days followed, and was attributed to the initial thermal energy from the outer, high-velocity,
 hot ejecta launched in the polar direction during the coalescence of the secondary's core onto the primary \citep[e.g.][]{met17}.

Although such high-cadence monitoring is not available for other LRNe, fragments of the sequence of the physical processes leading to an LRN outburst were 
observed for a number of extragalactic objects\footnote{Extensive datasets of bright LRNe were provided by 
\protect\citet{bla17,mau15,mau18,kan15,wil15,kur15,smi16,gor16,lip17,bla17,bla20,bla21,cai19,cai22,pasto19a,pasto19b,pasto21a,pasto21b,str20}.}. All of them are 
intrinsically more luminous and longer lasting than \object{V1309~Sco}, but the general morphology of the light curve is similar. In particular, for the closest 
events, we monitored the slow brightening phase after the common-envelope ejection, which lasted up to a few years (see Fig. \ref{Fig:21blu_lc}, top panel, 
for \object{AT~2021blu}), followed by a rapid brightening to the first maximum, and a subsequent luminosity decline to a plateau (or a new rise to a second, 
much broader and redder maximum). This particular morphology of the light curve was discussed in a number of studies. \citet{mcl17} proposed that the first
light-curve maximum is produced by a violent, merger-driven, high-velocity gas outflow. Then, a rapid luminosity decline is followed by a plateau or a second 
broad peak, usually explained by the recombination of the H-rich gas \citep{iva13b,mcl17}, with most of the LRN energy being radiated during the plateau phase. 
This interpretation is supported by the effective temperature showing a minor evolution in this phase, the H$\alpha$ emission component disappearing, and the 
spectrum becoming dominated by narrow absorption lines of metals. However, some scatter in the effective  plateau temperatures can be noticed among the 
objects, with $T_{\rm eff}$ spanning from 3000~K to 6000~K, suggesting that shell-shell interaction or even a further mass ejection \citep{iva13b} can also 
contribute to sustaining the light curve during this phase. 

%
\begin{sidewaystable*}
\centering
\tiny
\caption{\label{tab:param_LRNe} Photometric parameters for the complete LRN sample.}
\begin{tabular}{lccccccccccc}
\hline\hline
LRN name                   &  Host      & $t_{\rm host}$  & Distance       & $E(B-V)_{\rm tot}$  & $M_V$(prog)      & $M_V$(CE)        & $M_{\rm pk1}$        & $M_{\rm pk2}$       & $L_{\rm pk1}/L_{\rm pk2}$ & $\Delta~t_{1~{\rm dex}}$ &  $M_{\rm prog}$         \\
                           &  galaxy    &             & (Mpc)          &    (mag)         &     (mag)          &  (mag)             &    (mag)           &    (mag)          &  optical                &    (days)          &  (M$_\odot$)        \\ \hline
\object{AT~2021afy}        & UGC~10043  & $4.2\pm0.7$ & $49.2\pm8.0$     & $0.43\pm0.11$  & $>-11.66\pm0.56$ & $>-13.20\pm0.56$ & $-14.44\pm0.57$  & $-14.57\pm0.56$ &   1.0            &   120            & $46_{-25}^{+64}$      \\   
\object{UGC12307-2013OT1}  & UGC~12307  & $9.8\pm1.0$ & $39.7\pm2.8$     & $0.22\pm0.02$  & $>-11.88\pm0.17$ &  --              &    --            & $-15.03\pm0.42$ &   --             &    --            & $54_{-30}^{+67}$      \\   
\object{AT~2017jfs}        & NGC~4470   & $1.9\pm2.1$ & $35.2\pm2.7$     & $0.02\pm0.01$  & $>-11.26\pm0.17$ &  --              & $-15.46\pm0.46$  & $-14.38\pm0.17$ &   2.6            &   157            & $43_{-23}^{+50}$      \\   
\object{SNhunt248}         & NGC~5806   & $3.2\pm0.8$ & $22.5\pm3.8$     & $0.04\pm0.01$  & $-8.99\pm0.36$   & $-11.18\pm0.36$  & $-14.87\pm0.36$  & $-14.07\pm0.36$ &   2.1            &   169            & $58\pm2~(\star)$      \\
\object{AT~2020kog}        & NGC~6106   & $5.3\pm0.6$ & $22.5\pm4.7$     & $0.37\pm0.07$  & $>-9.82\pm0.50$  & $-11.17\pm0.53$  & $-13.17\pm0.51$  & $-12.68\pm0.51$ &   2.0            &   $>$100         & $23_{-11}^{+23}$      \\    
\object{AT~2018hso}        & NGC~3729   & $1.3\pm0.8$ & $21.3\pm0.6$     & $0.30\pm0.08$  & $-9.05\pm0.25$   &  --              & $-13.89\pm0.28$  & $-12.16\pm0.26$ &   3.7            &   201            & $18.6_{-9.1}^{+17.7}$ \\    
\object{NGC3437-2011OT1}   & NGC~3437   & $5.2\pm0.5$ & $20.9\pm4.2$     & $0.02\pm0.01$  & $>-9.98\pm0.43$  & $>-10.83\pm0.43$ & $-13.06\pm0.48$  & $-13.33\pm0.43$ &   0.9            &   174            & $29_{-15}^{31}$       \\    
\object{AT~2014ej}         & NGC~7552   & $2.4\pm0.7$ & $20.6\pm1.5$     & $0.31\pm0.15$  & $>-8.22\pm0.50$  &  --              & $>-14.70\pm0.50$ & $-14.36\pm0.50$ &  $>2.2$          &   $>98$          & $42_{-23}^{+49}$      \\    
\object{NGC4490-2011OT1}   & NGC~4490   & $7.0\pm0.4$ & $9.6\pm1.3$      & $0.32\pm0.32$  & $-7.32_{-1.03}^{+1.10}$ & $-9.18_{-1.10}^{+1.17}$ & $-14.35_{-1.00}^{+1.08}$& $-14.54_{-1.00}^{+1.08}$ & 0.9 & 200 & $30_{-22}^{+50}~(\star)$\\ 
\object{AT~1997bs}         & NGC~3627   & $3.1\pm0.4$ & $9.2\pm0.3$      & $0.21\pm0.04$  & $-7.61\pm0.21$   &  --              & $-13.34\pm0.15$  & $-11.51\pm0.17$ &   3.2            &   62             & $14.6_{-6.9}^{+13.1}$ \\    
\object{AT~2021blu}        & UGC~5829   & $9.8\pm0.6$ & $8.64\pm0.61$    & $0.02\pm0.01$  & $-6.72\pm0.30$   & $-8.33\pm0.43$   & $-13.06\pm0.15$  & $-12.26\pm0.15$ &   2.4            &   242            & $19.3_{-9.5}^{+18.6}$ \\    
\object{AT~2021biy}        & NGC~4631   & $6.5\pm0.7$ & $7.46\pm0.50$    & $0.27\pm0.02$  & $-7.93\pm0.17$   & $-8.78\pm0.20$   & $-13.81\pm0.16$  & $-12.65\pm0.16$ &   2.7            &   375            & $20.5\pm3.5~(\star)$ \\
\object{AT~2018bwo}        & NGC~45     & $7.8\pm0.7$ & $6.79\pm1.13$    & $0.02\pm0.01$  & $-5.92\pm0.36$   & $-7.47\pm0.36$   &  --              & $-10.14\pm0.45$ &   --             &   $>72$          & $13_{-2}^{+3}~(\star)$\\   
\object{AT~2015dl}         & M~101      & $5.9\pm0.3$ & $6.43\pm0.57$    & $0.01\pm0.01$  & $-7.19\pm0.36$   & $-10.10\pm0.47$  & $-12.70\pm0.21$  & $-11.46\pm0.31$ &   2.0            &   229            & $18\pm1~(\star)$     \\ 
\object{AT~2020hat}        & NGC~5068   & $6.0\pm0.4$ & $5.16\pm0.21$    & $0.09\pm0.01$  & $-2.99\pm0.09$   & $-8.87\pm0.76$   & $-10.72\pm0.27$  & $-10.08\pm0.26$ &   1.5            &   131            & $8.5_{-3.7}^{+6.6}$   \\    
\object{AT~2019zhd}        & M~31       & $3.0\pm0.4$ & $0.785\pm0.009$  & $0.055\pm0.005$& $0.17\pm0.14$    & $-5.74\pm0.28$   & $-9.08\pm0.13$   & $-7.59\pm0.32$  &   5.1            &   27             & $3.4_{-1.3}^{+2.0}$   \\    
\object{M31LRN2015}        & M~31       & $3.0\pm0.4$ & $0.785\pm0.009$  & $0.35\pm0.11$  & $-2.25\pm0.47$   & $-5.41\pm0.42$   & $-10.12\pm0.42$  & $-9.13\pm0.42$  &   2.4            &   62             & $4.0_{-1.0}^{+1.5}~(\star)$\\ 
\object{M31RV}             & M~31       & $3.0\pm0.4$ & $0.785\pm0.009$  & $0.12\pm0.02$  & $-5.04\pm0.32$   & $>-7.04\pm0.15$  & $-9.54\pm0.15$   & $-8.66\pm0.15$  &   2.0            &   110            & $5.0_{-2.0}^{+3.3}$   \\    
\object{CK~Vul}$(\dag$)    & Galaxy     & -- &$3.2_{-0.6}^{+0.9}\times10^{-3}$& $0.80\pm0.15$ & $>-9.0_{-2.4}^{+1.3}$ & -- & $-12.0_{-2.4}^{+1.3}$ & $-12.4_{-2.4}^{+1.3}$ & 0.7        &   400            & --                    \\   
\object{V838~Mon}          & Galaxy     & -- &$6.1(\pm0.6)\times10^{-3}$ & $0.85\pm0.02$  & $-1.29\pm0.22$   & $-6.67\pm0.22$   & $-9.76\pm0.22$   & $-9.43\pm0.22$  &   1.7            &    82            & $8\pm3~(\star)$         \\  
\object{V4332~Sgr}         & Galaxy&--&$3.85_{-1.57}^{+4.65}\times10^{-3}$& $0.32\pm0.10$ & $3.94_{-1.72}^{+1.14}$ & --         & --               & $-5.21_{-1.93}^{+1.33}$ &   --     &    --            & $1.0\pm0.5~(\star)$     \\  
\object{V1309~Sco}         & Galaxy     & --    &$3.5(\pm1.5)\times10^{-3}$& $0.70\pm0.15$& $3.33\pm1.04$    & $-1.39\pm1.04$   & $-7.02\pm1.04$   & $-5.48\pm1.04$  &   3.1            &    29            & $1.54\pm0.5~(\star)$    \\  
\object{OGLE-2002-BLG-360} & Galaxy     & -- &$8.20(\pm0.15)\times10^{-3}$&$1.0\pm0.2~(\ddag)$& $2.23\pm0.50$& $-1.43\pm0.54$   & $ 0.10\pm0.54$   & $ 0.79\pm0.65$  &   1.0            &   837            & $0.15_{-0.01}^{+0.01}$\\    
\object{OGLE-2002-BLG-360} & Galaxy     & -- &$8.20(\pm0.15)\times10^{-3}$&$1.0\pm0.2~(\ddag)$& $1.13\pm0.28$& $-3.54\pm0.28$   & $-4.56\pm0.28$   & $-4.65\pm0.30$  &   1.0            &   837            &  -- \\ 
\hline
\end{tabular}
\\
\tablefoot{The table reports the LRN name (Column 1); the host-galaxy name (Column 2) and its morphological type 
code (from Hyperleda; Column 3); the distance (Column 4); the total colour excess (Column 5); the $V$-band absolute magnitude of the quiescent progenitor (Column 6), 
the brightest $V$ absolute magnitude of the pre-outburst phase (Column 7), the first light-curve peak (Column 8), and the second light-curve peak (Column 9); the optical luminosity 
ratio of the first to the second light-curve peak (Column 10); time taken by the LRN to decrease its luminosity by one order of magnitude from the peak 
(Column 11); and the progenitor mass estimate using Eq.~\ref{eq1} or taken from \protect\citet{cai22} (Column 12).\\
 $(\star)$ Mass estimates obtained through the direct detection of the progenitor or via light-curve modelling, as adopted by \protect\citet{cai22}.
$(\ddag)$ The Milky Way reddening towards \object{OGLE-2002-BLG-360} follows a non-standard reddening law \protect\citep[$E(B-V)_{\rm MW} = 1$~mag, with 
$R_V=2.5$;][]{nat13}. While the parameters for the $I$-band photometry are robust, those inferred for  $V$ photometry are uncertain owing to the 
low-cadence follow-up observations in that filter. The $V$ magnitudes are obtained through an interpolation of the $V-I$ colour curve at the epochs of the $I$-band 
peaks. $(\dag$) In this table, the parameters for \object{CK~Vul} are taken from \protect\citet{ban20}.}
\end{sidewaystable*}
%

\citet{mat22} recently presented accurate one-dimensional models of LRN light curves which improve on previous studies based on \citet{pop93} approximations. 
The models of \citet{mat22} assume that the short-lasting initial blue peak is due to thermal energy release from the low-mass, fast outer ejecta 
dominated by radiation pressure, while the second long-duration red peak emission is powered by hydrogen recombination. This study offers a grid of light-curve
models showing two luminosity peaks, remarkably similar to those observed for LRNe. Following  \citet{mat22}, we can estimate the LRN parameters during 
the first peak. In particular, the ejected mass  ($M_{\rm ej}$) is inferred from

\begin{equation} \label{eq2}
$$ \frac{M_{\rm ej}}{10^{-2}~{\rm M}_\odot} \approx \frac{v_{\rm ej}}{500~{\rm km~s}^{-1}} \times \biggl(\frac{t_{\rm pk1}}{6.7~{\rm d}}\biggl)^{2} $$,
\end{equation}

\noindent where $t_{\rm pk1}$ is the duration of the first peak and $v_{\rm ej}$ is the velocity of the outer ejecta. The launching radius ($R_0$) is given by

\begin{equation} \label{eq3}
$$ \frac{R_0}{10~{\rm R}_\odot} \approx \frac{L_{\rm pk1}}{10^{39}~{\rm erg~s}^{-1}} \times \biggl(\frac{v_{\rm ej}}{500~{\rm km~s}^{-1}}\biggl)^{-2} $$,
\end{equation}

\noindent where $L_{\rm pk1}$ is the bolometric luminosity of the first light-curve peak. Finally, an upper limit to the energy radiated during the first peak ($E_{\rm pk1}$) 
can be obtained from

\begin{equation} \label{eq4}
$$ E_{\rm pk1} \approx L_{\rm pk1} \times t_{\rm pk1} $$.
\end{equation}

We use Eq.~\ref{eq2}-\ref{eq4}  to infer the early physical parameters of the \object{AT~2021blu} ejecta, adopting the following values for the observed parameters:
$v_{\rm ej} = 460$~km~s$^{-1}$ (see Sect. \ref{sect:2021blu_spec}, for the ejecta velocity at early phases), $L_{\rm pk1} = 6.5 \times 10^{40}$~erg~s$^{-1}$, and 
$t_{\rm pk2} \approx 30$~days. We obtain $M_{\rm ej} \approx 0.18$~M$_\odot$, $R_0 \approx 770$~R$_\odot$, and $E_{\rm pk1} \approx 1.7 \times 10^{47}$~erg. We note that 
the above launching-radius estimate is reasonably similar to that inferred from the blackbody fit to the pre-outburst SED in Sect. \ref{sect:TLR} (see 
Fig. \ref{Fig:TLR}, top-right panel).

The same calculation can be performed for  \object{AT~2021afy}, taking into account that the distance towards \object{UGC~10043} adopted in this paper is affected by 
a large uncertainty (Sect. \ref{Sect:hosts}). From the observed parameters $v_{\rm ej} = 560$~km~s$^{-1}$, $L_{\rm pk1} = 2.05 (\pm 0.61) \times 10^{41}$~erg~s$^{-1}$, and 
$t_{\rm pk1} \approx 30$~days, we infer $M_{\rm ej} \approx 0.22$~M$_\odot$ and $R_0 = 1640\pm490$~R$_\odot$, while 
the upper limit to the energy radiated during the first peak is $E_{\rm pk1} = 5.3 (\pm 1.6) \times 10^{47}$~erg. 
Hence, although the mass of the fast and hot ejecta is similar in the two objects, the energy radiated during the first peak is a least a factor of two (up to four) times  
higher in \object{AT~2021afy} than in \object{AT~2021blu}.

After the early peak, the light curve reaches a minimum before rising again to the second broad maximum, which is mostly powered by hydrogen recombination.
We still follow \citet{mat22} to describe the recombination phase. The mass of the recombining hydrogen shell ($M_{\rm rec}$) is obtained through the relation
 \citep[equivalent to Eq.~16 in][assuming that H recombines at a characteristic constant density of $\rho_{\rm rec} \approx 10^{-11}$~g~cm$^{-3}$]{mat22}

\begin{equation} \label{eq5}
$$ \frac{M_{\rm rec}}{{\rm M}_\odot} \approx \biggl(\frac{t_{\rm pk2}}{140~{\rm days}} \times \frac{v_{\rm rec}}{300~{\rm km~s}^{-1}}\biggl)^3 $$.
\end{equation}

For \object{AT~2021blu}, we assume a recombination phase lasting $t_{\rm pk2} = 200$~days and a luminosity $L_{\rm pk2} \approx 3.1 \times 10^{40}$~erg~s$^{-1}$
during the second peak. More tricky is the choice of the velocity of the recombining material ($v_{\rm rec}$). 
If $v_{\rm rec} \approx v_{\rm FWHM}({\rm H}\alpha) = 360$~km~s$^{-1}$ (at the time of the second peak; see Sect. \ref{sect:2021blu_spec} and Fig. \ref{Fig:Halpha_flux}), 
we obtain $M_{\rm rec} \approx 5$~M$_\odot$. If we instead adopt the photospheric velocity from the minimum of the absorption metal lines (250~km~s$^{-1}$), 
we infer a much smaller mass value of $M_{\rm rec} \approx 1.2$~M$_\odot$. We remark that the above mass estimates should be regarded as upper limits, obtained 
through the crude assumption that H recombination is the only source powering the light curve at this phase.

In the case of \object{AT~2021afy}, we adopt the following parameters during the second peak: $t_{\rm pk2} = 50$~days, $L_{\rm pk2} \approx 2.1 \times 10^{41}$~erg~s$^{-1}$,
and $v_{\rm rec} \approx 550$~km~s$^{-1}$. This last value is obtained from a weighted average of $v_{\rm FWHM}({\rm H}\alpha)$ measured in the spectra of 
\object{AT~2021afy} from +30 to +80~d after the first peak. We find that $M_{\rm rec} \approx 0.3$~M$_\odot$, about one order of magnitude smaller than the 
mass of the recombining material inferred for \object{AT~2021blu}.

Although we have poor constraints on the epoch of the \object{AT~2018bwo} outburst onset, we can tentatively estimate the mass of the recombining gas from the 
observed parameters during the plateau. We adopt $v_{\rm rec} \approx 500$~km~s$^{-1}$ (Sect. \ref{sect:2018bwo_spec}) and $L_{\rm pk2} \approx 10^{40}$~erg~s$^{-1}$ 
(Sect. \ref{sect:TLR}), while the minimum plateau duration is $t_{\rm pk2} = 40$~days. With these assumptions, we obtain $M_{\rm rec} > 0.1$~M$_\odot$. With a plateau duration 
of at least 60~days, the mass rises to $M_{\rm rec} \approx 0.4$~M$_\odot$, which is within the range of ejected mass extremes (0.02--2~M$_\odot$) determined 
by \citet{bla21} using different calibration methods (see their Sect.~5.3).

Finally, using the values of $t_{\rm pk2}$ and  $L_{\rm pk2}$ reported above, the upper limits to the total energy radiated during the second peak can be derived from

\begin{equation} \label{eq6}
$$ E_{\rm rec} \approx L_{\rm pk2} \times t_{\rm pk2} $$.
\end{equation}
\noindent
Applying Eq.~\ref{eq6} to \object{AT~2021blu}, we obtain  $E_{\rm rec} \approx 5.3 \times 10^{47}$~erg, which is equivalent to the energy radiated during the first peak, and it is smaller 
than the value inferred for \object{AT~2021afy} ($E_{\rm rec} \approx 9 \times 10^{47}$~erg)\footnote{If we account for the error in the luminosity at the second peak,
$L_{\rm pk2} = 2.1(\pm0.6) \times 10^{41}$~erg~s$^{-1}$, the upper limit to the total energy radiated by \object{AT~2021afy} in this phase ranges from $0.65$ to $1.15 \times 10^{48}$~erg.}. 
Assuming a plateau duration of 60~days, for \object{AT~2018bwo} we infer $E_{\rm rec} \approx 5 \times 10^{46}$~erg, which is over one order of magnitude less than that of \object{AT~2021blu}.

Once hydrogen has fully recombined, the luminosity abruptly declines, analogous to what is observed in Type IIP supernovae at the end of the plateau. Without radioactive 
material powering the light curve as happens in supernova explosions, we expect that the LRN bolometric light curve settles onto the nearly constant luminosity 
threshold of the resulting merger, although occasionally late-time humps can be observed, especially in the NIR domain, probably consequences of late-time 
interaction with confined circumstellar shells. This was observed in  \object{AT~2021blu}, \object{AT~2021biy}, \citep[][and Fig. \ref{Fig:LRN_absolute}]{cai22}, as well as earlier in 
\object{AT~2017jfs} \citep{pasto19b}.

%
\begin{table*}
\caption{\label{tab:param_LRNe2} Additional observed parameters for a sub-sample of LRNe followed during the early blue peak. }
\centering
\begin{tabular}{lccccc}
\hline\hline
LRN name & $L_{\rm bol,peak}$ & $ L_{+7~{\rm d}}({\rm H}\alpha)$ & $v_{{\rm FWHM},+7~{\rm d}}({\rm H}\alpha)$ & $T_{\rm eff,peak}$ & $R_{\rm ph,peak}$ \\
         &  ($10^{39}$~erg~s$^{-1}$) &  ($10^{37}$~erg~s$^{-1}$) & (km~s$^{-1}$) & (K) & (au) \\
\hline \hline
AT~2021afy       &  $205\pm61$        &   $53.0\pm6.8$   &  $560\pm110$ &   $6910\pm810$  &  $24\pm5$        \\   
AT~2017jfs       &  $552\pm172$       &  $390.6\pm36.9$  &  $<820$      &   $7190\pm260$  &  $36\pm6$        \\   
SNhunt248        &  $309\pm110$       &  $513.5\pm21.3$  &  $440\pm100$ &   $7310\pm70$   &  $26\pm5$        \\   
AT~2020kog       &   $89\pm13$        &   $28.7\pm2.9$   &  $460\pm50$  &  $10860\pm1090$ &   $6.3\pm1.0$    \\   
AT~2018hso       &  $109\pm27$        &   $38.5\pm9.2$   &  $<625$      &   $8060\pm280$  &  $12.7\pm1.3$    \\   
NGC3437-2011OT1  &   $61\pm27$        &   $95.7\pm12.5$  &  $<740$      &   $6090\pm300$  &  $17\pm4$        \\   
AT~2014ej        &  $300\pm20$        &  $138.6\pm15.5$  &  $690\pm10$  &   $8200\pm320$  &  $19\pm2$        \\   
NGC4490-2011OT1  &  $282\pm207$       &  $157.9\pm51.2$  &  $480\pm25$  &  $11830\pm3310$ &   $9.5\pm5.1$    \\   
AT~1997bs        &   $63\pm5$         &   $34.1\pm3.4$   &  $585\pm40$  &   $7500\pm640$  &  $11.5\pm1.5$    \\   
AT~2021blu       &   $65\pm9$         &   $47.2\pm1.6$   &  $500\pm100$ &   $8730\pm140$  &   $8.4\pm0.6$    \\   
AT~2021biy       &  $159\pm130$       &  $101.8\pm8.1$   &  $<500$      &  $11430\pm1410$ &   $7.7\pm3.1$    \\   
AT~2020hat       &    $7.8\pm0.7$     &  $0.41\pm0.15$   &  $<640$      &   $4630\pm120$  &  $10.3\pm0.6$    \\   
AT~2019zhd       &    $2.09\pm0.17$   &  $0.073\pm0.014$ &  $130\pm30$  &   $7030\pm210$  &   $1.08\pm0.07$  \\   
M31LRN2015       &    $3.81\pm0.90$   &  $0.21\pm0.05$   &  $<640$      &   $7940\pm1560$ &   $2.45\pm0.75$  \\   
V838~Mon         &    $3.15\pm0.30$   &  $0.081\pm0.008$ &  $230\pm20$  &   $7920\pm170$  &   $2.24\pm0.21$  \\   
V1309~Sco        &    $0.29\pm0.08$   & $0.0046\pm0.0012$&  $150\pm15$  &   $5970\pm260$  &   $1.00\pm0.29$  \\   
\hline
\end{tabular}
\tablefoot{The table reports the LRN name (Column 1); the bolometric luminosity at the first peak (Column 2); the H$\alpha$ luminosity (Column 3) and the FWHM velocity of H$_\alpha$ (Column 4) at phase $\sim+7$~d; the effective temperature (Column 5) and the photospheric radius (Column 6) at the first maximum.}
\end{table*}
%

The three objects discussed in this paper follow the general evolutionary framework of the best-studied events in the Milky Way and M~31. For this reason, we 
believe that most (or even all) of them are the outcome of merging events \citep[but see][]{gor20}. But the heterogeneity observed in the light-curve 
shape, luminosity, and duration suggests a wide range of the physical parameters involved.  In particular, the early-time sharp blue peak observed in 
\object{AT~2021blu} and its higher temperature at maximum brightness suggest a smaller photospheric radius than that of \object{AT~2021afy}. The interpretation of the 
observational properties of \object{AT~2018bwo} is more tricky, as the object was likely discovered a long time after maximum brightness. However, even if the object was 
older at discovery than assumed by \citet{bla21}, the very low photosperic temperature implies an initially larger photospheric radius than that of \object{AT~2021blu}.

While other considerations indicate that the progenitors of \object{AT~2021afy} and \object{AT~2021blu} were both massive systems (see Sect. \ref{sect:corr}), 
the above estimates suggest very different configurations for the two LRNe. A very expanded primary star was likely the progenitor of \object{AT~2021afy}, while 
a proportionally smaller amount of material was launched by this event. In contrast, the \object{AT~2021blu} precursor was characterised by a very compact 
initial configuration and more-massive ejecta. The parameters of  \object{AT~2018bwo} stay in the middle, although the progenitor system was likely less 
massive than the other two LRNe. The enormous difference between the inferred parameters for these three objects can be explained by a different fate of the 
system: while the large ejected mass of \object{AT~2021blu} can only result from the coalescence of massive stars,
two different scenarios can be invoked to explain the low ejected mass of \object{AT~2021afy}: the massive primary merged with a very low-mass companion, or the system survived as a binary 
system. However, as remarked by \citet{mat22}, the ejected mass and the radius strongly depend on the adopted velocity, and the presence of an extra heating 
source (such as shock interaction with circumbinary material) may severely affect the above estimates.

%
   \begin{figure*}
   \centering
   {\includegraphics[angle=270,width=15.6cm]{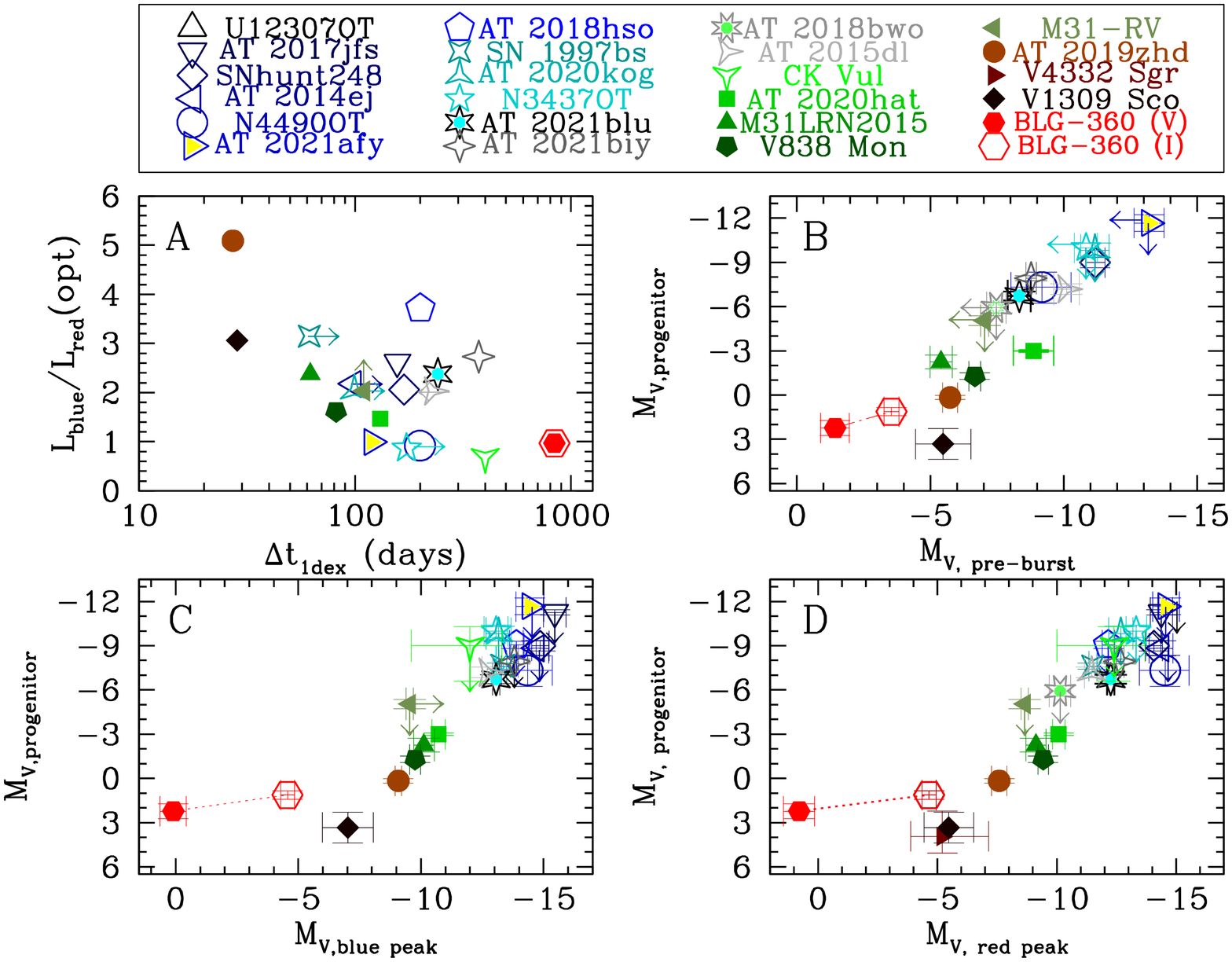}
\includegraphics[angle=270,width=15.6cm]{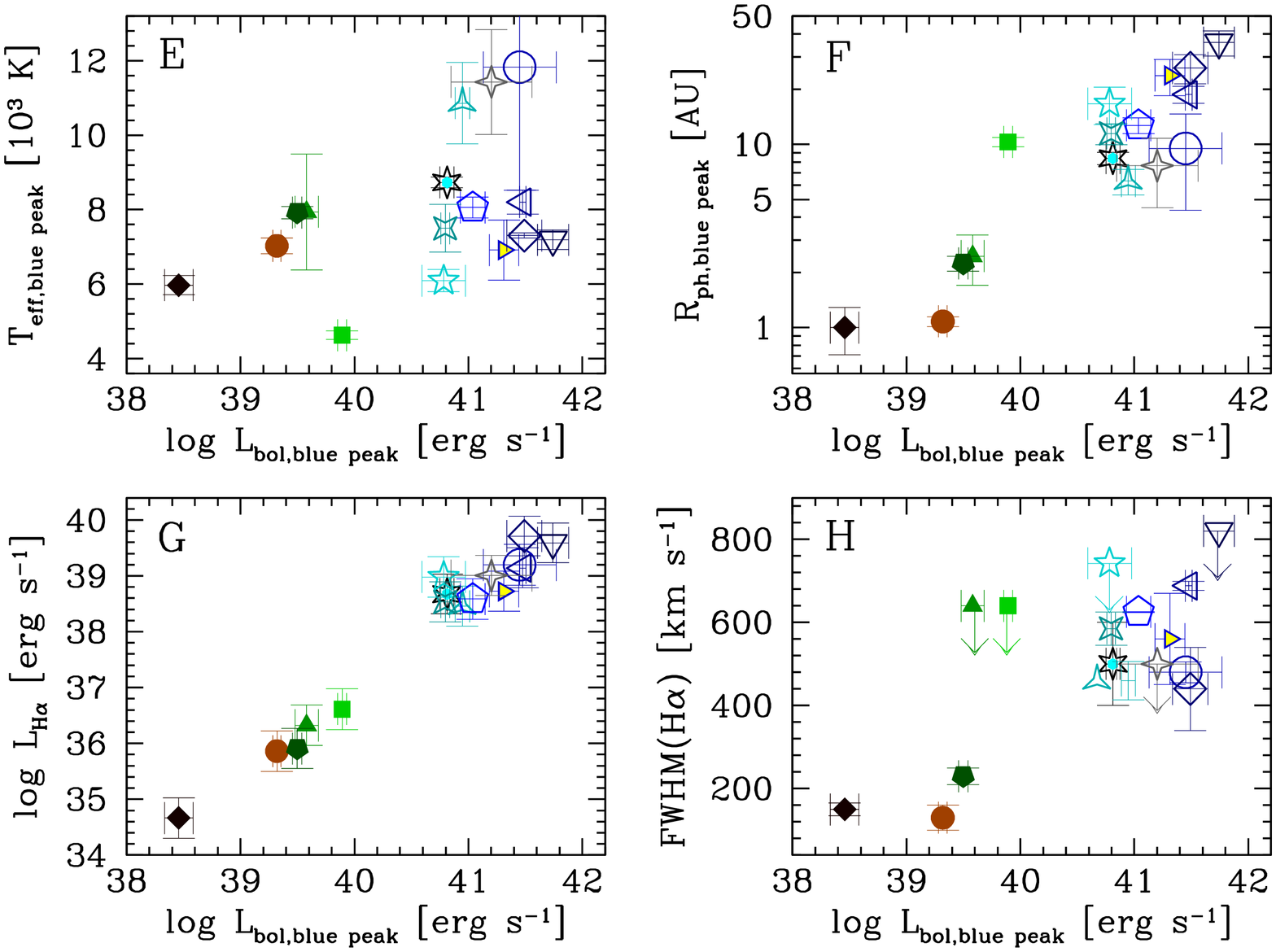}}
      \caption{Correlations among observational parameters for a sample of LRNe, as detailed in the text. The values of the physical parameters 
shown in the top panels are listed in Table \ref{tab:param_LRNe}, while those in the bottom panels are in Table \ref{tab:param_LRNe2}. 
}
         \label{Fig:correlations}
   \end{figure*}
%
   
\subsection{Correlations among physical parameters} \label{sect:corr}

With the inclusion of data presented in this paper \citep[plus AT~2021biy, studied in detail by][]{cai22}, we update with four new objects the diagrams showing possible correlations among the photometric parameters 
of LRNe presented by \citet{pasto21b}. The results are shown in Fig.~\ref{Fig:correlations} (top panels), while the parameters adopted for all objects are reported in Table \ref{tab:param_LRNe}. The new objects confirm the 
correlations between the absolute magnitude of the quiescent progenitor system with the absolute magnitude at the end of the slowly rising pre-outburst phase (panel B), the blue peak (panel C), and the broad 
red peak (or the plateau; panel D), with more-luminous LRN outbursts being produced by more-luminous (and, consequently, more-massive) stellar systems, as pointed out by \citet{koc17} and \citet{bla21}.

To quantify the strength of the correlations, we carried out a Pearson test\footnote{The parameters of \object{OGLE-2002-BLG-360}, a very peculiar object, were excluded in running the Pearson test.}, obtaining the following 
$p$-values: $1.3 \times 10^{-5}$, $1.1 \times 10^{-6}$, and $6.9 \times 10^{-8}$ for panels B, C, and D (respectively). 
We also note a weak correlation ($p$-value = 0.02) between the luminosity ratio of the two LRN maxima, and the time during which the luminosity stays between $L_{\rm peak}$ and $0.1\,L_{\rm peak}$ (panel A). As noticed by \citet{bla21}, 
dimmer objects seem to have a shorter duration, although \object{OGLE-2002-BLG-360} \citep{tyl13} appears to be an outlier, as it does not follow the general observational LRN trends. However, this object was observed 
mostly in the Johnson-Cousins $I$ band, had a limited colour information, and showed a very peculiar, triple-peaked light curve which is challenging to interpret. We also remark that for \object{CK~Vul}, quite limited 
photometric information is available\footnote{The object erupted in June 1670 and brightened again in April 1671. Only uncertain visual observations are documented from historical records, collected by 
\protect\citet{sha85}.}; hence, the inferred quantities should be regarded as simply indicative.

We inspect other possible correlations of physical parameters computed at the time of the early blue peak (Fig. \ref{Fig:correlations}, bottom panels; the values for individual objects are reported in 
Table \ref{tab:param_LRNe2}). In this analysis, we do not consider LRNe discovered after the early peak, such as AT~2018bwo and AT~2014ej, or whose photometric information is not accurate enough for a reliable 
estimate of $T_{\rm eff}$ and $R_{\rm ph}$ at that phase. In Fig.~\ref{Fig:correlations}, we report the bolometric luminosity at the first peak (obtained through a blackbody fit to the SED) versus $T_{\rm eff}$ (panel E) 
and $R_{\rm ph}$  (panel F) computed at the same phase. Again, there is a general trend, with dimmer LRNe having lower temperatures and smaller radii at the photosphere. Following the same approach as above, we performed 
a Pearson test to verify the robustness of the correlations in panels E and F, and obtained $p$-values of 0.02 and $2.6 \times 10^{-5}$, respectively. 

Finally, we inspect possible correlations of the bolometric luminosity at the blue peak with the
H$\alpha$ luminosity $L_{+7~{\rm days}}$(H$\alpha$) (panel G) and the velocity of the
expanding material $v_{\rm FWHM}$(H$\alpha$)\footnote{This value was computed for all objects through a Lorenzian fit to the line profile.}
(panel H) inferred from spectra obtained $\sim 1$~week after the first maximum. We made this choice 
because only a very limited number of LRN spectra are available at maximum brightness. Panel G shows a clear trend between $L_{\rm bol,peak}$ and $L_{+7~{\rm days}}$(H$\alpha$) ($p$-value = $9.2 \times 10^{-5}$), with dimmer LRNe having fainter H$\alpha$ luminosity. 
On the other hand, when the FWHM velocity is considered (panel H), the Pearson test does not reveal a significant correlation ($p$-value = 0.48), although  a correlation between peak luminosity and outflow velocity
was predicted for LRNe by \citet[][see their Fig.~21]{pej16b}.

The above correlations resemble those proposed by \citet{bla21} with slightly different parameters, but giving very similar outcomes: globally, the most-luminous LRNe are longer-duration events, often showing an 
 early light-curve peak. Hence, they are expected to have initially a hotter, more expanded, and higher-velocity photosphere, producing more-luminous H$\alpha$ spectral lines. \citet{bla21} suggested that 
the presence or the absence of an early blue peak in the light curve is due to different ionisation states of the gas shell where the photosphere is located. This shell would be initially fully ionised 
in 'hot'  events, and only marginally ionised in 'cold' LRNe. Hot LRNe are usually high-luminosity events produced by more-massive progenitor systems. \citet{bla21} proposed that energetic outflows 
generated by such massive binaries can efficiently ionise the circumstellar shell generated during a previous mass-loss event. However, while the spectral appearance during the first photometric peak may support 
this explanation, it cannot be comfortably applied to all objects, including the faint-but-hot \object{AT~2019zhd} \citep[Fig. \ref{Fig:TLR}, and][]{pasto21a}.

While \citet{sana12} predicted that mergers are a common endpoint of the evolution of massive stars in binary systems, precise rate estimates are still not available. We may expect that the LRN rates in 
different luminosity bins depend on the systemic mass. Using ZTF data, a volumetric rate of  $7.8^{+6.5}_{-3.7} \times 10^{-5}$~Mpc$^{-1}$~yr$^{-1}$ has been recently computed by \citet{kar22} in the absolute magnitude range $-16 \leq M_r \leq -11$~mag, hence considering intrinsically luminous events only. This is consistent with the larger
volumetric rate of $8 \times 10^{-4}$~Mpc$^{-1}$~yr$^{-1}$ estimated in the local Universe by \citet[][including intrinsically faint LRNe]{how20}, with 1--2 events per decade being 
expected in the Galaxy. This estimate is consistent with the discovery of a handful of LRNe in the Galaxy over the past 30~yr. The rates of LRNe are broadly dominated by the dimmest events, as discussed by 
\citet{koc14}, implying that mergers of low-mass binaries are more common by 2--3 orders of magnitude than those of massive binaries. We also note that a number of 
low-mass Galactic contact binaries have been proposed to be in the pathway to become mergers \citep[see, e.g.][]{wad20}.

\section{Conclusions} \label{Sect:conclusions}

We presented photometric and spectroscopic datasets for three new objects (\object{AT~2018bwo}, \object{AT~2021afy}, and \object{AT~2021blu}) that belong to the high-luminosity population of LRNe. 
All of them are most likely the outcome of merging events involving massive stars. However, they exhibit different properties, with  \object{AT~2021blu} being initially hotter and having a smaller photospheric
radius than \object{AT~2021afy} (and probably also \object{AT~2018bwo}, despite its epoch of outburst onset being poorly constrained). In addition, the duration of the \object{AT~2021blu} 
outburst is twice as long as that of the other two objects, suggesting a larger outflowing mass. 

Comparisons among observed parameters suggest that the three objects discussed here belong to the bright LRN population, with \object{AT~2021afy} being one of the most luminous events discovered to date.
Making use of the correlation between the absolute magnitude of the outburst and the progenitor's mass presented by \citet{cai22}, we estimate that the progenitor of  \object{AT~2021afy} has a mass likely 
exceeding 40~M$_\odot$ (although with a very large uncertainty). The binary progenitor system of \object{AT~2021blu} is characterised by a primary star of 13--18~M$_\odot$, slightly more massive than that of 
\object{AT~2018bwo} (11--16~M$_\odot$) reported by \citet{bla21}.

Our study supports previous evidence \citep[e.g.][]{pasto19a} that LRNe span a very wide range of physical properties, and that most observational parameters are somewhat correlated. In particular, the peak 
luminosity of LRN light curves appears to be correlated with the outburst duration, the H$\alpha$ luminosity, the photospheric radius, the effective temperature, and (most importantly)
the luminosity and mass of the progenitor stellar systems, as advocated by \citet{koc17} and \citet{bla21}. 

To increase our ability to characterise LRN variety, we need to expand the sample of events with excellent spectral and photometric coverage, and with available information regarding the quiescent progenitors. 
This will enable us to fine-tune the above correlations, making them a valuable tool for estimating the parameters of LRNe when only incomplete datasets are available, as well as for inferring the luminosity and 
mass of LRN binary progenitors without the need for a  direct detection of the progenitor flux from archival pre-outburst images obtained with high-spatial-resolution facilities.

\begin{acknowledgements}

We thank Jorge Anais Vilchez, Abdo Campillay, Nahir Mu\~noz-Elgueta, Natalie Ulloa, and Jaime Vargas-Gonz\'alez for performing the observations on the Swope Telescope at Las Campanas Observatory, Chile; Takashi Nagao for his help with the NOT observations; 
and WeiKang Zheng for his help with Keck observations. We also thank Jun Mo  for his help with the TNT data reduction.  We acknowledge the support of the staffs of the various observatories where data were obtained.\\

MF is supported by a Royal Society -- Science Foundation Ireland University Research Fellowship. 
AR acknowledges support from ANID BECAS/DOCTORADO NACIONAL 21202412.
GP and  AR acknowledge support from the Chilean Ministry of Economy, Development, and Tourism’s Millennium Science Initiative through grant IC12009, awarded to the Millennium Institute of Astrophysics.
EC, NER, and LT acknowledge support from MIUR, PRIN 2017 (grant 20179ZF5KS) ``\textit{The new frontier of the Multi-Messenger Astrophysics: follow-up of electromagnetic transient counterparts of gravitational wave sources}.''
NER also acknowledges partial support from the Spanish MICINN grant PID2019-108709GB-I00 and FEDER funds, and from the program Unidad de Excelencia María de Maeztu CEX2020-001058-M.
TMR acknowledges the financial support of the Jenny and Antti Wihuri and the Vilho, Yrj{\"o} and Kalle V{\"a}is{\"a}l{\"a} Foundations.
Research by SV is supported by NSF grants AST-1813176 and AST-2008108.
Time-domain research by the University of Arizona team and DJS is supported by NSF grants AST-1821987, 1813466, 1908972, $\&$ 2108032, and by the Heising-Simons Foundation under grant \#2020-1864.
KAB acknowledges support from the DIRAC Institute in the Department of Astronomy at the University of Washington. The DIRAC Institute is supported through generous gifts from the Charles and Lisa Simonyi Fund for Arts and Sciences, and the Washington Research Foundation.
The LCO team is supported by NSF grants AST-1911225 and AST-1911151.
JB is supported by NSF grants AST-1911151 and AST-1911225, as well as by
National Aeronautics and Space Administration (NASA) grant 80NSSC19kf1639.
YZC is funded by China Postdoctoral Science Foundation (grant 2021M691821).
RD acknowledges funds by ANID grant FONDECYT Postdoctorado No. 3220449.
LG acknowledges financial support from the Spanish Ministerio de Ciencia e Innovaci\'on (MCIN), the Agencia Estatal de Investigaci\'on (AEI) 10.13039/501100011033, and the European Social Fund (ESF) ``Investing in your future'' under the 2019 Ram\'on y Cajal program RYC2019-027683-I and the PID2020-115253GA-I00 HOSTFLOWS project, from Centro Superior de Investigaciones Cient\'ificas (CSIC) under the PIE project 20215AT016, and the program Unidad de Excelencia Mar\'ia de Maeztu CEX2020-001058-M.
RK acknowledges support from the Academy of Finland (340613).
KM acknowledges BRICS grant DST/IMRCD/BRICS/Pilotcall/ProFCheap/2017(G) for this work.
AMG acknowledges financial support from the 2014–2020 ERDF Operational Programme and by the Department of Economy, Knowledge, Business and University of the Regional Government of Andalusia through the FEDER-UCA18-107404 grant.
MDS is supported by grants from the VILLUM FONDEN (grant 28021) and the Independent Research Fund Denmark (IRFD; 8021-00170B).
AVF's group at UC Berkeley has received support from the Miller Institute for Basic Research in Science (where AVF was a Miller Senior Fellow), the Christopher R. Redlich Fund, and numerous individual donors.
LW is sponsored (in part) by the Chinese Academy of Sciences (CAS), through a grant to the CAS South America Center for Astronomy (CASSACA) in Santiago, Chile.
This work was supported in part by NASA Keck PI Data Award 2021A-N147 (PI: Jha), administered by the NASA Exoplanet Science Institute. 
This work is supported by National Natural Science Foundation of China (NSFC grants 12033003, 11633002), the Scholar Program of Beijing Academy of Science and Technology (DZ:BS202002), and the Tencent Xplorer Prize.
This work is partially supported by China Manned Space Project (CMS-CSST-2021-A12).\\

This work is based on observations made with the Nordic Optical Telescope (NOT), owned in collaboration by the University of Turku and Aarhus University, and operated jointly by Aarhus University, 
the University of Turku and the University of Oslo, representing Denmark, Finland and Norway, the University of Iceland and Stockholm University at the Observatorio del Roque de los Muchachos, 
La Palma, Spain, of the Instituto de Astrofisica de Canarias; the 10.4\,m Gran Telescopio Canarias (GTC), installed in the Spanish Observatorio del Roque de los 
Muchachos of the Instituto de Astrof\'isica de Canarias, in the Island of La Palma;  the 2.0\,m Liverpool Telescope operated on the island of La Palma by Liverpool John Moores University 
at the Spanish Observatorio del Roque de los Muchachos of the Instituto de Astrof\'isica de Canarias with financial support from the UK Science and Technology Facilities Council; 
the 3.58\,m Italian Telescopio Nazionale Galileo (TNG) operated the island of La Palma by the Fundaci\'on Galileo Galilei of the Istituto Nazionale di Astrofisica (INAF) at the Spanish 
Observatorio del Roque de los Muchachos of the Instituto de Astrof\'isica de Canarias; the 4.1\,m Southern African Large Telescope (SALT) of the South African Astronomical Observatory, Southerland, 
South Africa; the 6.5\,m Magellan-Baade Telescope located at the Las Campanas Observatoy, Chile; the Southern Astrophysical Research Telescope and the Panchromatic Robotic Optical Monitoring and 
Polarimetry Telescopes (PROMPT) of the Cerro Tololo Inter-American Observatory (CTIO), on Cerro Pach\'on, Chile; the 3.05\,m Shane telescope of the Lick Observatory, University of California 
Observatories, USA; the 1.82\,m Copernico and the 67/92\,cm Schmidt telescopes of INAF --- Osservatorio Astronomico di Padova, Asiago, Italy; the 3.6\,m Devasthal Optical Telescope, 
the 1.3\,m Devasthal Fast Optical Telescope (DFOT), and the 1.04\,m Sampurnanand Telescope (ST) of the Aryabatta Research Institute of Observational Sciences (ARIES), Manora Peak, Naintal, Uttarakhand, India; 
the 0.8\,m Tsinghua-NAOC Telescope at Xinglong Observatory (China); the 0.6\,m Rapid Eye Mount (REM) INAF telescope, hosted at ESO La Silla Observatory, Chile, under program ID 43308;
and the 2.3\,m Bok Telescope, operated by Stewart Observatory, Kitt Peak National Observatory, Arizona, USA.
This work makes also use of observations from the Las Cumbres Observatory global telescope network (including the Faulkes North Telescope, and the 2.54\,m Isaac Newton telescope operated by the Isaac Newton Group of Telescopes (ING),
Roque de los Muchachos, La Palma, Spain. 
The NIRES data presented herein were obtained at the W. M. Keck Observatory from telescope time allocated to NASA through the agency's scientific partnership with the California 
Institute of Technology and the University of California. The Observatory was made possible by the generous financial support of the W. M. Keck Foundation.
A major upgrade of the Kast spectrograph on the Shane 3\,m telescope at Lick Observatory was made possible through generous gifts from William and Marina Kast as well as the Heising-Simons Foundation. Research at Lick Observatory is partially supported by a generous gift from Google.
The paper is also based on observations obtained at the international Gemini Observatory, a program of NSF's NOIRLab, which is managed by AURA, Inc., 
under a cooperative agreement with the NSF on behalf of the Gemini Observatory partnership: the National Science Foundation (United States), National Research Council (Canada), 
Agencia Nacional de Investigaci\'on y Desarrollo (Chile), Ministerio de Ciencia, Tecnolog\'ia e Innovaci\'on (Argentina), Minist\'erio da Ciência, Tecnologia, Inova\c{c}\~oes e Comunica\c{c}\~oes (Brazil), 
and Korea Astronomy and Space Science Institute (Republic of Korea).
The data presented herein were obtained in part with ALFOSC, which is provided by the Instituto de Astrofisica de Andalucia (IAA) under a joint agreement with the University of Copenhagen and NOT.\\

This study is also based on observations made with the NASA/ESA {\it Hubble Space Telescope}, obtained from the Data Archive at the Space Telescope Science Institute (STScI), which is operated by AURA, Inc., under NASA contract NAS 5-26555. This work made use of data from the All-Sky Automated Survey for Supernovae (ASAS-SN), obtained through
the Sky Patrol interface (\url{https://asas-sn.osu.edu/}).
This research is based on observations made with the mission, obtained from the MAST data archive at STScI, which is operated by AURA, Inc., under NASA contract NAS 5–26555. These observations are associated with programs with IDs 8645, 15922, and 16691.
This work has made use of data from the Asteroid Terrestrial-impact Last Alert System (ATLAS) project. ATLAS is primarily funded to search for near-Earth objects through NASA 
grants NN12AR55G, 80NSSC18K0284, and 80NSSC18K1575; byproducts of the NEO search include images and catalogs from the survey area. The ATLAS science products have been made 
possible through the contributions of the University of Hawaii Institute for Astronomy, the Queen's University Belfast, STScI, and the South
 African Astronomical Observatory, and The Millennium Institute of Astrophysics (MAS), Chile.\\

The Pan-STARRS1 Surveys (PS1) and the PS1 public science archive have been made possible through contributions by the Institute for Astronomy, the University of Hawaii, the 
Pan-STARRS Project Office, the Max-Planck Society and its participating institutes, the Max Planck Institute for Astronomy, Heidelberg and the Max Planck Institute for 
Extraterrestrial Physics, Garching, The Johns Hopkins University, Durham University, the University of Edinburgh, the Queen's University Belfast, the Harvard-Smithsonian Center for 
Astrophysics, the Las Cumbres Observatory Global Telescope Network Incorporated, the National Central University of Taiwan, STScI, NASA
under grant NNX08AR22G issued through the Planetary Science Division of the NASA Science Mission Directorate, NSF
grant AST-1238877, the University of Maryland, Eotvos Lorand University (ELTE), the Los Alamos National Laboratory, and the Gordon and Betty Moore Foundation.\\

This publication is partially based on observations obtained with the Samuel Oschin 48-inch Telescope at the Palomar Observatory as part of the Zwicky Transient Facility (ZTF) project. 
ZTF is supported by the NSF under grant AST-1440341 and a collaboration including Caltech, IPAC, the Weizmann Institute for Science, the Oskar Klein 
Center at Stockholm University, the University of Maryland, the University of Washington, Deutsches Elektronen-Synchrotron and Humboldt University, Los Alamos National Laboratories, 
the TANGO Consortium of Taiwan, the University of Wisconsin at Milwaukee, and Lawrence Berkeley National Laboratories. Operations are conducted by COO, IPAC, and UW.\\

We acknowledge ESA Gaia, DPAC, and the Photometric Science Alerts Team (\url{http://gsaweb.ast.cam.ac.uk/alerts}).
We acknowledge the use of public data from the {\it Swift} data archive.
This research made use of the WISeREP database (\url{https://wiserep.weizmann.ac.il}).
This research has made use of the NASA/IPAC Extragalactic Database (NED) which is operated by the Jet Propulsion Laboratory, California Institute of Technology, under contract 
with NASA.
This publication used data products from the Two Micron All-Sky Survey, which is a joint project of the University of Massachusetts and the Infrared Processing and Analysis Center/California Institute of Technology, funded by NASA and the NSF. \\
 This work used the Binary Population and Spectral Synthesis (BPASS)
models as last described by Eldridge, Stanway, et al. (2017) and by Stanway, Eldridge, et al. (2018).\\

The authors wish to recognise and acknowledge the very significant cultural role and reverence that the summit of Maunakea has always had within the indigenous Hawaiian community.  
We are most fortunate to have the opportunity to conduct observations from this mountain. 

\end{acknowledgements}

\begin{appendix}
  
\section{Instruments used in the photometric campaigns} \label{Appendix:A}

\object{AT~2018bwo} was monitored in the optical bands using the 0.41~m Prompt~3 (with Sloan $g$ and $r$ filters) and the 0.41~m Prompt~6
(with Johnson $V$ and $R$ filters), both hosted by the Cerro Tololo Inter-American Observatory (CTIO, Chile). Additional unfiltered photometry
was obtained with the 0.41~m Prompt~5 (CTIO) telescope and the 0.41~m Prompt-MO-1 telescope (Meckering Observatory, south-western Australia), 
both operating in the framework of the DLT40 survey. Other data were obtained with the 1~m Swope Telescope of the Las Campanas Observatory, and 
several 1~m telescopes equipped with Sinistro cameras and $UBVgri$ filters, operating within the Las Cumbres Observatory global telescope network, 
in the framework of the Global Supernova Project. In particular, the telescopes used for the \object{AT~2018bwo} campaign are hosted at the Siding Spring 
Observatory (SSO; Australia), the South African Astronomical Observatory (SAAO, South Africa), and CTIO. Single-epoch observations were also obtained
with the 3.56~m New Technology Telescope (NTT) equipped with EFOSC2, hosted at ESO-La Silla Observatory (Chile), and the 10.4~m GTC with OSIRIS, 
hosted at La Palma (Canary Islands, Spain).
A small number of photometric data were provided by the {\it Gaia} survey in the {\it Gaia} $G$ band, and by ZTF. Four epochs taken by ZTF with the Sloan $g$ and $r$
filters\footnote{The images were retrieved from \url{https://www.ztf.caltech.edu/}; \protect\cite{mas19}.} during the late-time light-curve 
decline were measured by us using the template-subtraction technique. Finally, photometry in the orange ($o$) and cyan ($c$) bands 
are provided by the two 0.5~m ATLAS telescopes (ATLAS~1 is on Haleakal\"a and ATLAS~2 on Maunaloa, Hawaii, USA).

\object{AT~2021afy} was followed in $g$ and $r$ by ZTF. Although ZTF publicly provides forced photometry, we re-analysed the
ZTF images obtained through the Public Data Release 3
without applying a template subtraction. We made this choice because the object was located in a remote position of the host-galaxy
centre, and the ZTF templates were not optimal (in particular, in the $g$ band).
 We complemented the ZTF data with multi-band observations obtained with
the following instruments: the 2~m Liverpool Telescope (LT) equipped with IO:O; the 2.56~m Nordic Optical Telescope (NOT) equipped with 
the Alhambra Faint Object Spectrograph and Camera (ALFOSC) and the NOT near-infrared Camera and spectrograph (NOTCam); 
the 10.4~m GTC with OSIRIS; 
the 1.82~m Copernico Telescope with the Asiago Faint Object Spectrograph and Camera (AFOSC) and  the 0.67/0.92~m Schmidt Telescope with a Moravian 
camera of the Padova Observatory (Istituto Nazionale di Astrofisica, INAF, hosted at Mt. Ekar, near Asiago, Italy); and the 0.6~m Rapid Eye Mount (REM) 
telescope with the ROSS2 and REMIR cameras, hosted by the European Southern Observatory (ESO) in La Silla (Chile).

\object{AT~2021blu} was extensively observed by the following public surveys:  the All-Sky Automated Survey for Supernovae 
\citep[ASAS-SN;][in the $g$ band]{sha14,koc17}\footnote{ASAS-SN photometry is publicly released through the Sky Patrol ASAS-SN interface 
(\url{https://asas-sn.osu.edu}).} which works through a network of small (0.14~m) telescopes in different world-wide sites; 
the 1.22~m Samuel Oschin Telescope at the Palomar Observatory (California, USA) serving the ZTF survey\footnote{In this case, we used public
ZTF forced photometry, which is released through the Lasair 
(\url{https://lasair.roe.ac.uk/}) and ALeRCE (\url{https://alerce.online/}) brokers, and already shown by \protect\citet{sora22}.}; 
the two 0.5~m ATLAS telescopes,
and the two 1.8~m Pan-STARRS\footnote{The acronym stands for Panoramic Survey Telescope $\&$ Rapid Response System.} \citep[PS1 and PS2;][]{cha19,fle20,mag20} 
telescopes at Haleakal\"a (Hawaii, USA).
Multi-Band data were also obtained with the same facilities used for the \object{AT~2021afy} campaign 
(except REM, owing to the northern declination of \object{AT~2021blu}),
plus 0.4~m to 1~m telescopes of the Las Cumbres Observatory global telescope network equipped with SBIG STL6303 and Sinistro cameras, respectively, 
and hosted at the McDonald Observatory (Texas, USA) and the Teide Observatory (Tenerife, Canary Islands, Spain).
Other telescopes used in the monitoring campaign of \object{AT~2021blu} are
the 0.8~m Tsinghua-NAOC Telescope \citep[TNT;][]{wang08,hua12} with a PIXIS back-illuminated 1300B CCD camera
at Xinglong Observatory (China); the 3.6~m Devasthal Optical Telescope (DOT) equipped with ADFOSC, the 1.3~m Devasthal Fast Optical 
Telescope (DFOT) and the 1.04~m Sampurnanand Telescope (ST) operated by Aryabhatta Research Institute of Observational Sciences 
(ARIES; India) with optical imagers.
Data in the optical (with $u$, $b$, and $v$ filters) and UV (in the $uvw2$, $uvm2$, and $uvw1$ bands) domains were obtained by the 
{\it Neil Gehrels Swift Observatory} spacecraft \citep{geh04} equipped with UVOT \citep{rom05}.

\subsection{Photometry tables}  \label{Appendix:A.1}

Tables A1, A2, and A3 are available in electronic form at the CDS, and contain the following information: 
the epoch and the MJD of the observation (Columns 1 and 2, respectively); the filter (Column 3); the magnitude and the error (Columns 4 and 5, respectively); the instrumental configuration (Column 6); additional notes (Column 7).

\begin{table*}
\setcounter{table}{3}
\caption{\label{Table:A1.4} Information on the PS1 and ZTF stacked images obtained in the decade before the outburst of \object{AT~2021afy}, and detection limits.}
\centering
\begin{tabular}{cccccccc}
\hline\hline
Initial date & Initial MJD & Final date & Final MJD & Average MJD & Filter & magnitude & CCD code\\
\hline 
2011-05-31 & 55712.31 & 2012-05-16 & 56063.55 & 55888.13 & $g$ & $>$23.05 & PS1 \\
2010-07-01 & 55378.42 & 2013-07-03 & 56476.40 & 56069.75 & $r$ & $>$23.20 & PS1 \\
2011-05-19 & 55700.48 & 2014-07-11 & 56849.38 & 56396.19 & $i$ & $>$23.40 & PS1 \\
2010-03-01 & 55256.57 & 2014-09-17 & 56917.23 & 55928.69 & $z$ & $>$22.81 & PS1 \\
2018-03-26 & 58203.42 & 2018-09-08 & 58369.23 & 59286.82 & $g$ & $>$20.95 & ZTF-c15 \\
2018-03-09 & 58186.48 & 2018-05-20 & 58258.25 & 59222.37 & $r$ & $>$22.05 & ZTF-c15 \\
2018-04-20 & 58228.51 & 2018-09-06 & 58367.14 & 59297.82 & $i$ & $>$21.33 & ZTF-c15 \\
2020-12-11 & 59194.55 & 2020-12-13 & 59196.57 & 59195.64 & $r$ & $>$21.11 & ZTF-c15 \\
2020-12-15 & 59198.56 & 2020-12-18 & 59201.57 & 59199.56 & $r$ & $>$21.11 & ZTF-c15 \\
\hline
\end{tabular}
\tablefoot{The table reports the epoch and MJD of the first image (Columns 1 and 2), the epoch and MJD of the last image (Columns 3 and 4), the average MJD of the stacked image (Column 5), 
the filter (Column 6), the detection magnitude limits (Column 7), and the identificatione code of the stacked images (Column 8).}
\end{table*}

\begin{table*}
\setcounter{table}{4}
\caption{\label{Table:A1.5} Information on the ZTF stacked images and photometry of the pre-outburst source at the location of \object{AT~2021blu}.}
\centering
\begin{tabular}{cccccccc}
\hline\hline
Initial date & Initial MJD & Final date & Final MJD & Average MJD & Filter & magnitude & CCD code\\
\hline 
2018-04-25 &  58233.22 &  2018-06-23 &  58292.20 &  58260.72 & $g$ &  $>$21.98     &  ZTF-c12 \\ 
2018-03-25 &  58202.26 &  2018-06-24 &  58293.18 &  58248.47 & $g$ &  $>$21.85     &  ZTF-c09 \\            
2018-10-31 &  58422.50 &  2019-06-28 &  58662.21 &  58520.01 & $g$ &  22.20 (0.54) &  ZTF-c09 \\
2019-10-09 &  58765.51 &  2019-12-29 &  58846.52 &  58802.75 & $g$ &  $>$22.17     &  ZTF-c12 \\
2019-10-02 &  58758.51 &  2019-12-29 &  58846.51 &  58790.41 & $g$ &  $>$21.74     &  ZTF-c09 \\
2020-01-04 &  58852.52 &  2020-01-29 &  58877.38 &  58865.24 & $g$ &  22.10 (0.46) &  ZTF-c12 \\ 
2020-01-01 &  58849.48 &  2020-01-31 &  58879.27 &  58869.97 & $g$ &  22.04 (0.36) &  ZTF-c09 \\
2020-02-01 &  58880.42 &  2020-02-27 &  58906.29 &  58891.04 & $g$ &  $>$22.09     &  ZTF-c09 \\
2020-02-01 &  58880.42 &  2020-02-27 &  58906.31 &  58895.65 & $g$ &  $>$22.00     &  ZTF-c12 \\
2020-03-01 &  58909.28 &  2020-03-28 &  58936.27 &  58916.00 & $g$ &  $>$21.55     &  ZTF-c12 \\
2020-03-04 &  58912.26 &  2020-03-31 &  58939.31 &  58925.76 & $g$ &  $>$21.60     &  ZTF-c09 \\
2020-04-15 &  58954.17 &  2020-04-29 &  58968.24 &  58962.46 & $g$ &  22.04 (0.42) &  ZTF-c09 \\
2020-04-15 &  58954.28 &  2020-04-29 &  58968.27 &  58963.53 & $g$ &  22.07 (0.33) &  ZTF-c12 \\
2020-05-02 &  58971.25 &  2020-05-29 &  58998.18 &  58983.74 & $g$ &  $>$21.90     &  ZTF-c12 \\
2020-05-03 &  58972.23 &  2020-06-05 &  59005.17 &  58987.19 & $g$ &  $>$21.84     &  ZTF-c09 \\
2020-06-01 &  59001.17 &  2020-06-27 &  59027.20 &  59016.53 & $g$ &  $>$21.69     &  ZTF-c12 \\
2020-10-18 &  59140.51 &  2020-11-05 &  59158.48 &  59149.21 & $g$ &  $>$21.46     &  ZTF-c09 \\ 
2020-10-18 &  59140.51 &  2020-11-06 &  59159.52 &  59150.39 & $g$ &  $>$21.74     &  ZTF-c12 \\
2020-11-12 &  59165.53 &  2020-11-29 &  59182.56 &  59174.27 & $g$ &  21.83 (0.49) &  ZTF-c09 \\ 
2020-11-12 &  59165.52 &  2020-12-01 &  59184.45 &  59175.00 & $g$ &  21.83 (0.50) &  ZTF-c12 \\
2020-12-02 &  59185.51 &  2020-12-22 &  59205.43 &  59196.32 & $g$ &  21.62 (0.66) &  ZTF-c09 \\ 
2020-12-05 &  59188.47 &  2020-12-22 &  59205.43 &  59198.29 & $g$ &  21.72 (0.46) &  ZTF-c12 \\
2021-01-01 &  59215.49 &  2021-01-17 &  59231.42 &  59222.69 & $g$ &  21.64 (0.48) &  ZTF-c12 \\
2021-01-08 &  59222.39 &  2021-01-18 &  59232.43 &  59227.44 & $g$ &  21.60 (0.32) &  ZTF-c09 \\
2018-04-08 &  58216.22 &  2018-06-15 &  58284.18 &  58243.86 & $r$ &  21.62 (0.51) &  ZTF-c09 \\
2018-04-06 &  58214.20 &  2018-06-15 &  58284.17 &  53245.90 & $r$ &  21.74 (0.46) &  ZTF-c12 \\
2018-11-07 &  58429.52 &  2019-06-24 &  58658.18 &  58524.29 & $r$ &  $>$21.72     &  ZTF-c12 \\
2018-10-31 &  58422.53 &  2019-07-05 &  58669.18 &  58509.83 & $r$ &  21.75 (0.54) &  ZTF-c09 \\ 
2019-09-25 &  58751.53 &  2019-12-29 &  58846.46 &  58800.39 & $r$ &  $>$21.58     &  ZTF-c09 \\
2019-10-20 &  58776.51 &  2019-12-29 &  58846.46 &  58812.89 & $r$ &  $>$21.61     &  ZTF-c12 \\
2020-01-01 &  58849.45 &  2020-01-29 &  58877.34 &  58861.95 & $r$ &  $>$21.40     &  ZTF-c12 \\
2020-01-01 &  58849.44 &  2020-01-31 &  58879.34 &  58867.73 & $r$ &  21.74 (0.36) &  ZTF-c12 \\
2020-02-01 &  58880.32 &  2020-02-27 &  58906.23 &  58890.86 & $r$ &  21.75 (0.50) &  ZTF-c09 \\
2020-02-01 &  58880.34 &  2020-02-20 &  58899.34 &  58891.63 & $r$ &  21.71 (0.47) &  ZTF-c12 \\
2020-03-01 &  58909.32 &  2020-03-31 &  58939.21 &  58916.04 & $r$ &  $>$21.68     &  ZTF-c12 \\  
2020-04-15 &  58954.24 &  2020-04-29 &  58968.19 &  58961.56 & $r$ &  21.82 (0.33) &  ZTF-c12 \\
2020-04-15 &  58954.22 &  2020-04-29 &  58968.17 &  58962.46 & $r$ &  21.83 (0.33) &  ZTF-c09 \\
2020-05-01 &  58970.21 &  2020-05-29 &  58998.25 &  58981.72 & $r$ &  21.91 (0.42) &  ZTF-c12 \\
2020-05-03 &  58972.17 &  2020-06-27 &  59027.18 &  58990.65 & $r$ &  21.93 (0.52) &  ZTF-c09 \\
2020-06-04 &  59004.25 &  2020-06-27 &  59027.17 &  59017.06 & $r$ &  $>$21.34     &  ZTF-c12 \\
2020-10-14 &  59136.52 &  2020-11-05 &  59158.53 &  59147.65 & $r$ &  21.74 (0.53) &  ZTF-c09 \\
2020-10-18 &  59140.53 &  2020-11-06 &  59159.49 &  59149.61 & $r$ &  21.61 (0.47) &  ZTF-c12 \\   
2020-11-12 &  59165.49 &  2020-11-28 &  59181.49 &  59172.69 & $r$ &  21.48 (0.39) &  ZTF-c12 \\
2020-11-12 &  59165.49 &  2020-12-02 &  59185.47 &  59174.79 & $r$ &  21.26 (0.31) &  ZTF-c09 \\ 
2020-12-01 &  59184.51 &  2020-12-17 &  59200.52 &  59193.48 & $r$ &  21.30 (0.38) &  ZTF-c12 \\
2020-12-10 &  59193.52 &  2020-12-27 &  59210.43 &  59200.32 & $r$ &  21.29 (0.29) &  ZTF-c09 \\
2021-01-04 &  59218.45 &  2021-01-04 &  59218.45 &  59218.45 & $r$ &  21.32 (0.30) &  ZTF-c09 \\
2021-01-01 &  59215.44 &  2021-01-05 &  59219.36 &  59219.42 & $r$ &  21.32 (0.42) &  ZTF-c12 \\  
2021-01-01 &  59215.44 &  2021-01-17 &  59231.47 &  59222.68 & $r$ &  21.29 (0.37) &  ZTF-c12 \\
2021-01-07 &  59221.40 &  2021-01-11 &  59225.41 &  59223.43 & $r$ &  21.31 (0.30) &  ZTF-c12 \\  
2021-01-08 &  59222.44 &  2021-01-12 &  59226.49 &  59224.47 & $r$ &  21.34 (0.26) &  ZTF-c09 \\
2021-01-13 &  59227.40 &  2021-01-17 &  59231.47 &  59229.43 & $r$ &  21.36 (0.34) &  ZTF-c12 \\
2021-01-14 &  59228.47 &  2021-01-18 &  59232.42 &  59230.45 & $r$ &  21.35 (0.28) &  ZTF-c09 \\
2018-04-24 &  58232.30 &  2018-05-28 &  58266.22 &  58251.60 & $i$ &  $>$20.98     &  ZTF-c12 \\
2018-04-24 &  58232.30 &  2018-05-28 &  58266.22 &  58251.97 & $i$ &  $>$21.06     &  ZTF-c09 \\
\hline
\end{tabular}
\tablefoot{The table reports the epoch and MJD of the first image (Columns 1 and 2), the epoch and MJD of the last image (Columns 3 and 4), the average MJD of the stacked image (Column 5), 
the filter (Column 6), the magnitude (Column 7), and the CCD chip identification code of the ZTF images (Column 8).}
\end{table*}

\section{Instruments used in the spectroscopic campaigns} \label{Appendix:B}

The spectra of \object{AT~2018bwo}, which cover four months of the LRN evolution, were taken with the $11.1\times9.8$~m Southern African Large Telescope (SALT) with the Robert Stobie Spectrograph 
(RSS; hosted near Sutherland, South Africa); the 8.1~m Gemini South Telescope equipped with FLAMINGOS2 and  the 4.1~m Southern Astrophysical Research (SOAR) Telescope plus 
the Goodman spectrograph (both located on Cerro Pach\'on, Chile); the 6.5~m Magellan-Baade Telescope with the Folded-port InfraRed Echelette (FIRE)\footnote{FIRE data were reduced 
following the prescriptions detailed by \protect\citet{hsi19}.} spectrometer at the Las Campanas Observatory (Chile); and the 10~m Keck-I Telescope with the Low Resolution Imaging Spectrograph \citep[LRIS;][]{oke95} 
on Maunakea (Hawaii, USA). GTC, equipped with OSIRIS, was used for a late-time spectrum of \object{AT~2018bwo}, and all spectra of \object{AT~2018bwo}.

 The following instruments were used in the 
spectroscopic campaign of \object{AT~2021blu}: the 2.0~m Faulkes North Telescope (FNT) with FLOYDS of the Las Cumbres Observatory node on Haleakal\"a (Hawaii, USA); the 3.05~m Shane telescope 
equipped with the Kast spectrograph (hosted at Lick Observatory, near San Jose, California, USA); Keck-I plus LRIS and Keck-II with NIRES;
 the 1.82~m Copernico Telescope plus AFOSC; the DOT plus ADFOSC; the GTC with OSIRIS; the NOT with ALFOSC; and the 3.58~m Telescopio Nazionale Galileo (TNG) 
with the Device Optimized for the LOw RESolution (DOLORES, or LRS).

\end{appendix}

\end{document}